%% file: EOI-NuSOnG.tex
\pdfoutput=1 
\documentclass[12pt]{article}
\usepackage{graphicx}
\usepackage{amssymb}
\usepackage{epstopdf}
\usepackage[dvips]{epsfig} 
\def\Journal#1#2#3#4{{#1} {\bf #2}, #3 (#4)}

\def\NPA{{Nucl. Phys.} A}

\def\PLB{{Phys. Lett.}  B}
\def\PRL{Phys. Rev. Lett.}

\def\PRD{{Phys. Rev.} D}

\title{Expression of Interest for\\ Neutrinos Scattering on Glass:
  NuSOnG}

\begin{document}
\date{September 13, 2007}
\maketitle
\thispagestyle{empty}
\begin{centering}
{ \large
T. Adams$^4$, L. Bugel$^2$, J.M. Conrad$^2$, P.H. Fisher$^6$,
  J.A. Formaggio$^6$,\\ A. de Gouv\^ea$^9$, W.A. Loinaz$^1$,
  G. Karagiorgi$^2$, T.R. Kobilarcik$^3$, S. Kopp$^{13}$,\\ G. Kyle$^8$, D.A. Mason$^3$,  R. Milner$^6$, 
  J.~G.~Morf\'{\i}n$^3$,
   M. Nakamura$^7$,\\ D. Naples$^{10}$, 
  P. Nienaber$^{11}$, F.I Olness$^{12}$, J.F. Owens$^4$, 
  W.G. Seligman$^2$, \\
  M.H. Shaevitz$^2$,  H. Schellman$^9$,
  M.J. Syphers$^3$, C.Y. Tan$^3$, \\ R.G. Van de Water$^5$, 
R.K. Yamamoto$^6$, G.P. Zeller$^5$\\

~~~\\

}

$^1$\/Amherst College, Amherst, MA 01002 \\
$^2$\/Columbia University, New York, NY 10027  \\
$^3$\/Fermi National Accelerator Laboratory, Batavia IL 60510 \\
$^4$\/Florida State University, Tallahassee, FL 32306  \\
$^5$\/Los Alamos National Accelerator Laboratory, Los Alamos, NM 87545  \\
$^6$\/Massachusetts Institute of Technology, Cambridge, MA 02139  \\
$^7$\/Nagoya University, 464-01, Nagoya, Japan \\
$^8$\/New Mexico State University, Las Cruces, NM 88003 \\
$^9$\/Northwestern University, Evanston, IL 60208  \\
$^{10}$\/University of Pittsburgh, Pittsburgh, PA 15260  \\
$^{11}$\/Saint Mary's University of Minnesota, Winona, MN 55987\\
$^{12}$\/Southern Methodist University, Dallas, TX 75205 \\
$^{13}$\/University of Texas, Austin TX 78712\\
\end{centering}

\begin{figure}[h]
\centering
\includegraphics[width=4.in]{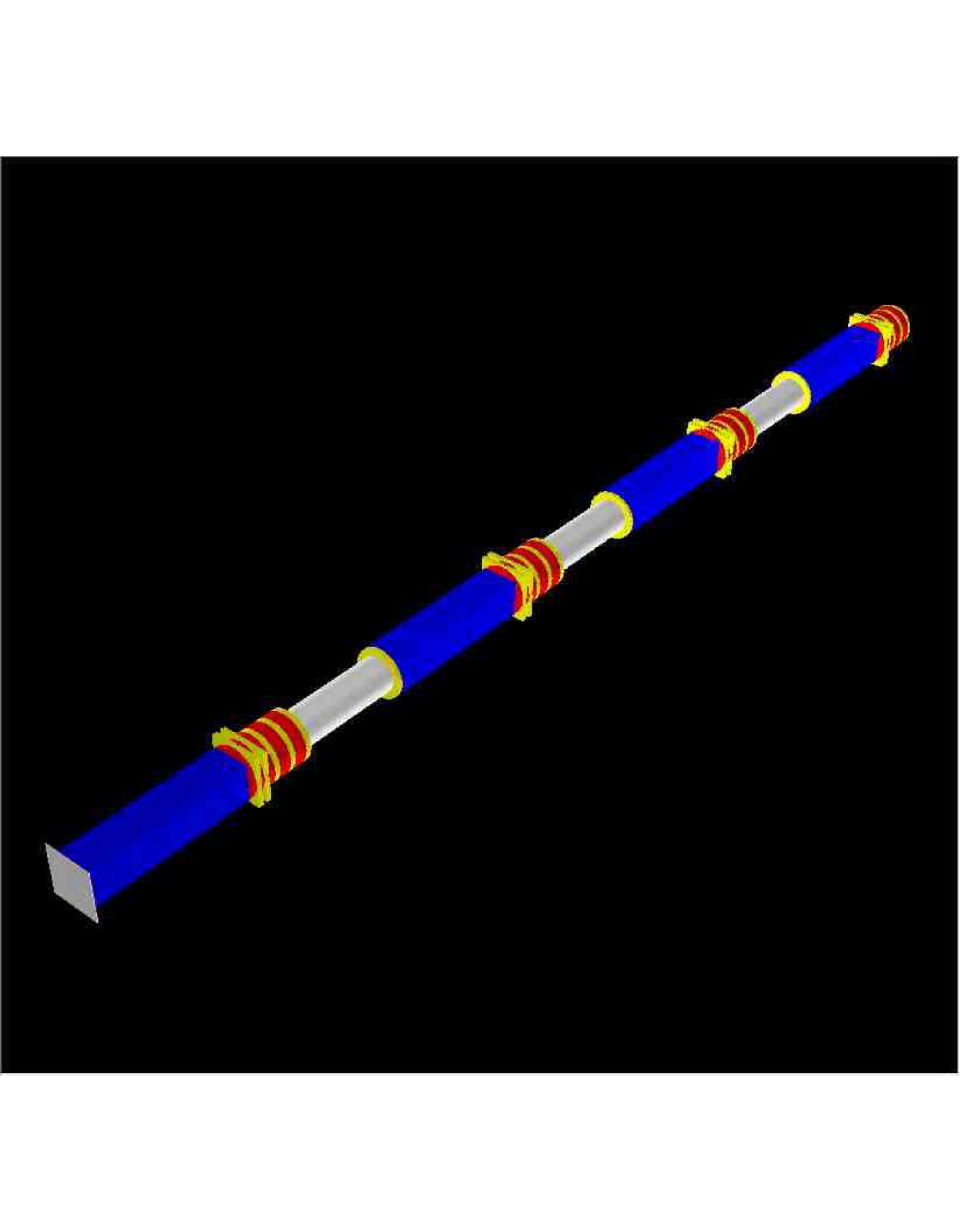}
\end{figure}

\newpage

\abstract{ We propose a 3500 ton (3000 ton fiducial volume) 
  $\mathrm{SiO}_2$ neutrino detector with sampling calorimetry,
  charged particle tracking, and muon spectrometers to run in a
  Tevatron Fixed Target Program.  Improvements to the Fermilab
  accelerator complex should allow substantial increases in the
  neutrino flux over the previous NuTeV quad triplet beamline.  With 4
  $\times$ 10$^{19}$ protons on target/year, a 5 year run would
  achieve event statistics more than 100 times higher than NuTeV. With 100 times
  the statistics of previous high energy neutrino experiments, the
  purely weak processes $\nu_{\mu}+e^- \rightarrow \nu_{\mu}+ e^-$ and
  $\nu_{\mu}+ e^- \rightarrow \nu_e + \mu^-$ (inverse muon decay) can
  be measured with high accuracy for the first time.  The inverse muon
  decay process is independent of strong interaction effects and can
  be used to significantly improve the flux normalization for all
  other processes.  The high neutrino and antineutrino fluxes also
  make new searches for lepton flavor violation and neutral heavy
  leptons possible.  In this document, we give a first look at the
  physics opportunities, detector and beam design, and calibration
  procedures.  }

\newpage

\tableofcontents

\newpage

\section{Introduction}
\input{intro_v2A.tex}

\newpage

\section{Physics Opportunities}
\label{se:physics}

The physics opportunities of the experiment arise from NuSOnG's uniquely high 
statistics:  $>$20k neutrino-electron scatters and 
$>$100M neutrino-quark scatters.
Roughly equal statistics will be obtained from antineutrino scattering.
More information on the event rates for various processes is 
given in Sec.~\ref{se:rates}.
These rates present a wide range of physics opportunities including
precision electroweak measurements, direct searches for new physics, 
and parton distribution studies.

\subsection{Electroweak Precision Measurements}
\label{EWsection}

\input{EWsection_v13}

\subsection{Direct Searches for New Physics}

\input{DirectSearches_v4A}

\subsection{Measurement of Parton Distribution Functions}

\input{DIS_PDF_NucA}

\newpage

\section{Neutrino Flux and Event Rates}
\label{se:rates}

\input{FluxContent_v1A}

\input{EventContent_v3A}

\input{precisionflux_v3A}

\newpage

\section{Preliminary Design}
\label{se:design}

This report focuses upon the determination of the physics goals
of the experiment.  In order to maintain realistic goals, we
have developed a preliminary design for a beam and detector based on
existing technology.  There are two particularly challenging 
aspects of the design.  The first is the high Tevatron
intensity discussed in sec.~\ref{protdeliv}.  
The second is the high precision required for the detector calibration
discussed in sec.~\ref{detcalib}. 

The 2007 Fermilab Steering Group Report considers the Tevatron-based 
neutrino beam described here.    The preliminary concept
for the facility recieved an endorsement \cite{FSG}.

\input{ProtondeliveryMJSA}

\label{protdeliv}

\subsection{Neutrino Beam Design}
\label{se:beam}

\input{SsqtA}

\subsection{Detector Design}
\label{se:detector}

\input{NusOnGDetector_v2A}

\input{mag_v2A}

\input{cal_v2A}

\input{where_is_itA}

\newpage

\newpage

\section{Summary}
\label{se:summary}

\input{summaryA}

\input{bib_v1A}
 \end{document}

%% file: intro_v2A.tex
The Neutrino Scattering on Glass (NuSOnG) experiment will consist of four
detector modules, each composed of a finely segmented calorimeter
followed by a muon spectrometer. The detector will be illuminated by a
neutrino or antineutrino beam from the Tevatron.  In its five-year data
acquisition period, NuSOnG will make precise measurements of three types of 
neutrino scattering and will accumulate the world's largest sample of 
electron-neutrino scatters.  These data will provide unique opportunities to 
discover physics beyond the Standard Model (including, {\it inter alia}, lepton flavor 
violation and new particles) as well as determine structure functions over a
wide range of $x$ and $Q^2$.  The breadth of anticipated
measurements makes NuSOnG a program rather than an experiment;
the design heritage ensures that the approach is low-risk and cost-effective.  

This Expression of Interest arises from our view that an experiment
probing the high energy interactions of neutrinos is a necessary
complement to the LHC and an important lead-in to the ILC.  
In the next few years, the LHC will reveal the nature of
electroweak symmetry breaking; the Higgs mass will cease 
being a prediction of the electroweak theory and will become 
an input to the theory.  Without the Higgs mass as a fit parameter, precision 
electroweak data, including neutrino scattering data,
will be much more powerful as a tool for constraining
that physics beyond the Standard Model which directly influences 
the electroweak sector.  More important still, precision
neutrino scattering will probe areas of phenomenology that may be 
inaccessible to the LHC and ILC.  
NuSOnG is not a precision test of the Standard
Model; NuSOnG is a discovery experiment aimed at the terrain not covered
by the collider experiments.

This Expression of Interest presents the physics case and initial
design for NuSOnG.  The detector draws on the heritage of FMMF, CDHS,
CHARM and CCFR/NuTeV.  
The design uses an SiO$_2$ target in 
one-quarter radiation length panels interleaved
with active detector elements (proportional tubes and/or
scintillator).  This will provide the very high segmentation needed to ensure good
separation between different classes of events.  We will develop
these ideas in the coming months and submit a proposal to the Fermilab
Directorate.

Our report is organized as follows: the physics opportunities follow
in Section~\ref{se:physics}; Section~\ref{se:rates} describes the flux
and expected event rates; and Section~\ref{se:design} describes our
preliminary design for the NuSOnG beam and apparatus.  We summarize in
Section~\ref{se:summary}.

%% file: EWsection_v13.tex
NuSOnG's considerable discovery potential derives from its ability to
do precision electroweak tests through two independent channels:
electron scattering and quark scattering. These measurements probe
for new particles and new neutrino properties beyond the present
Standard Model. As examples, NuSOnG will be sensitive to extra $Z$
bosons with masses beyond the 1 TeV scale (depending on the model), and to
compositeness scales above 5 TeV.  Thus the energy scales explored by
this experiment overlap the LHC, and we present the discovery
potential for the new physics we will explore within this context.
This experiment also directly addresses questions raised by the ``NuTeV
anomaly,'' an electroweak precision measurement in
disagreement with the Standard Model.

\subsubsection{Electroweak Measurements in Neutrino Scattering}
\label{numeasuresubsec}

NuSOnG is sensitive to new physics through neutral current (NC)
scattering. 
The exchange of the $Z$ boson between the neutrino $\nu$ and fermion $f$
leads to the effective interaction:
\begin{eqnarray}
\mathcal{L}
& = & -\sqrt{2}G_F
\Bigl[\, \bar{\nu}\gamma_\mu\bigl(g_V^\nu - g_A^\nu \gamma_5\bigr)\nu \,\Bigr]
\Bigl[\, \bar{f}\gamma^\mu\bigl(g_V^f - g_A^f \gamma_5\bigr)f \,\Bigr] \cr
& = & -\sqrt{2}G_F
\Bigl[\, g_L^\nu\,\bar{\nu}\gamma_\mu(1-\gamma_5)\nu
     + g_R^\nu\,\bar{\nu}\gamma_\mu(1+\gamma_5)\nu \,\Bigr] \cr
& & \qquad\qquad
\times
\Bigl[\, g_L^f \,\bar{f}\gamma^\mu(1-\gamma_5)f
     + g_R^f \,\bar{f}\gamma^\mu(1+\gamma_5)f \,\Bigr] \;, \cr
& & 
\end{eqnarray}
where the Standard Model values of the couplings are:
\begin{eqnarray}
g_L^\nu & = & \sqrt{\rho}\left(+\frac{1}{2}\right) \;,\cr
g_R^\nu & = & 0\;, \cr
g_L^f & = & \sqrt{\rho}\left(I_3^f - Q^f\sin^2\theta_W \right) \;,\cr
g_R^f & = & \sqrt{\rho}\left(-Q^f\sin^2\theta_W\right) \;,
\end{eqnarray}
or equivalently,
\begin{eqnarray}
g_V^\nu \;=\; g_L^\nu + g_R^\nu & = & \sqrt{\rho}\left(+\frac{1}{2}\right)\;,\cr
g_A^\nu \;=\; g_L^\nu - g_R^\nu & = & \sqrt{\rho}\left(+\frac{1}{2}\right)\;,\cr
g_V^f   \;=\; g_L^f + g_R^f & = & \sqrt{\rho}\left(I_3^f - 2Q^f\sin^2\theta_W\right) \;,\cr
g_A^f   \;=\; g_L^f - g_R^f & = & \sqrt{\rho}\left(I_3^f\right) \;.
\end{eqnarray}
Here, $I_3^f$ and $Q^f$ are the weak isospin and electromagnetic
charge of fermion $f$, respectively.  In these formulae, $\rho$ is the
relative coupling strength of the neutral to charged current
interactions ($\rho=1$ at tree level in the Standard Model). The weak
mixing parameter, $\sin^2 \theta_W$, is related (at tree level) to to
$G_F$, $M_Z$ and $\alpha$ by
\begin{equation}
\sin^2 2 \theta_W=\frac{4 \pi \alpha  }{ \sqrt{2} G_F M_Z^2} .
\end{equation}

NuSOnG is unique in its ability to test the NC couplings by studying
scattering of neutrinos from both electrons and quarks. A deviation
from the Standard Model predictions in both the electron and quark
measurements will present a compelling case for new physics.

\vspace{0.1in}
{\bf Neutrino Electron Scattering}
\label{nuesubsub}
\vspace{0.1in}

The differential cross section for muon neutrino and antineutrino
scattering from electrons, defined using the coupling constants
described above, is:
\begin{eqnarray}
d\sigma & = &
\frac{2G_F^2 m_e E_\nu}{\pi}
\Biggl[ 
(g_L^\nu g_V^e \pm g_L^\nu g_A^e)^2 \,\frac{dT}{E_\nu} 
\Biggr.\cr
& & \qquad\qquad + 
(g_L^\nu g_V^e \mp g_L^\nu g_A^e)^2 
\left(1-\frac{T}{E_\nu}\right)^2\frac{dT}{E_\nu} \cr
& & \qquad\qquad - 
\Biggl. 
\Bigl\{ (g_L^\nu g_V^e)^2   - (g_L^\nu g_A^e  )^2 \Bigr\}
\frac{m_e T}{E_\nu^2}\frac{dT}{E_\nu}
\Biggr] \;.
\end{eqnarray}
The upper and lower signs corresponding to the neutrino and
anti-neutrino cases, respectively.  In this equation, $E_{\nu}$ is the
incident ${\nu}_{\mu}$ energy and $T$ is the electron recoil kinetic
energy.

More often in the literature, the cross section is defined in terms of
the parameters $(g_{V}^{\nu e},g_{A}^{\nu e})$, which are defined as
\begin{eqnarray}
g_V^{\nu e} 
& \equiv & (2g_L^\nu g_V^e)
\;=\; \rho\left(-\frac{1}{2}+2\sin^2\theta_W\right) \;,\cr
g_A^{\nu e}
& \equiv & (2g_L^\nu g_A^e)
\;=\; \rho\left(-\frac{1}{2}\right) \;,
\end{eqnarray}
In terms of these parameters, we can write:
\begin{eqnarray}
d\sigma & = &
\frac{G_F^2 m_e E_\nu}{2\pi}
\Biggl[ 
(g_V^{\nu e} \pm g_A^{\nu e})^2 \,\frac{dT}{E_\nu} 
\Biggr. + 
(g_V^{\nu e} \mp g_A^{\nu e})^2 
\left(1-\frac{T}{E_\nu}\right)^2\frac{dT}{E_\nu} \cr
& & \qquad\qquad - 
\Biggl. 
\Bigl\{ (g_V^{\nu e})^2   - (g_A^{\nu e}  )^2 \Bigr\}
\frac{m_e T}{E_\nu^2}\frac{dT}{E_\nu}
\Biggr] \;,
\end{eqnarray}
When $m_e \ll E_\nu$, the third terms in these expressions can be neglected.
If we introduce the variable $y=T/E_\nu$, then
\begin{eqnarray}
\frac{d\sigma}{dy}
& = & \frac{G_{F}^{2}m_e E_\nu}{2\pi}
\left[ 
  \left( g_{V}^{\nu e} + g_{A}^{\nu e} \right)^{2}
+ \left( g_{V}^{\nu e} - g_{A}^{\nu e} \right)^{2}
  \left( 1 - y \right)^{2} 
\right]\;.
\end{eqnarray}
Integrating over the region $0\le y\le 1$, we obtain the total cross sections
which are
\begin{eqnarray}
\sigma & = & 
\frac{G_{F}^{2}m_e E_\nu}{2\pi}
\left[ 
  \left( g_{V}^{\nu e} \pm g_{A}^{\nu e} \right)^{2}
+ \frac{1}{3}\left( g_{V}^{\nu e} \mp g_{A}^{\nu e} \right)^{2}
\right]\;.
\label{sigma_enu}
\end{eqnarray}
Note that
\begin{eqnarray}
\left( g_{V}^{\nu e} + g_{A}^{\nu e} \right)^{2}
& =&  \rho^2\left(-1+4\sin^2\theta_W\right)^2  
\;=\; \rho^2\left(1-2\sin^2\theta_W+4\sin^4\theta_W\right) \;,\cr
\left( g_{V}^{\nu e} - g_{A}^{\nu e} \right)^{2}
& = & \rho^2\left(2\sin^2\theta_W\right)^2
\;=\; \rho^2\left(4\sin^4\theta_W\right) \;.
\end{eqnarray}
Therefore,
\begin{eqnarray}
\sigma(\nu_{\mu}\, e) & = & \frac{G_F^2 m_e E_\nu}{2\pi}
\,\rho^2
\Biggl[ 1 - 4\sin^2\theta_W + \frac{16}{3}\sin^4\theta_W \Biggr] \;,\cr
\sigma({\bar \nu_{\mu}}\, e) & = & \frac{G_F^2 m_e E_\nu}{2\pi}
\,\frac{\rho^2}{3}
\Biggl[ 1 - 4\sin^2\theta_W + 16\sin^4\theta_W \Biggr] \;.
\end{eqnarray}

\begin{figure}
\vspace{-5cm}
\centering
\scalebox{0.8}{\includegraphics{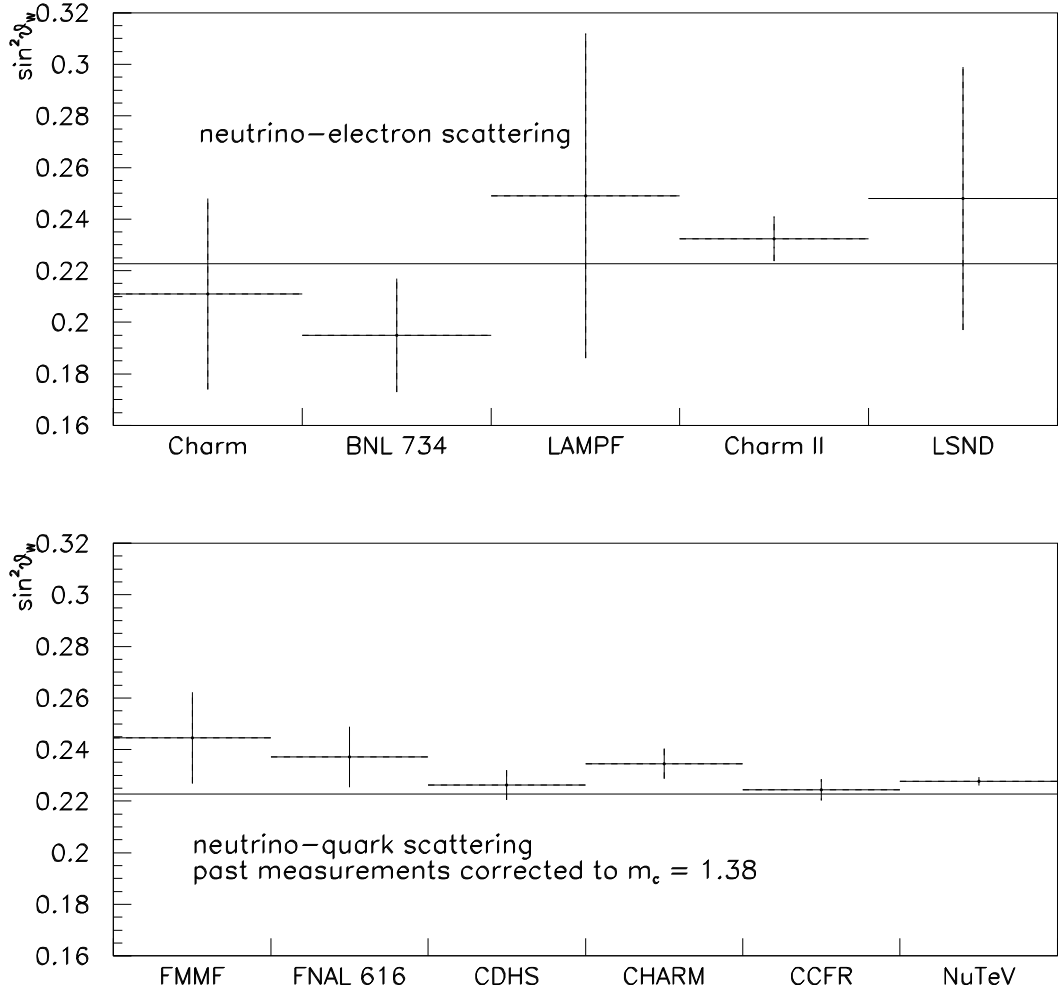}}
\vspace{-4cm}
\caption{\label{pastnu} Measurements of $\sin^2 \theta_W$ from past
  experiments.  Top: neutrino-electron elastic scattering experiments.
  Bottom: neutrino DIS experiments.  All DIS results are adjusted to
  the same charm mass (relevant for experiments not using P-W method).
  The Standard Model value, indicated by the line, is $0.2227$.}
\end{figure}

The ratio of the integrated cross sections for neutrino to antineutrino
electron scattering is
\begin{equation}
R_{e}
\;=\; \frac{\sigma(\nu_{\mu L}\, e)}{\sigma^({\bar \nu_{\mu L}} e)}
\;=\; 3\;
\frac{1-4\sin^2\theta_W+{{16}\over{3}}\sin^4\theta_W}
      {1-4\sin^2\theta_W+16\sin^4\theta_W} \;.
\end{equation}
Many systematics, including flux errors, cancel in this ratio, as does the $\rho$ dependence.  
Fig. \ref{pastnu}(top) shows the results for $\sin^2 \theta_W$ 
from many past experiments.

NuSOnG will make independent measurements of the electroweak
parameters for both $\nu_\mu$ and $\bar \nu_\mu$-electron scattering.
We can achieve this via ratios or by direct extraction of the cross
section.    In the case of $\nu_\mu$-electron scattering, we
will use the ratio of the number of events in 
neutrino-electron elastic scattering to inverse muon decay:
\begin{equation} 
{{N(\nu_\mu e^- \rightarrow \nu_\mu e^-)}
\over{N(\nu_\mu e^- \rightarrow \mu^- \nu_e)}}
=
\frac{\sigma^{\nu e}_{NC} \times \Phi^\nu}{\sigma^{IMD} \times \Phi^{\nu}}.
\end{equation}
Because the cross section for IMD events is well determined by the
standard model, this ratio should have low errors and will isolate the
EW parameters from NC scattering.  In the case of $\bar \nu_\mu$
running, the ratio is more complex because there is no equivalent
process to inverse muon decay (since there are no positrons in the
detector).  In this case, we use the fact that, for low exchange
energy in Deep Inelastic Scattering, the cross sections in neutrino
and antineutrino scattering approach the same constant, $A$, as is
explained in Sec.~\ref{subsubfixednu}.  Thus, for Deep Inelastic
events with low energy transfer and hence low hadronic energy ($5
\lesssim E_{had} \lesssim 10$ GeV), $N^{low~E_{had}}_{\nu DIS} =
\Phi^\nu A$ and $N^{low~E_had}_{\bar\nu DIS} = \Phi^{\bar \nu} A$.
The result is that we can extract the electroweak parameters to high
precision using the ratio:
\begin{equation}
\frac {{N^{low~E_{had}}_{\nu DIS}}} {N^{low~E_{had}}_{\bar\nu DIS}}  \times
\frac {N(\bar \nu_\mu e^- \rightarrow \bar \nu_\mu e^-)}
{N(\nu_\mu e^- \rightarrow \mu^- \nu_e)}
=
\frac {\Phi^\nu} {\Phi^{\bar \nu}} \times
\frac {\sigma^{\bar \nu e}_{NC} \times \Phi^{\bar \nu}}
{\sigma^{IMD} \times \Phi^\nu}.
\end{equation}
The first ratio cancels the DIS cross section, leaving the
energy-integrated $\nu$ to $\bar \nu$ flux ratio.  The IMD events in
the denomenator of the second term cancel the integrated $\nu$ flux.
The NC elastic events cancel the integrated $\bar \nu$ flux.
Alternatively, because we will have accurate knowledge of the flux as
a function of the energy (see Sec.~\ref{subprecisionflux}) we could
directly measure the cross sections.

An important point is that the two
independent measurements, one in neutrino and the other in
antineutrino mode, will in turn allow independent extraction of
$g_A^{\nu e}$ and $g_V^{\nu e}$.
The previous best measurement from $\nu_\mu$ and $\bar \nu_\mu$
cross-section measurements is from CHARM II, 
which used 2677$\pm$82 events
in neutrino mode and 2752$\pm$88 events in antineutrino mode  \cite{CHARMIIsin2thw} to find
\begin{eqnarray}
g_V^{\nu e} & = &-0.035 \pm 0.012 {\rm (stat)} \pm 0.012 {\rm (sys)} \\
g_A^{\nu e} & = &-0.503 \pm 0.006 {\rm (stat)} \pm 0.016 {\rm (sys)}. 
\end{eqnarray}
This can be compared to electroweak measurements from LEP provide a 
very precise prediction of these parameters \cite{LEPgagv}:
\begin{eqnarray}
g_V^{\nu e} & = &-0.0397 \pm 0.0003 \\
g_A^{\nu e} & = &-0.5065 \pm 0.0001.
\end{eqnarray}
The CHARM II results are in agreement with LEP, but with large errors.
Errors on the neutrino measurement must be
substantially reduced in order to meaningfully probe for physics beyond the Standard Model.   
The goal of NuSOnG is to measure the neutrino-electron and
antineutrino-electron cross sections to $0.7\%$.

\paragraph{\it  Neutrino Quark Scattering}
\label{PWsection}

~~~\\
~~~

Substantially higher precision has been obtained using neutrino-quark
scattering, which compares neutral-current (NC) to charged-current (CC)
scattering to extract $\sin^2 \theta_W$.  However, these experiments are 
subject to issues of modeling in the quark sector.  
Fig.~\ref{pastnu}(bottom) reviews the history of these measurements.

The lowest systematic errors come from 
implementing a ``Paschos-Wolfenstein style'' \cite{PW} analysis, which 
would be the technique 
used by NuSOnG.  This requires separated 
$\nu$ and $\bar \nu$ beams, for which 
the following ratios could be formed:
\begin{eqnarray}
R^\nu &=& \frac{\sigma_{NC}^\nu}{\sigma_{CC}^\nu} \\
R^{\bar \nu} &=& \frac{\sigma_{NC}^{\bar \nu}}{\sigma_{CC}^{\bar \nu}}. \\
\end{eqnarray}
Paschos and Wolfenstein \cite{PW} recast these as:
\begin{equation}
R^- = \frac{\sigma_{NC}^\nu - \sigma_{NC}^{\bar \nu}}{\sigma_{CC}^\nu - \sigma_{CC}^{\bar \nu}} = \frac{R^\nu - r R^{\bar \nu}}{1-r},
\end{equation}
where $r=\sigma_{CC}^{\bar \nu}/\sigma_{CC}^{\nu}$.  In $R^-$ many systematics cancel to first order, including the effects of the quark and
antiquark seas for $u, d, s$, and $c$.  Charm production only enters
through $d_{valence}$ (which is Cabbibo suppressed) and at high $x$;
thus the error from the charm mass is greatly reduced.  
The cross section ratios can be written in terms of the effective 
neutrino-quark coupling parameters $g_L^2$ and $g_R^2$ as
\begin{eqnarray}
R^\nu &=& g_L^2+rg_R^2  \\
R^{\bar \nu} &=& g_L^2 + {1 \over r} g_R^2\\
R^- &=& g_L^2-g_R^2 = \rho^2 ({1 \over 2} - \sin^2\theta_W),
\end{eqnarray}
in which
\begin{eqnarray}
g_L^2 & = (2 g_L^\nu g_L^u)^2 + (2 g_L^\nu g_L^d)^2~= & \rho^2 ({1 \over 2} - \sin^2 \theta_W + {5 \over 9} \sin^4 \theta_W) \\
g_R^2 & = (2 g_L^\nu g_R^u)^2 + (2 g_L^\nu g_R^d)^2~= &\rho^2({5 \over 9} \sin^4 \theta_W).
\end{eqnarray}

NuTeV fit for $R^\nu$ and $R^{\bar \nu}$ simultaneously to
extract $\sin^2 \theta_W,$ obtaining the value $\sin^2\theta_W=0.2277\pm 0.00162$.   
The goal of NuSOnG is to improve on this error by 
a factor of two.  
Table~\ref{NuTeVerrs} lists the errors which NuTeV identified and
indicates those for which NuSOnG expects improvement.  Many of the
largest experimental systematics of NuTeV came from the method of
separating CC and NC events, which relied on length.  NuSOnG will have
a more sophisticated model for differentiating CC and NC events, using
shower shape and identification of Michel-electron followers from low energy pion decays. 

\begin{table}[tbp]
\begin{center}
  \begin{tabular}{|c|c|c|} \hline

 Source & Error & Reduction in
    NuSOnG \\ \hline \hline
 Statistics & 0.00135 & 100 times the
    statistics \\ \hline \hline $\nu_e$, $\bar \nu_e$ flux prediction
    & 0.00039 & see Sec.~\ref{eflux}\\ \hline
    Interaction vertex position & 0.00030 & Better detector segmentation and \\
    & & more sophisticated shower identification. \\ \hline
    Shower length model & 0.00027 & Better segmentation and \\
    & & more sophisticated shower identification. \\ \hline Counter
    efficiency and noise & 0.00023 & Better, Minos-style counter
    design \\ \hline Energy Measurement & 0.00018 & Likely to be at a
    similar level. \\ \hline \hline Charm production, strange sea &
    0.00047 & See Sec. \ref{subsec:strangesea} \\ \hline $R_L$ &
    0.00032 & Likely to be at a similar level. \\ \hline $\sigma^{\bar
      \nu}/\sigma^{\nu}$ & 0.00022 & See Sec. \ref{subsubxsec} \\
    \hline Higher Twist & 0.00014 & Likely to be at a similar
    level. \\ \hline Radiative Corrections & 0.00011 & Likely to be at
    a similar level. \\ \hline Charm Sea & 0.00010 & Under study \\
    \hline Non-isoscalar target & 0.00005 & Glass is isoscalar \\
    \hline
\end{tabular}

\caption{Source and value of NuTeV error on $\sin^2 \theta_W$, and
reason why the error will be reduced in the PW-style analysis of NuSOnG. }
\label{NuTeVerrs}
\end{center}
\end{table}

From Fig.~\ref{pastnu}, it is apparent that the NuTeV measurement
is in agreement with past neutrino scattering results, although
these have much larger errors.
However,
the NuTeV result is in disagreement with the
global fits to the electroweak data which give a Standard Model
value of $\sin^2\theta_W =0.2227$ \cite{NuTeVanomaly}.  
Expressed in terms of the couplings,
NuTeV measures:
\begin{eqnarray}
g_L^2 = 0.30005 \pm 0.00137 \\
g_R^2 = 0.03076 \pm 0.00110, 
\end{eqnarray}
which can be compared to the Standard Model values of $g_L^2=0.3042$ and 
$g_R^2=0.0301$, respectively.
Sec. \ref{nutevsection} (below)
considers possible sources for this disagreement, both within and
outside the Standard Model.

\subsubsection{NuSOnG and New Physics}

\label{NewPhyssub}

NuSOnG will provide important probes of physics beyond the Standard Model
distinct from and complementary to those of the LHC.  NuSOnG will seek 
indirect evidence for new physics by addressing
anomalies in the precision electroweak data, and by providing unique
information about neutrino coupling to the $Z$.  In addition, precision 
measurements from NuSOnG will help to disentangle the complicated set of 
observations that will be present at the LHC and, in doing so, elucidate the mechanism
of electroweak symmetry breaking.  NuSOnG and the LHC provide distinct probes of
new physics because new physics enters collider and 
neutrino scattering processes differently:  neutrino physics measures 
different combinations of couplings to light quarks; neutrino scattering probes
new physics at space-like momentum transfer (versus the time-like scattering
at colliders); and systematics are very different between low and high energy
experiments.  Finally, NuSOnG will directly search for new particles and interactions
in the lepton sector that might be missed by the LHC and must otherwise await 
discovery by the ILC.

\paragraph{\it New Physics Observed through Coupling to the $Z$}
\label{gllgrrsection}

~~~\\
~~~

NuSOnG is unique among experiments in its ability to address the
nature of the neutrino couplings to the $Z$ boson in the near 
future.  In the
Standard Model, the neutrino coupling to the $Z$- and $W$-bosons is
purely left-handed. Indeed, the fact that the neutrino coupling to the
$W$-boson and an electron is purely left-handed is, experimentally, a
well-established fact (evidence includes precision measurements of
pion and muon decay, nuclear processes, etc.). By contrast, the
nature of the neutrino coupling to the $Z$ boson is, experimentally,
far from being precisely established \cite{Carena:2003aj}.

The best measurement of the neutrino coupling to the $Z$-boson is
provided by indirect measurements of the invisible $Z$-boson width at
LEP. In units where the Standard Model neutrino--$Z$-boson couplings are
$g_L^{\nu}=0.5$, $g_R^{\nu}\equiv 0$, the LEP measurement \cite{invZ} 
translates
into $(g^{\nu}_L)^2+(g^{\nu}_R)^2=0.2487\pm 0.0010$. Note that
this result places no meaningful bound on $g_R^{\nu}$.

Precise, model-independent information on $g^{\nu}_L$ can be obtained by combining 
$\nu_{\mu}+e$ scattering data from CHARM II and LEP and SLD data. Assuming model-independent 
couplings of the fermions to the $Z$-boson, $\nu_{\mu} +e$ scattering measures $g_L^{\nu}=2\rho$, while LEP and SLD measure the 
left and right-handed couplings of the electron to the $Z$. The CHARM II result translates 
into $|g_L^{\nu}|=0.502\pm 0.017$ \cite{Carena:2003aj}, assuming that the charged-current weak 
interactions produce only left-handed neutrinos. In spite of the good precision of the CHARM II 
result (around 3.5\%), a combination of all available data allows $|g_R^{\nu}/g_L^{\nu}|\sim 0.4$ at 
the two $\sigma$ confidence level \cite{Carena:2003aj}. 

Significant improvement in our understanding of $g_R^{\nu}$ can only be obtained with more 
precise measurements of $\nu+e$ scattering, or with the advent of a new high 
intensity $e^+e^-$ collider, such as the ILC. By combining ILC running at the $Z$-boson pole mass 
and at $\sqrt{s}=170$~GeV, $|g_R^{\nu}/g_L^{\nu}|\lesssim 0.3$ could be constrained 
at the two $\sigma$ level after analyzing $e^+e^-\to\gamma+$missing energy events \cite{Carena:2003aj}. 

\begin{figure}
\centering
\scalebox{0.8}{\includegraphics[clip=true]{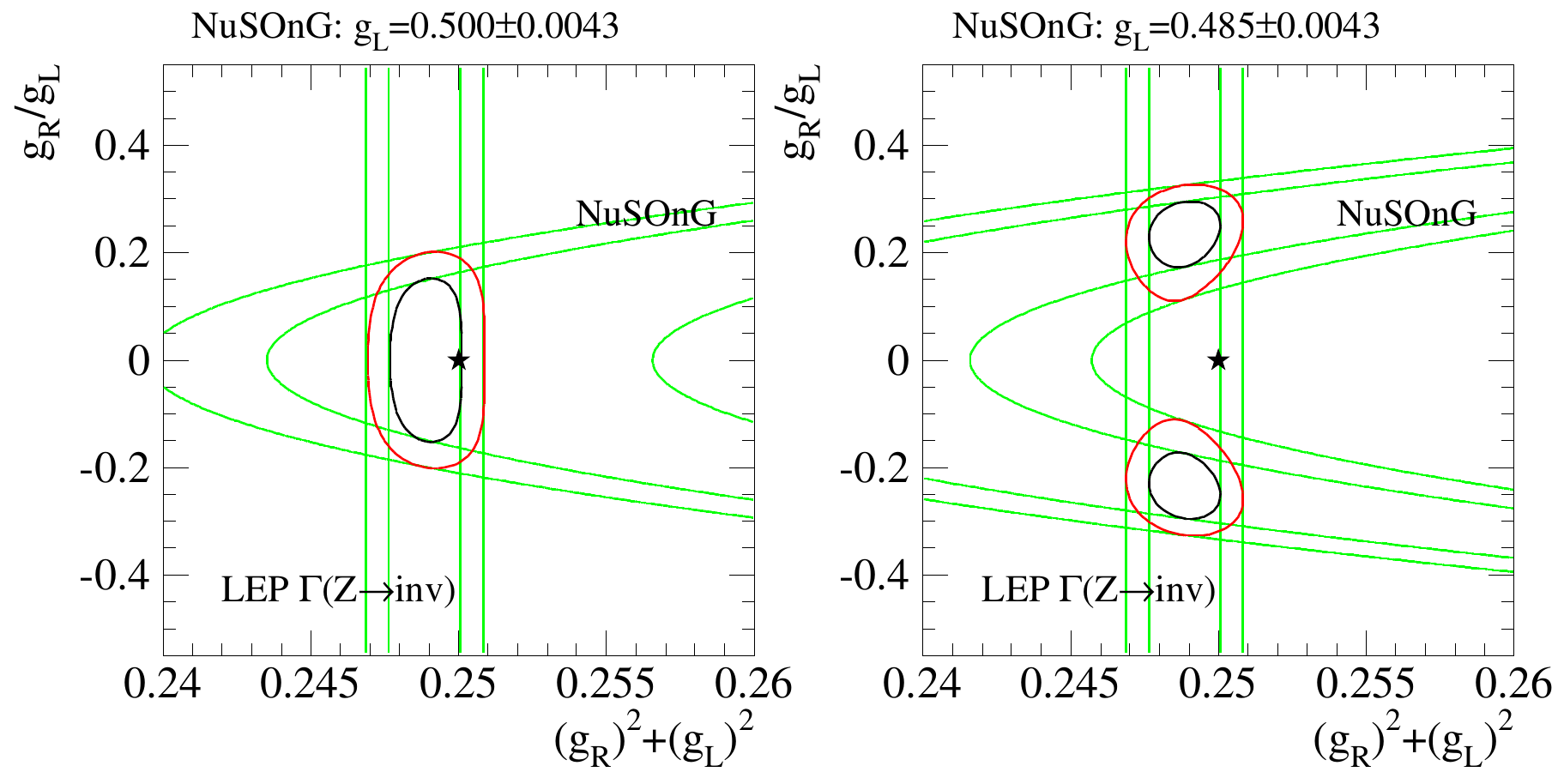}}
\vspace{1mm}
\caption{Precision with which the right-handed neutrino--$Z$-boson coupling can be determined by combining NuSOnG measurements of $g_L^{\nu}$ with the indirect determination of the invisible $Z$-boson width at LEP. In the left panel, we assume that the $\nu+e$ scattering measurement is consistent with the Standard Model prediction $g_L^{\nu}=0.5$, while in the right panel we assume that the $\nu+e$ scattering measurement is significantly lower, $g_L^{\nu}=0.485$, but still in agreement with the CHARM II measurement (at the one sigma level). Contours (black, red) are one and two sigma, respectively, while the star indicates the Standard Model expectation. See  \cite{Carena:2003aj} for more details.}
\label{glgr}
\end{figure}

At NuSOnG, we estimate that $g_L^{\nu}$ can be measured  
at around the 0.86\% level. This estimate is obtained by
combining the statistical uncertainty (20,000 $\nu+e$ elastic
scattering events) with an estimated 0.5\% systematic uncertainty from
the flux estimate.  Fig.~\ref{glgr} (left) depicts an estimate of how
precisely $g_R^{\nu}$ could be constrained if the NuSOnG result,
assumed to agree with the Standard Model prediction, is combined with
the indirect LEP constraints.  One can clearly see that this
measurement ($|g_R^{\nu}/g_L^{\nu}|\lesssim 0.2$ at the two sigma
level) compares favorably with the ILC capabilities described
above. If the NuSOnG result is incompatible with Standard Model
expectations but still in agreement with the CHARM II experiment, a
combined NuSOnG--LEP analysis should be able to establish that
$g_R^{\nu}\neq 0$, as depicted in Fig.~\ref{glgr} (right).

\paragraph{\it  New Physics Observed through Oblique Corrections}
\label{SandTsubsub}

~~~\\
~~~

Precision neutrino scattering measurements made at NuSOnG can reveal
new physics even when new particles are not created in the
final state, through the effects of these particles in loops.
For models of new physics in which the dominant
loop corrections are vacuum polarization corrections to the gauge boson 
propagators (``oblique'' corrections), the $ST$ parameterization introduced by 
Peskin and Takeuchi \cite{PT} provides a convenient framework in which
to describe the effects of the new physics.

\begin{figure}
\vspace{5mm}
\centering
\scalebox{0.8}{\includegraphics{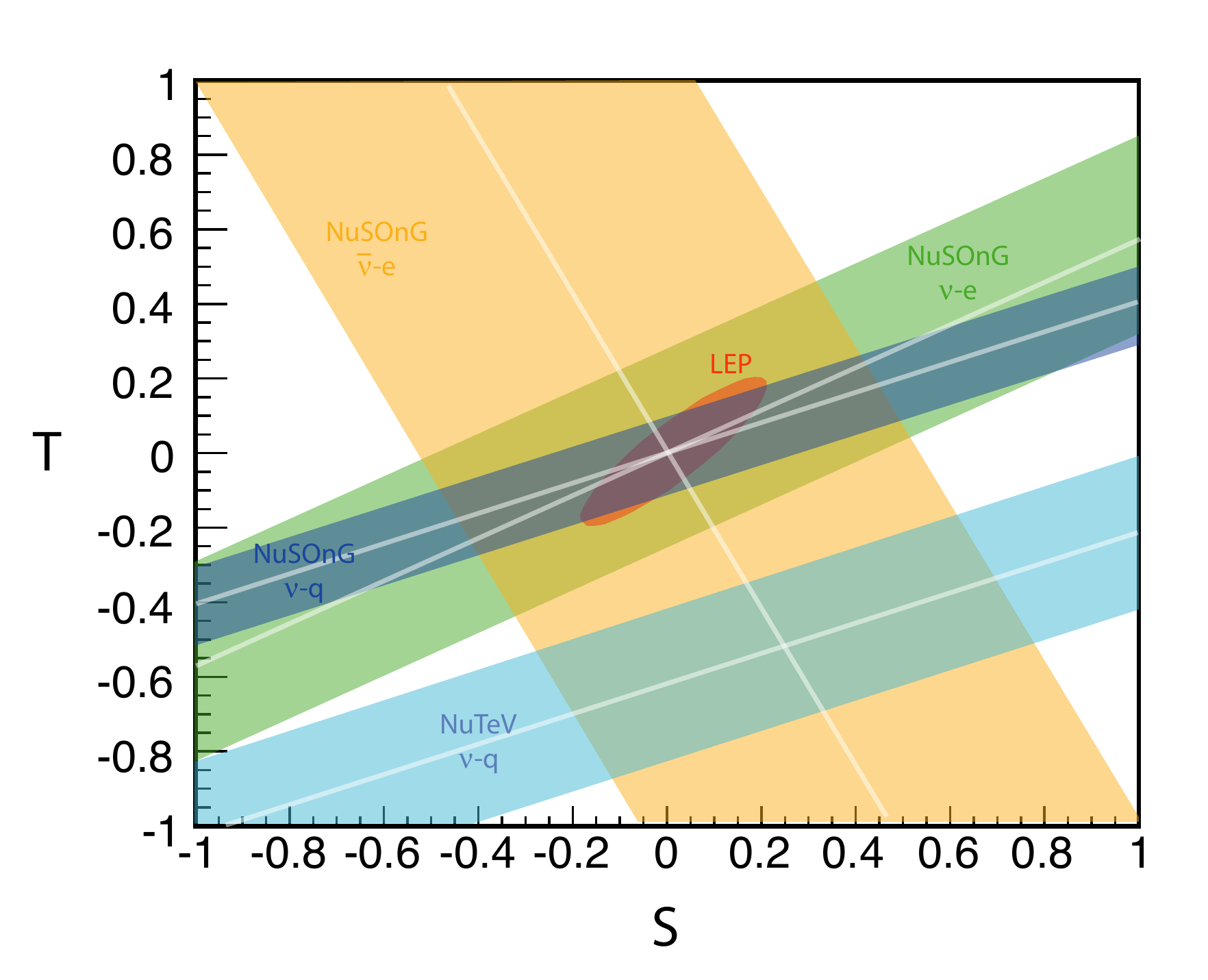}}
\vspace{1mm}
\caption{\label{SandT}{Three projected electroweak measurements from NuSOnG in S-T plane.  LEP/SLD error ellipse is shown in red and the current NuTeV $\nu-q$ measurement is shown as a light blue band.  The ochre band shows NuSOnG $\overline{\nu}- e$, the dark blue band shows NuSOnG $\nu-q$ and the green shows NuSOnG $\nu-e$.  The width of the bands correspond to 68\% confidence level for statistics as described in the text.  The NuSOnG measurements assume $(S,T)=(0,0)$.}}
\end{figure}

The $ST$ parameterization begins with a reference Standard Model,
including reference values for the Higgs and top masses, and predictions for
observables in this reference Standard Model.  Differences between
predicted and experimental values of the observables are
then parameterized by and used to fit for $S$ and $T$, which can then
be compared to predictions from new physics.  
The full set of precision 
electroweak data can then be used to constrain $S$ and $T$, as shown
in Fig.~\ref{SandT}.
The $T$ parameter is 
sensitive to new physics that violates isospin 
and is zero for new physics that conserves isospin.  Isospin-breaking
new physics such as heavy non-degenerate
fermion doublets or scalar multiplets would affect the $T$ parameter.
The $S$ parameter is sensitive to isospin-conserving physics,
such as heavy degenerate fermion doublets.

The status of electroweak measurements
are shown in Fig.~\ref{SandT} \cite{rosner}.  The combined analysis of the LEP and
SLD data by the LEP Electroweak Working Group (EWWG) \cite{EWWG}
indicates an allowed region
shown by the small oval, centered at 
$S=0.05\pm0.10$ and $T=0.07\pm0.11$.  
A different choice of reference Higgs or top mass changes Standard
Model predictions for observables and thus shifts the center
of the $ST$ plot \cite{PeskinWells}; setting the Higgs mass to 1000 GeV would shift the center of the oval to roughly $(S,T)=(0.12,-0.36)$.  Measurements of the $W$ mass, which are not shown, are also consistent with the LEP measurements.
The highest precision neutrino result
comes from $\nu q$ and $\bar \nu q$ scattering by the NuTeV
experiment.  This result clearly disagrees with the other
measurements, as discussed in Sec. \ref{nutevsection}.

The goal of NuSOnG is to make measurements which are competitive with
or better than past electroweak measurements.  These goals are
indicated by the magenta ellipse and orange band on Fig.~\ref{SandT}.
The magenta ellipse shows the area in $ST$ space which can be probed
if a 0.7\% measurement of the $\nu$ and $\bar \nu$ NC
electron-scattering cross sections is achieved.  The orange band shows
the improvement in the neutrino-quark, ``Paschos-Wolfenstein''-style
measurement which is expected from NuSOnG.

Disregarding the NuTeV offset for the moment, one can now ask: how
will this plot look in the era of LHC and what will NuSOnG add?  We
consider this question in light of three scenarios:
\begin{enumerate}
\item  a light Higgs (115-200
GeV)
\item a heavy Higgs (200-1000 GeV)
\item no Higgs signal.
\end{enumerate}

\paragraph{\it NuSOnG Impact for  a Light Higgs (115-200 GeV) Scenario}
\label{lighthiggs}

~~~\\
~~~

A light Higgs is consistent with LEP/SLD and $W$ mass data.  The
fit to the electroweak data excluding NuTeV indicates a mass less than
144 GeV at 95\% CL.  This is also consistent with the current best
direct-search limit which finds $m_H >114$ GeV \cite{EWWG}.  In the case of the
lightest Higgs masses, where the cleanest signal may be in $H
\rightarrow \gamma \gamma$, a clear observation above background will
be experimentally difficult and may take some time.

Once the LHC measurement of the Higgs mass is made, the center of the
$ST$ ellipse (Fig.~\ref{SandT}) will be fixed at a point (modulo any remaining
uncertainty in the top mass).  
Our experiment is especially interesting if the
NuSOnG result disagrees with this LEP+SLD+LHC point.  If the LHC
measurement is high, {\it i.e.} $m_H\sim 200$ GeV, the result would be
marginally inconsistent with the $M_W$
analysis, which is $85^{+39}_{-28}$ GeV\cite{EWWG}.
In this case, comparison with the $\nu_\mu$ scattering results from
NuSOnG could resolve the question of a discrepancy between these
measurements.

If all other electroweak results are in good agreement, but disagree
with NuSOnG, this would indicate new properties associated exclusively
with the neutrino.  An example would be decreased coupling of the
neutrino to the $Z$ boson,
where suppression of the coupling comes from intergenerational mixing
of the light neutrino with a moderately heavy neutrino:
\begin{equation}
\label{mixeq}
\nu_\mu = (\cos \alpha) \nu_{{\rm light}} + (\sin \alpha) \nu_{{\rm heavy}}.
\end{equation} 
The $Z\nu_\mu\nu_\mu$ coupling is modified by $\cos^2 \alpha$ and
the $W\mu\nu_\mu$ coupling is modified by $\cos \alpha$.  
This model,
inspired by the NuTeV anomaly (see Sec. \ref{nutevsection}), would
yield a measurement in NuSOnG with a low NC-to-CC ratio in both
the case of electron and quark scattering.  

These moderately heavy right-handed states, dubbed ``neutrissimos''
\cite{hep-ph/0304004}, could have masses as low as 
just above the current bound of
the $Z$ mass.  They may well be within the reach of the LHC and may
appear as missing energy in events \cite{hep-ph/0304004}. Some models
allow for neutrissimos as light as $\sim 100$ GeV
\cite{0706.1732v1.pdf}.  The neutrissimos decay very quickly, but not
always invisibly.  For example, in the reaction $N \rightarrow \ell + W$,
the $W$ may decay to either two jets or a neutrino--charged-lepton
pair; only the latter case has missing energy.  This may make
recognition of the neutrissimo at LHC rather difficult.  In the case of $m_H
< 130$ GeV, a dominant decay mode of the Higgs (along with $b \bar b$)
could be into $\nu N$, where the neutrissimo subsequently decays.
Reconstructing the Higgs in this case may be difficult at LHC; if
neutrissimos exist, the result from NuSOnG may significantly improve
our understanding of LHC results.

With a large tuning among the neutrino Yukawa
couplings \cite{0706.1732v1.pdf}
neutrissimos could be the seesaw right-handed neutrinos.
 Relatively ``large" mixing is
marginally consistent with other constraints, including neutrinoless
double-beta decay, which constrains $|U_{e4}|^2$ to be less than a few
$\times 10^{-5}$ for a 100 GeV right-handed neutrino, and rare pion and
tau decays, which constrain $|U_{\mu 4}|^2$ to be less than, most
conservatively, 0.004 and $|U_{\tau 4}|^2$ to be less than 0.006. Other
bounds come from $\mu \rightarrow e$ conversion in nuclei and other
charged-lepton-flavor violation.  A new experiment to search for $\mu
\rightarrow e$ has been proposed at Fermilab \cite{Prebys} should also
be sensitive to neutrissimos.  The combination of NuSOnG and this
experiment will be powerful in identifying the existence of these
particles.

If the neutrissimo is a Majorana particle, it could be instrumental in
elucidating the mechanism for leptogenesis.  The present models of
leptogenesis require very high mass scales for the neutral lepton, but 
theorists are pursuing ways to
accommodate lower masses \cite{hep-ph/0410075v2}.  There also may be a
wide mass spectrum for these particles, with one very heavy state required by 
standard leptogenesis models and others with masses
in the range observable at LHC \cite{hep-ph/0608147}.

\paragraph{\it NuSOnG Contribution in a Heavy Higgs ($200-1000$ GeV) Scenario}
\label{heavyhiggs}

~~~\\
~~~

While present electroweak data excluding NuTeV
favor a light Higgs ($\lesssim 200$ GeV), as indicated in Fig.~\ref{SandT}, 
the Higgs mass can extend up to about 1000 GeV without violating unitarity
\cite{quigguni}.  Thus, if LHC finds that the Higgs is between
$\sim$200 and 1000 GeV and the LEP+SLD ellipse has no major systematic error,
then new physics must explain the discrepancy.  
Candidate models of new physics may well affect the neutrino scattering
and $e^+e^-$ scattering differently, so the 
high-precision neutrino scattering measurements from NuSOnG will
provide an important piece of the puzzle if the Higgs mass found at LHC is
genuinely inconsistent with  LEP+SLD predictions.

Introduction of a fourth family would compensate for a modestly heavy
($\sim 300$ GeV) Higgs by shifting the LEP+SLD allowed region back up
in $S$ and $T$ \cite{Tait}.  This family would need to exist above the
bounds of direct searches, which is $\gtrsim 300$ GeV.  Mixing must be
confined within the allowed bounds of the CKM matrix measurements
\cite{PDGCKM}.  A nice feature of this model is that a
fourth-generation Majorana neutrino could play the role of dark
matter.  Depending on the underlying physics, evidence of a fourth
family would be apparent in a shift of the NuSOnG result on the
$ST$ plot.  This could be especially important if the physics
introducing the fourth family is from a mechanism like ``Top See Saw''
\cite{DobrescuHill}, which will not be observable at LHC.  The impact
of this particular model on neutrino scattering is not yet thoroughly
explored, but could prove interesting \cite{hill}.

A classic method for masking a heavy Higgs is to introduce heavy $Z$
bosons \cite{additionalZ}, which, as shown in ref.
\cite{PeskinWells}, tend to move the LEP-SLD ellipse upward in $T$,
compensating for the heavy Higgs.  Introduction of a $Z^{'}$ tends to
increase NC rate in neutrino scattering and also to move the neutrino
result upward on the $ST$ plot (although with a different
dependence than the LEP-SLD result).  

There are good theoretical reasons for considering the existence
of additional neutral heavy gauge bosons.  Extra $Z$ bosons appear in
various GUT and string-motivated extensions to the Standard Model
\cite{LangackerErler04}.  For example, the $E(6)$ breakdown to $SO(10)
\times U(1)_\psi$ results in the $Z_{\psi}$.  The $SO(10)$ break down
to $SU(5) \times U(1)_\chi$ yields the $Z_{\chi}$.  Thus the new
exchange boson could be: $Z^{'}=Z_{\chi}cos\beta+Z_{\psi}sin\beta$,
where the mixing angle $\beta$ is an arbitrary parameter.  Extra $Z$
bosons also appear in other beyond Standard Model theories, including
extra dimensions with gauge fields in the bulk \cite{Zextradim};
little Higgs theories \cite{ZlittleH}, which use heavy $Z$s to cancel
divergences in the Higgs mass; and topcolor in which they drive
electroweak symmetry breaking \cite{Ztopcolor}.  Heavy $Z$s provide a
mechanism for new SUSY theories to evade the LEP bound of $m_H=114$ GeV
\cite{Zsusy}.  These models all produce new physics signatures at LHC.
The precision measurement from NuSOnG can aid in differentiating
models.

Models which introduce new physics to mask a heavy Higgs may seem
contrived until one looks at the LEP+SLD data more closely.  Up to
this point we have considered the LEP+SLD measurements as a single
result, however, many measurements enter this fit, and larger than 
expected inconsistencies  between these measurements exist \cite{chanowitz}.
For example, there is a 3.2$\sigma$ discrepancy between the
forward-backward ($A_{FB}$) and left-right ($A_{LR}$) asymmetry
measurements.  Excluding the $A_{FB}$ result, the LEP+SLD fit yields
$m_H<115$ GeV at 95\%, with the best fit at 42 GeV -- {\it i.e.} a
range already excluded by direct searches, which require $m_H>114$ GeV
at 95\% CL.   

There are several ways to interpret this deviation.
It may simply be that there are systematics involved
in the $A_{FB}$ measurement which have yet to be identified
and which would bring this result into agreement with the others.
In this case, we are in the dramatic situation of having 
already ruled out the Higgs.    The scenario of no Higgs
is considered in the next section.    Alternatively, new 
physics is involved.
This result is dominated by purely leptonic measurements.
On the other hand, the fit to the hadronic asymmetries, dominated by
$A_{FB}^b$ has two $\chi^2$ minima, at 450 and 3000 GeV. 
Thus, one may either introduce new physics
which produces a 20\% shift on $A_{FB}^b$ alone; or introduce new
physics which would indicate apparently low values of $m_H$ in the
lepton-based measurements, when actually the value is large.
Within any of these scenarios, new precision results from 
NuSOnG will be valuable for understanding the underlying physics.

\paragraph{\it NuSOnG and the Case of No Higgs}
\label{nohiggs}

~~~\\
~~~

Higgsless models do not employ the Higgs mechanism to
render the Standard Model renormalizable \cite{ref:birkedal};
instead they introduce some other scheme.
The Higgs mechanism enforces unitarity in the scattering amplitudes of
longitudinally polarized gauge bosons, $W_L^{\pm}+Z_L^o \rightarrow
W_L^{\pm} Z_L^o$, for example.  A requirement that the transition
probability remains less than one gives the energy scale $\Lambda$ at
which a new mechanism must come into play,
\begin{equation}
\Lambda \sim \frac{4\pi M_W}{g}\sim 1.8 \mathrm{ TeV}.
\end{equation}
Higgsless theories generally contain new mass bosons $V_i$ with masses
on the TeV scale that act to cancel the divergences in gauge boson
scattering.  Cancelling the amplitudes while respecting bounds from
current electroweak couplings typically give small couplings: 
\begin{equation}
g_{WZV} <\frac{g_{wwz}M_Z^2}{\sqrt{3}M_1^{\pm}M_W}=0.04
\end{equation}
for $M_1^{\pm}$=700 GeV.  

At the LHC, the typical cross sections for $V_i$ are hundreds of
femtobarns, so, after cuts, the LHC experiments will record tens to
hundreds of events in the first years of data taking.  Since the $V_i$
resonances serve the same purpose as the Higgs boson, additional
information will be necessary to determine whether these resonances
originate from spontaneous symmetry breaking or from strong coupling
between the known gauge bosons. The electroweak measurements 
from NuSOnG will play a role in understanding the origin of
such events, en route to a more complete explanation provided by the ILC.

\subsection{The NuTeV Anomaly}
\label{nutevsection}

\begin{figure}
\vspace{5mm}
\centering
\scalebox{0.75}{\includegraphics[clip=true]{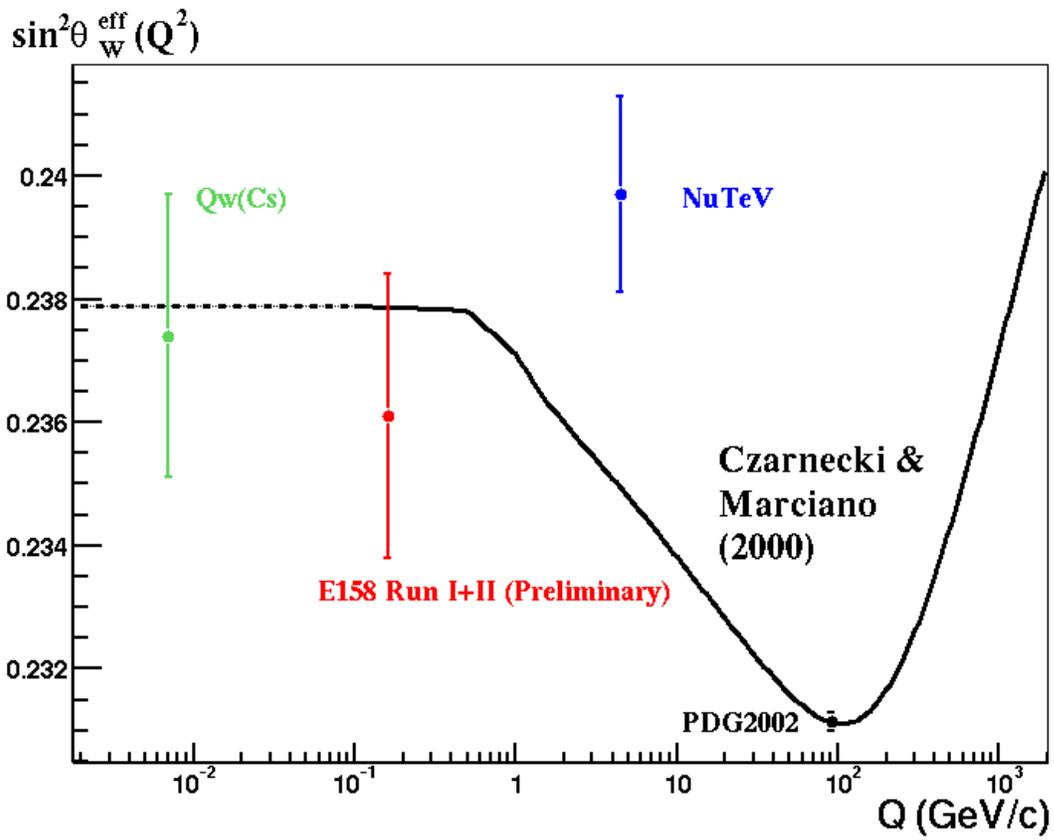}}
\vspace{-2.cm}
\caption{Measurements of $\sin^2 \theta_W$ as a function of $Q$; 
from ref. \cite{E158web}.   The curve shows 
the Standard Model expectation. }
\label{Marciano}
\end{figure}

The NuTeV anomaly is a $3 \sigma$ deviation of $\sin^2 \theta_W$ from
the Standard Model prediction \cite{NuTeVanomaly}.  NuTeV employed the
PW-inspired method discussed in Sec. \ref{PWsection}, which resulted
in a 0.75\% measurement of the weak mixing angle (see
Tab.~\ref{NuTeVerrs}).  Two systematic adjustments to the NuTeV result
have been identified since the result was published.  The first is the
new measurement of the $K_{e3}$ branching ratio from KTeV, which does
not significantly reduce the error, but introduces a correction moving
the result away from the Standard Model.  The second is the final
measurement of the difference between the strange and antistrange seas
(called ``the strange sea asymmetry'', see Sec. \ref{subsec:strangesea}),
which will pull the NuTeV result toward the Standard Model. A new
analysis of the NuTeV data which will include these two corrections is
expected be available in late summer, 2007 \cite{SamPrivateComm}.
It should be noted that while an error from the strange sea
appeared in the NuTeV analysis, no error on a strange sea {\it
  asymmetry} appeared in the original NuTeV analysis; this will be
included in the upcoming re-analysis.

NuTeV is one of a set of $Q^2 \ll m_Z^2$ experiments measuring
$\sin^2\theta_W$.  It was performed at $Q^2 =$ 1 to 140 GeV$^2$,
$\langle Q^2_\nu \rangle=26$ GeV$^2$, $\langle Q^2_{\bar \nu} \rangle
=15$ GeV$^2$, which is also the expected range for NuSOnG.  Two
other precision low $Q^2$ measurements are from atomic parity
violation\cite{APV} (APV), which samples $Q^2 \sim 0$; and SLAC E158,
a M{\o}ller scattering experiment at average $Q^2=0.026$ GeV$^2$
\cite{E158}.  Using the measurements at the $Z$-pole with $Q^2=M_z^2$
to fix the value of $\sin^2 \theta_W$, and evolving to low $Q^2$,
Fig.~\ref{Marciano}, from ref. \cite{E158web}, shows that APV and
SLAC E158 are in agreement with the Standard Model.  However, the
radiative corrections to neutrino interactions allow sensitivity to
high-mass particles which are complementary to the APV and
M{\o}ller-scattering corrections.  Thus, these results may not be in conflict
with NuTeV.
The NuSOnG measurement will provide valuable additional information on
this question.

Since the NuTeV result was published, more than 300 papers have been
written which cite this result.  Various Beyond-the-Standard-Model
explanations have been put forward; those which best explain the
result require a follow-up experiment which probes the neutral weak
couplings specifically with neutrinos, such as NuSOnG.  Several
``within-Standard-Model'' explanations have also been put forward,
based on the inherent issues involving scattering off quarks.  NuSOnG
can address these criticisms in two ways.  First, we will provide
better constraints of the quark-related distributions at issue.
Second, we perform the measurement of the weak mixing angle in both a
purely leptonic mode (scattering from electrons) and via the PW
method.  Agreement between the two results would address the questions
which have been raised.

\subsubsection{Explanations Within the Standard Model}
\label{smexplain}

\begin{figure}
\vspace{5mm}
\centering
\includegraphics[height=0.75\textheight]{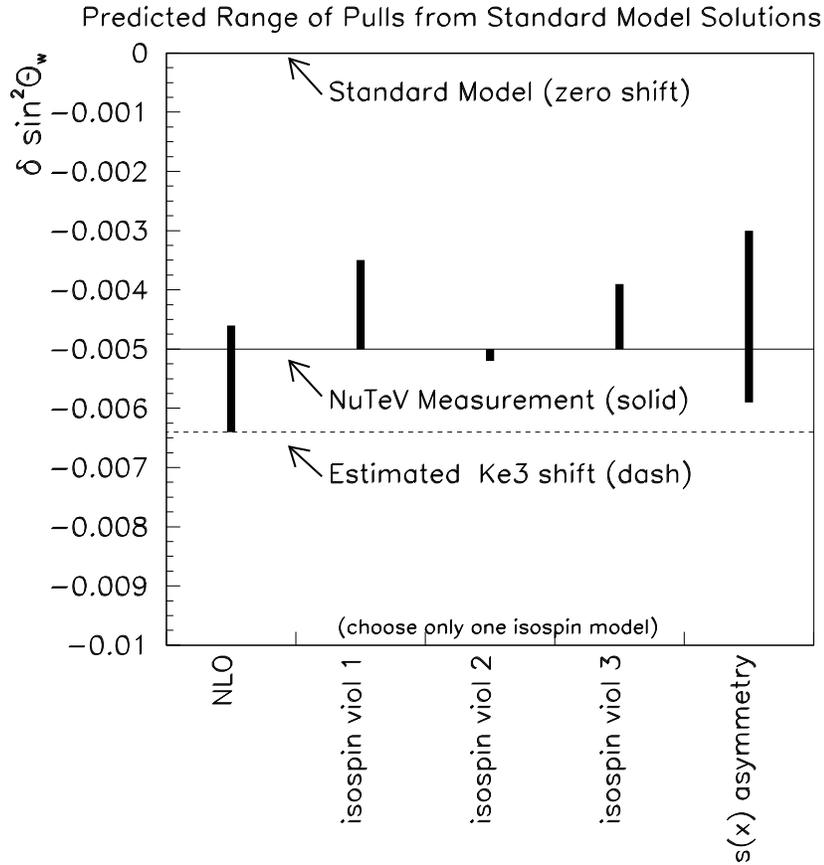}
\vspace{-1cm}
\caption{\label{pulls} Effect of various ``Standard Model''
explanations on the NuTeV anomaly. The $y$-axis is the deviation from
the Standard Model.    The solid line is the NuTeV deviation.  The dashed line
is an estimate of the effect of correcting for the new $K_{e3}$ branching 
ratio. Thick black lines extending from the NuTeV deviation show the range
of possible pulls from the various suggested sources, as described in the 
text.}
\end{figure}

Four explanations for the NuTeV anomaly that are ``within the Standard
Model'' have been proposed.  These are: electromagnetic radiative
corrections; higher order QCD corrections; isospin (or charge symmetry)
violation; and the strange sea asymmetry.  The radiative
corrections will be disregarded here, since the results of this paper
\cite{diener}
are not reproducible.

The effect of the possible explanations is illustrated in
Fig.~\ref{pulls}.  On this plot, the solid horizontal line indicates
the deviation of NuTeV from the Standard Model.  The thick vertical
lines, which emanate from the NuTeV deviation, show the range of pulls
estimated for each explanation, as discussed below.  The dashed
horizontal line shows the estimated shift due the new $K_{e3}$
branching ratio.  We do not yet have an estimated shift due to the new
NuTeV strange sea measurement, but it is expected that this
will move the dashed line toward the Standard Model
\cite{SamPrivateComm}.

Three ``Standard Model'' explanations may be considered next
\cite{McFarland:2003jw,Davidson:2001ji}.  First, the NuTeV analysis
was not performed at a full NLO level; NuSOnG will need to undertake 
a full NLO analysis.  But the effect of going to NLO
on NuTeV can be estimated \cite{nlomodels}, and the expected pull is
away from the Standard Model, as shown on Fig.~\ref{pulls}.  Second,
the NuTeV analysis assumed isospin symmetry, that is, $u(x)^p =
d(x)^n$ and $d(x)^p = u(x)^n$.  Isospin violation can come about from
a variety of sources and is interesting in its own right. NuSOnG's
contribution to this study is discussed in Section \ref{subsubsec:isospin}.
Various models for isospin violation have been studied and their pulls
range from less than $1\sigma$ away from the Standard Model to $\sim 1
\sigma$ toward the Standard Model \cite{isomodels}. We have chosen
three examples \cite{isomodels} for illustration on Fig.~\ref{pulls}:
the full bag model, the meson cloud model, and the isospin QED model.
These are mutually exclusive models, so only one of these can affect
the NuTeV anomaly.  Third, variations in the predicted strange sea
asymmetry can either pull the result toward or away from the Standard
Model expectation \cite{Zeller:2002du, Olness:2003wz, Barone:1999yv}.
This issue is considered in detail in Sec.~\ref{subsec:strangesea}.

\subsubsection{Beyond Standard Model Interpretations}

Chapter 14 of the APS
Neutrino Study White Paper on Neutrino Theory \cite{APSwhitepaper} is
dedicated to ``the physics of NuTeV'' and provides an excellent
summary.  The discussion presented here is drawn from this
source.

The NuTeV measurements of $R^\nu$ and $R^{\bar \nu}$, the NC-to-CC
cross sections, are low.  If one is assumes that the
Higgs is light, then this must be interpreted as Beyond-Standard-Model
physics that suppresses the NC rate with respect to the CC rate.
Two types of models produce this effect and remain consistent with the
other electroweak measurements: 1) models which affect only the $Z$
couplings, {\it e.g.},  the introduction of a heavy $Z^\prime$ boson which
interferes with the Standard Model $Z$; or 2) models which affect only
the neutrino couplings, {\it e.g.},  the introduction of moderate mass
neutral heavy leptons which mix with the neutrino.

As discussed in Sec.~\ref{heavyhiggs}, introduction of $Z^\prime$
bosons tend to increase the NC rate rather than suppress it.   
Thus there is only a small subset of models which produce the 
destructive interference needed to explain the NuTeV result.
Models which introduce a $Z^\prime$ which selectively suppresses
neutrino scattering, without significantly affecting the other
electroweak measurements, include cases where the 
$Z^\prime$ couples to $B-3L_\mu$ \cite{hep-ph/209316}
or to $L_\mu -L_\tau$ \cite{hep-ph/0110146}.
In the former case, fitting the NuTeV anomaly requires that 
$M_{Z^\prime}/g_{Z^\prime} \sim 3$ TeV.  
From the bounds from direct searches, this sets a limit on 
$M_{Z^\prime} > 600$ GeV if the coupling is on the order of unity,
but as low as 2 to 10 GeV if the coupling is $\sim 0.1\%$.
The latter case is an example which improves the agreement between
NuTeV and other results, but does not entirely address the problem.
Its effectiveness in solving the NuTeV anomaly is limited by the
data constraining lepton universality. This model addresses 
more than just the NuTeV anomaly.   It is inspired by 
attempts to address bimaximal mixing in the neutrino sector.  
It has the nice features of also  
addressing the muon $(g-2)$ measurement and producing a distinctive
dimuon signature at LHC.

The case of models involving moderate-mass neutral heavy leptons, {\it a.k.a.}
neutrissimos, have been discussed in the Sec. \ref{lighthiggs} and
examples of viable models appear in ref. 
\cite{hep-ph/040201v1}. Eq.~\ref{mixeq} described how the muon
neutrino couplings might be modified by mixing.  This idea can be
extended to all three flavors, leading to a suppression factor for the
$Z$ coupling which is expressed as $(1-\epsilon_\ell)$ and for the $W$ by
$(1-\epsilon_\ell/2)$, where $\ell=e,\mu,$ or $\tau$.  This addresses
the NuTeV anomaly and at the same time suppresses the invisible width
of the $Z$, describing the LEP I data.

\begin{figure}
\vspace{5mm}
\centering
\scalebox{0.75}{\includegraphics{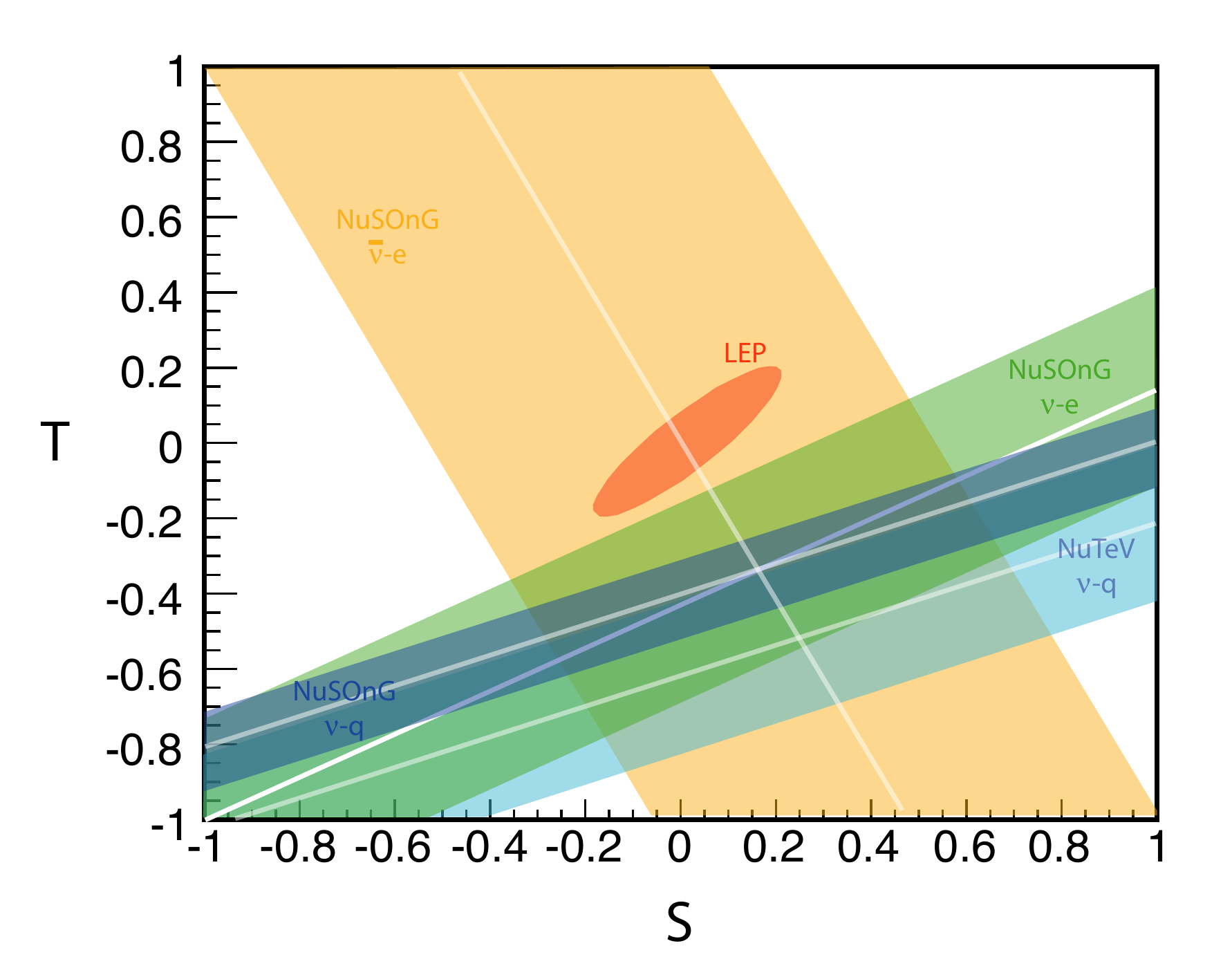}}
\vspace{1mm}
\caption{\label{SandTanomaly}{Three projected electroweak measurements from NuSOnG in S-T plane for a model model with a heavy Higgs inspired by the NuTeV measurement \cite{hep-ph/040201v1}.  In this model, $(S,T)=(0.12,-0.36)$.  The labeling is as in Fig.~\ref{SandT}.}}
\end{figure}

If the NuTeV anomaly is due to Beyond Standard Model physics, then
the effect will be visible in the neutrino-electron elastic scattering
measurement also.   Thus, if the NuTeV anomaly is borne out, NuSOnG would observe an $ST$ 
plot similar to~Fig.~\ref{SandTanomaly}.

%% file: DirectSearches_v4A.tex
\subsubsection{Light Neutrino Properties}

Evidence for three light neutrino masses has now been
established through neutrino oscillations in solar, atmospheric, and
reactor experiments (see references \cite{Cleveland:1998nv} through \cite{kamland}). 
Furthermore, although the
MiniBooNE experiment recently refuted the LSND two-neutrino
oscillation scenario at $\Delta m^2\sim1$ eV$^2$ \cite{MiniBooNE}, 
the question of the
existence of multiple light sterile neutrinos still remains open
\cite{hep-ph/0705.0107}. These observations already require beyond-the-Standard-Model physics, 
and consequently raise phenomenological
questions, such as: what are the mass and mixing parameters still allowed in
sterile neutrino models? What do sterile neutrinos imply about
neutrino mixing? Is the neutrino mixing matrix unitary, or is there
effective freedom of mixing parameters? As we illustrate in the
following sections, these are some of the questions that NuSOnG can
potentially address.

\paragraph{\it Matrix Freedom}

~~~\\
~~~

Perhaps the most interesting study of light neutrino properties which
can be performed at NuSOnG is the search for evidence of ``matrix
freedom'' or ``nonunitarity.''  For example, in the case of existence of
sterile neutrinos, the neutrino mixing matrix is extended to an
$N\times N$ matrix, where $N>$3. Under that assumption, it has been
suggested that the 3$\times$3 part of the matrix describing the three
active (SM) neutrinos is not necessarily unitary; or, equivalently,
the three flavor eigenstates are non-orthogonal (the 3$\times$3
neutrino mixing matrix is \textit{free}) \cite{BorisPrivateComm}.

This introduces striking changes to the probability formula for
neutrino flavor transitions.  Assuming unitarity, the survival
probability formula for a neutrino produced as flavor $\alpha$ is
\begin{equation}
\label{eq:unitaryprob}
P_{\alpha \alpha}^{unitary} = 1 - 4 |U_{\alpha 3}|^2[1-|U_{\alpha 3}|^2] \sin^2{\Delta_{31}},
\end{equation}
where one has made use of $\Delta_{31} = \Delta m^2_{31} {{L}\over{4E}}$, 
and $\Delta m^2_{21}{L\over{4E}}\ll1$.
 In the case of matrix freedom, the mixing matrix is no longer unitary. 
The level at which unitarity is violated can be defined as $X_{\alpha}$, where 
\begin{equation}
\label{eq:unitarityviol}
\sum_j |U_{\alpha j}|^2 = 1-X_\alpha,
\end{equation}
with $X_{\alpha}$ being small. 
Under that assumption, the survival probability formula is then found to be
\begin{equation}
\label{eq:nonunitaryprob}
P_{\alpha \alpha}^{general}  =  P_{\alpha \alpha}^{unitary} - 2 X_\alpha [1 - 2 |U_{\alpha 3}|^2 \sin^2 \Delta_{31}] + X_\alpha^2.
\end{equation}

As implied by Eq.~\ref{eq:nonunitaryprob} one of the main consequences of
such scenario is instantaneous ($L=$0) flavor transitions in
a neutrino beam.   This occurs regardless of the size of 
the mass splitting between
the mostly sterile and mostly active states, and thus allows for
a full-mass-range search for evidence of sterile neutrinos.
A recent study \cite{hep-ph/0607020} suggests that
current experimental data limit such an effect to up to the order of a few
percent. 


As a result, several interesting and potentially observable phenomena
can occur.  Extending the argument of ref.~\cite{hep-ph/0607020}, for
instance, the non-orthogonality of $\nu_{\mu}$ and $\nu_e$ that matrix
freedom introduces, results in an instantaeous transition at $L=0$
from $\nu_\mu$ to $\nu_e$ \cite{BorisPrivateComm}.  Thus one could
observe an excess of $\nu_e$ events in a pure $\nu_{\mu}$ beam.

\begin{figure}
\includegraphics[width=16cm]{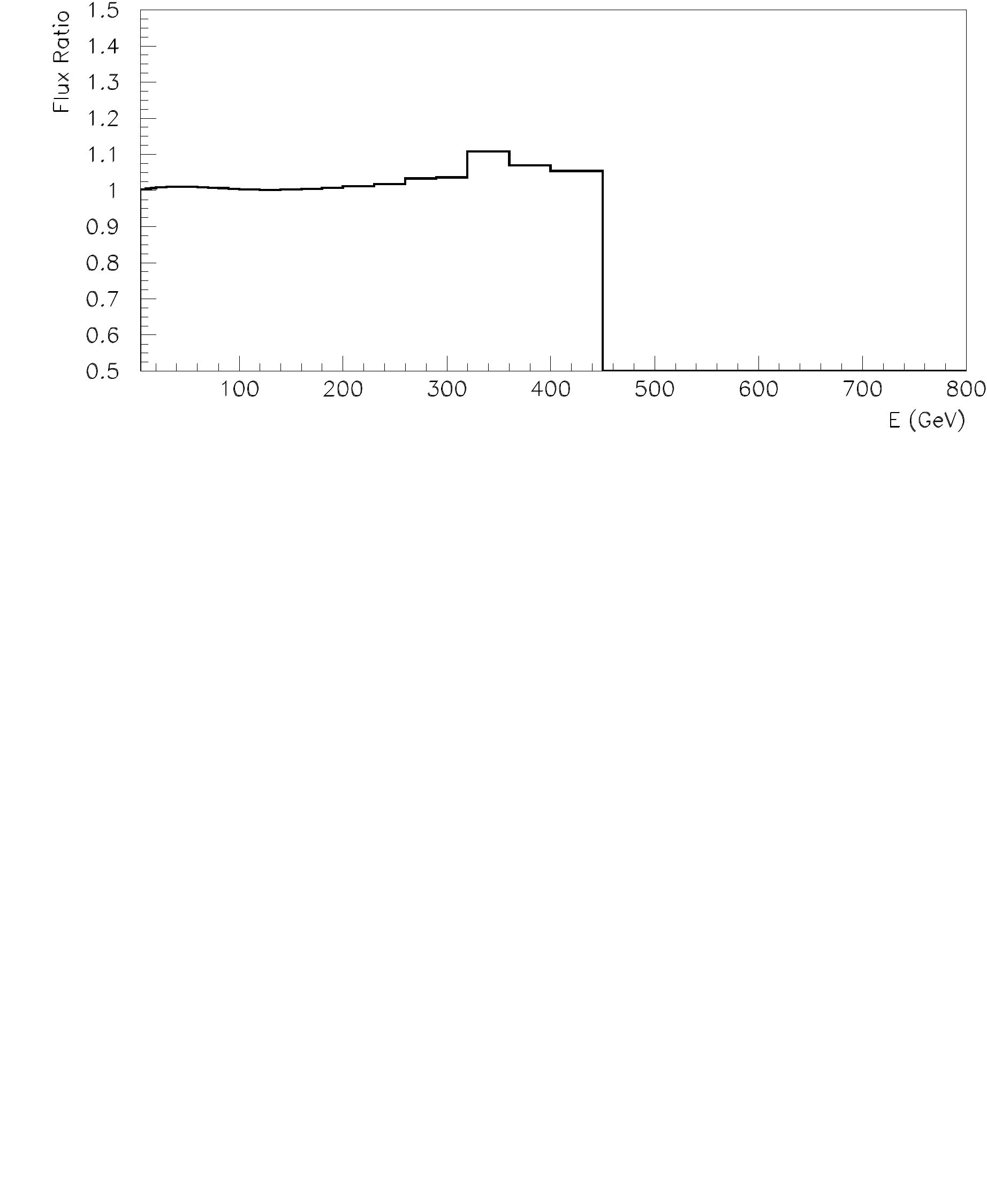}
\vspace*{-12cm}
\caption{Ratio of enhanced $\nu_e$ flux due to $\nu_\mu$ transitions to $\nu_e$ flux assuming no transitions. Obtained assuming 100M $\nu_\mu$ deep inelastic scattering events.} 
\label{epsfig:nueratio}
\end{figure}

The trick to searching for this instantaneous transition is to focus
on an energy range where the $\nu_e$ background is low and well
constrained.  In the case of NuSOnG, this is on the high energy tail
of the flux, above E$\gtrsim 250$ GeV.  For the limits on $\nu_\mu$
transformation to $\nu_e$\cite{hep-ph/0607020}, which are at the $\sim
1\times 10^{-4}$ level, NuSOnG would see an excess of $\sim200$
$\nu_e$ events in this high energy region.  Fig.
\ref{epsfig:nueratio} shows the ratio of $\nu_e$ flux with
$\nu_\mu$ transitions to $\nu_e$ flux without transitions.  The abrupt
cutoff is due to Monte Carlo statistics; higher energies can be
explored.  Assuming that such transitions indeed happen at the
$10^{-4}$ level, one would expect up to a 10\% increase in flux for
E$\sim350$ GeV.  In that high energy region, the $\nu_e$ flux is mainly
from $K^+$ decay, which is well constrained by the $\nu_\mu$ events.
Such an excess should therefore be measurable.

Other interesting effects of matrix freedom \cite{BorisPrivateComm}
include the oscillatory behavior in the total (flavor-summed) CC event
rate as a function of $L/E$, and (fake) CP-violating effects in the
$\nu$ and $\bar{\nu}$ neutral-current event rates (the two rates
oscillate differently with $L/E$). Potential observation of those
effects at NuSOnG has not been explicitly considered at this stage,
although it would be interesting to address this and we are planning
to do so in the near future. Regardless of that, evidence of $\nu_e$
contamination in a $\nu_\mu$ beam above expected background levels,
something for which NuSOnG can search, would strongly support
the matrix freedom hypothesis.

\paragraph{\it Sterile Neutrino Oscillations}

~~~\\
~~~

Direct observation of sterile neutrino oscillations may also be
possible in NuSOnG, depending on the mass and mixing parameters.
Oscillations of active to light sterile neutrinos have been introduced
to explain the LSND anomaly, as dark matter candidates, and in
describing the supernova collapse models.  These ideas span a wide
range of $\Delta m^2$ values.  The LSND anomaly requires a sterile
neutrino in the range of $\sim 1$ eV$^2$ with moderate mixing
($\lesssim 1\%$), while dark matter candidates and supernova collapse
models require $(\gtrsim 1~{\rm keV})^2$.  These models also require tiny
mixing ($10^{-13} < \sin^2 2\theta < 10^{-7}$)\cite{astro-ph/0101524}.
NuSOnG probes an intermediate range of $\Delta m^2$, between 
the LSND and astrophysical allowed regions. 
However, since sterile neutrinos may come in
families, it is worth exploring this previously uncharted territory.

The NuSOnG experimental design consists of a 30-600 GeV muon neutrino
beam, peaked at $\sim$100 GeV, incident on a $\sim 200$-meter long
detector located at L$\sim$1.5km from the neutrino
source. This detector design allows for $\nu_{\mu}$ disappearance
studies across the detector length by examining the
$\nu_{\mu}$ scattering rate variation across the detector. Such searches
would be limited by the detector energy resolution.
Preliminary studies have shown that, assuming a 10\%
energy resolution, $\Delta m^2\sim$ 600eV$^2$ regions with mixing of 
$\lesssim 0.1$ can be probed easily.  NuSOnG may also be able to explore smaller mixings and higher 
$\Delta m^2$s, depending on the
final experimental design.

NuSOnG can also probe for $\nu_\mu$ and $\nu_e$ disappearance in the
range of $L/E = (1.5 {\rm km}/ 100~{\rm GeV}) = 0.015$, thus in the
range of $\Delta m^2 \sim 50$ eV$^2$.  This is a range which has been
covered by past experiments including CCFR \cite{ccfr84}, CHDS \cite{cdhs}, and
NOMAD \cite{NOMAD}.  However, the improved quality of the first
principles prediction due to the new SPY secondary production data
\cite{SPY}, discussed in sec.~\ref{subprecisionflux}, 
should allow improvement of these limits.

\subsubsection{New Interactions}

\paragraph{\it Lepton Number Violation Searches}

~~~\\
~~~

The NuSOnG experiment possesses two valuable characteristics for the
search for lepton number violation.  First, it relies upon a high purity,
high intensity beam as its source of neutrinos; secondly, it employs an
instrumented detector optimized to measure inverse muon decay
with high accuracy.  An experiment with these two features naturally
lends itself to searches for the process:

\begin{equation}
\bar{\nu}_\mu + e^- \rightarrow \mu^- + \bar{\nu}_e . 
\end{equation}

This interaction is forbidden by the Standard Model since it
violates lepton family number conversation ($\Delta L_e = -\Delta
L_\mu = 2$).  As such, observation of this reaction would
immediately constitute direct observation of physics beyond the
Standard Model.

A number of theories beyond the Standard Model predict that lepton
number is not a true conserved quantum number; this means that processes that
violate lepton number are allowed to occur.  Theories which
incorporate multiplicative lepton number conservation~\cite{Feinberg,
  Ibarra:2004pe}, left-right symmetry~\cite{Herczeg:1992pt}, or the
existence of bileptons~\cite{Godfrey:2001xb} fall under this category.

The differential cross-section for lepton-violating processes can be
parametrized in the following form:

\begin{equation}
\frac{d \sigma}{dy} = \lambda \frac{G_F^2 s}{\pi} (A_V \cdot y(y-r) + A_S \cdot (1-r)),
\end{equation}

\noindent where $y$ is the fractional energy carried by the outgoing
lepton, $G_F$ the weak coupling constant, $s$ the square of the center of mass energy
of the system, and $r$  the threshold factor, defined as
$m_\mu^2/s$.  The parameters $\lambda, A_V,$ and $A_S$ describe the
strength of the reaction and whether the process is vector or scalar
in nature.  It is typical to compare this process to that of inverse
muon decay:

\begin{equation}
\frac{\sigma(\bar{\nu}_\mu e^- \rightarrow \mu^- \bar{\nu}_e)}{\sigma(\nu_\mu e^- \rightarrow \mu^- \nu_e)}
= \lambda \cdot (A_V \cdot (\frac{1+r/2}{3}) + A_S).
\end{equation}

The signature for such a reaction is the tagging of an $\mu^-$ during
antineutrino running with the same signature as expected from inverse
muon decays.  The main backgrounds to this reaction include (a)
$\nu_\mu$ contamination, (b) $\nu_e$ contamination, and (c) charge
misidentification of candidate events.  Our current estimates place
a very small beam contamination during antineutrino running: about
0.4\% contamination of $\nu_\mu$s and a 2.3\% contamination of
$\nu_e$ and $\bar{\nu}_e$ neutrinos (See Sec.~\ref{se:fluxcontent}).
Charge misidentification is expected to be very small, on the order
of 10$^{-5}$.  If we assume a conservative knowledge of the
backgrounds at the 5\% level, this would imply a limit on the lepton
number violation cross-section ratio of better than 0.2\% (at 90\%
C.L.) for V-A couplings and less than 0.06\% for scalar couplings.
Previous searches, based on $1.6 \times 10^{18}$ protons on target and
smaller target masses, have placed limits on this cross-section ratio
to less than 1.7\% at 90\% C.L. for V-A couplings and less than 0.6\%
for scalar couplings~\cite{Formaggio:2001jz}.  The NuSOnG experiment
can therefore reach an improvement of over an order of magnitude compared to
previous searches.  This limit can be improved if further selection
criteria are used in removing unwanted beam impurities or the
quasi-elastic background contamination.

\paragraph{\it Inverse Muon Decay}

~~~\\
~~~

The study of inverse muon decay, $\nu_{\mu} + e^- \rightarrow \mu^- +
\nu_e$ provides access to the helicity structure of the weak
interaction distinct from muon decay experiments.  The weak
interaction polarizes the incident $\nu_{\mu}$, making inverse muon
decay an excellent place to study departures from $V-A$ couplings.
For inverse muon decay,  $\sigma \propto (g_{L}^{V,e} g_L^{V,\mu})^2
(1-\epsilon)$ \cite{Fetscher:1986rf} where $\epsilon=h-(-1)$ and $h$ is the helicity of the
incident muon.  Ref. \cite{muonhelicity} has measured  $\epsilon < 4.1
\times 10^{-3}$ and the current limit on $g_{LL}^V = (g_{L}^{V,e}
g_L^{V,\mu})>0.96$ \cite{muonlimit}.  For a measurement of the total
cross section scaled to the predicted cross section, the uncertainty
on the coupling is $g_{LL}^V=(1/2)\sigma_{\sigma}/\sigma_{SM}$.

For NuSOnG, we expect $>200$k inverse muon decay events, which would
give a statistical uncertainty of 0.002 on $g_{LL}^V$. However, we
will need to determine the neutrino flux.  Taking the $\nu_{\mu} + e^-
\rightarrow \nu_{\mu} +e^-$ cross section as {\em known} gives the
neutrino flux to 0.7\%.  Since we plan to use the inverse muon decay
events for determining the flux for the electroweak measurements,
NuSOnG will need to measure the efficiency and fiducial volume for
both processes to better than 0.7\%.  Combined with other systematics,
we should be able to achieve an total uncertainty of about 1-2\% on
$g_{LL}^V$, an improvement by a factor of four.

The key background will come from CCQE events that have small hadronic
energy.  We expect our high granularity will allow us to keep the
systematic error from this source well below 1\%, but this needs
study. 

Obviously, the manner of analysis described above is somewhat questionable.  Ultimately, one
would want to carry out a combined analysis of both neutrino elastic
scattering on electrons and quarks and of inverse muon decay in the
context of a specific model which relates the charged and neutral
current coupling constants. For such an analysis, 1-2\% uncertainty
should still be achievable.

\subsubsection{New Particles}

\paragraph{\it Long-lived, Light Neutral Heavy Leptons}

~~~\\
~~~


Another interesting NuTeV result arose from the search for long-lived,
light ($<15$ GeV) neutral heavy leptons.  This was performed in a
helium-filled decay region located upstream of the calorimeter.  In
the mass region of 2.2-15 GeV, NuTeV has a small expected background
(0.07 $\pm$ 0.01 events), but observed three events.  All events had
two muons originating from a vertex within the helium decay region and
missing energy.  \cite{NuTeV3ev}.

Since publication in 2001, no widely accepted explanation has been
found.  In 2006, D$0$ published a search for a similar decay signature
in proton-antiproton interactions \cite{D0search}.  No events were
found and some production models were excluded.  The most viable
remaining model is by Dedes {\it et al.}, which hypothesizes that the
events are from decay of long-lived neutralinos. These are produced in
the NuTeV beam dump through $B$ hadron decays \cite{Dedes}.  No other
experiment has been able to match NuTeV's running conditions to
further explore this intriguing result.

NuSOnG can address the question by including a low-mass
(helium-filled) decay region between the calorimeter segments.  Assuming
parameters similar to those of NuTeV (except for a 20-fold increase in the number
of protons on target), NuSOnG would expect to see 60 events with
an expected background of 1-2 events.  The sensitivity would scale
directly with the decay volume, so the increased length compared to
NuTeV (26 $m$ $\rightarrow$ $\approx$40 $m$) would increase this to 90
signal events over a 2-3 event background.  Observing no signal would finally 
settle this outstanding question.

These decay regions allow exploration for a signal from a beyond-the-Standard-Model 
particle in other decay modes as well; other interesting
modes include $\mu \pi$, $\mu e$, $e \pi$ and $ee$.  NuSOnG's sensitivity
to other new particles is similarly improved over NuTeV by the
increase in beam intensity and decay volume, allowing us to study
new regions of phase space.

\paragraph{\it Muonic Photons}

~~~\\
~~~

In the mid-1990's there was interest in searching for ``leptonic
photons'' -- massless vector particles that couple according to
flavor.  Electronic, muonic, and tauonic photons, $\gamma_e$,
$\gamma_\mu$, and $\gamma_\tau$ were introduced \cite{hep-ph/9512435}.
Production occurs in secondary meson decays such as $\pi \rightarrow
\nu_\mu \mu \gamma_\mu$, and detection can proceed through $\gamma_\mu
+ Z \rightarrow \mu^+ + \mu^- + Z$, where $Z$ is the charged nucleus.
These events have small missing $p_T$ compared to the ``trident''
background,   $\nu + Z \rightarrow
\nu + \mu^+ \mu^-+ Z$.
The search by CHARM II sets the best limit at $1.6\times10^{-6}$
\cite{CHARMIImuonic}.

Since this time, neutrino oscillations have been confirmed (see
references \cite{Cleveland:1998nv} through \cite{kamland}).  This
complicates the theory of ``muonic photons,'' since, in this case,
lepton flavor-charge is not conserved.  As pointed out in reference
\cite{hep-ph/9512435}, a theory with a non-conserved charge cannot have
massless vector particles and a Coulomb-like potential.  It appears
very difficult to evade this problem.

Nevertheless, NuSOnG should search for these events.  With higher rate
and better segmentation than CHARM II, NuSOnG should have sensitivity
in the range of $\sim 10^{-7}$.  A significant excess would be quite
startling.

%% file: DIS_PDF_NucA.tex

The Deeply Inelastic Scattering (DIS) process provides crucial information
about the structure of the proton which is used to determine the Parton
Distribution Functions (PDFs). For example, in the recent CTEQ6HQ
analysis, DIS data accounted for more than two-thirds of the data
points used in the analysis.\footnote{Specifically, 
there were 1333 DIS data points used out of the 1925
total.\cite{Kretzer:2003it}} 
As such, the DIS measurements form the foundation for 
the many calculations which make use of the PDFs. 

In the basic DIS process, leptons scatter from hadrons via the exchange
of an intermediate vector boson: $\{\gamma,W^{\pm},Z\}$. Different
boson probes couple to the hadrons with different factors, and it
is important to combine data from these different probes to separate
the different flavor components in the hadron. Unfortunately, three
of the four DIS probes $\{ W^{\pm},Z\}$ have a (relatively) large
mass and couple only weakly; this introduces a number of complications: 

\begin{itemize}
\item  The statistics for these weak processes are limited as
compared with the photon-exchange processes. 
\item To compensate for the weak cross section, typically heavy nuclear
targets ({\it e.g.}, Fe and Pb) are used; this introduces nuclear corrections
when the results are scaled from the heavy target back to proton or
isoscalar targets.
\end{itemize}

The NuSOnG experiment will generate high statistics ($>100$M DIS events)
measurements on an intermediate atomic-weight nuclear target (SiO$_{2}$).
This will provide precise information on the linear
combinations of PDFs which couple to the weak charged currents
($W^{\pm}$), which can significantly improve the parton distribution fits.
In this section, we first introduce the basics of DIS and 
the connection to parton distribution functions.   
Then we concentrate on three aspects of parton distribution
studies where NuSOnG can make a unique contribution to the physics:
\begin{itemize}
\item Improved understanding of nuclear effects in neutrino scattering.
\item Study of Charge Symmetry Violation
\item Measurement of the Strange Sea
\item Measurement of $\sigma^\nu$ and $\sigma^{\bar \nu}$
\end{itemize}
The latter two items are directly relevant to the electroweak studies
proposed for NuSOnG (see Sec. \ref{smexplain}).

\subsubsection{Deep Inelastic Scattering and Parton Distribution Functions}
\label{subsubdis}

The differential cross
section for neutrino DIS depends on three structure functions:
$F_2$, $xF_3$ and $R_L$.  It is given by:
\begin{eqnarray}
\frac{d^2\sigma ^{\nu (\overline{\nu })N}}{dxdy} &=&\frac{G_F^2ME_\nu }{\pi
\left( 1+Q^2/M_W^2\right) ^2}  \nonumber \\
&&\left[ F_2^{\nu {(\overline{\nu })}N}(x,Q^2)\left( \frac{y^2+(2Mxy/Q)^2}{%
2+2R_L^{\nu {(\overline{\nu })}N}(x,Q^2)}+1-y-\frac{Mxy}{2E_\nu }\right)
\right.   \nonumber \\
&&\left. \pm xF_3^{\nu {(\overline{\nu })}N}y\left( 1-\frac y2\right)
\right] ,  \label{eq:sigsf}
\end{eqnarray}
where the $\pm $ is $+(-)$ for $\nu (\overline{\nu })$ scattering.  In
this equation, $x$ is the Bjorken scaling variable, $y$ the
inelasticity, and $Q^2$ the squared four-momentum transfer.

The function $xF_3(x,Q^2)$ is unique to the DIS cross section for the weak
interaction.  It originates 
from the parity-violating term in the product of the leptonic and hadronic
tensors.    For an isoscalar target, in the quark-parton model,
\begin{eqnarray}
xF_{3}^{\nu N}(x) &=&x\left( u(x)+d(x)+2s(x)\right. \\    \nonumber
&&\left. -\bar{u}(x)-\bar{d}(x)-2\bar{c}(x)\right) ,  \\
xF_{3}^{\bar{\nu}N}(x) &=&xF_{3}^{\nu N}(x)-4x\left(
s(x)-c(x)\right) .  \label{nubarxF3}
\end{eqnarray}
Defining $xF_3=\frac 12(xF_3^{\nu N}+xF_3^{\bar{\nu}N})$, 
at leading order in QCD, 
\begin{equation}
xF_{3,LO}=\sum_{i=u,d..}xq(x,Q^2)-x\overline{q}(x,Q^2).
\label{DIS xf3 definition}
\end{equation}
To the level that the sea quark distributions have the same $x$
dependence, and thus cancel, $xF_3$ can be thought of as probing the
valence quark distributions. 
The difference between the neutrino and
antineutrino parity violating structure functions, $\Delta
(xF_3)=xF_3^{\nu N}-xF_3^{\bar{\nu}N}$, probes the strange and 
charm seas.

Analogous functions for $F_2(x,Q^2)$ and $R_L(x,Q^2)$ appear in both the
cross section for charged lepton ($e$ or $\mu$) DIS  and the cross section
for $\nu$ DIS.  At leading order, 
\begin{equation}
F_{2,LO}=\sum_{i=u,d..}e^2 (xq(x,Q^2)+x\overline{q}(x,Q^2)),
\label{DIS F_2 definition}
\end{equation}
where $e$ is the charge associated with the interaction.  In the weak
interaction, this charge is unity. For charged-lepton scattering
mediated by a virtual photon, the fractional electromagnetic charge of
each quark flavor enters.  Thus $F_2^{\nu N}$ and $F_2^{e(\mu)N}$ are
analogous but not identical and comparison yields useful information
about specific parton distributions \cite{Arneodo:1996qe}.
$R_L(x,Q^2)$ is the longitudinal to transverse virtual boson
absorption cross-section ratio.  The best measurements for this come
from charged lepton scattering rather than neutrino scattering.  In
the past, neutrino experiments have used the charged lepton fits to
$R_L$ as an input to the measurements of $xF_3$ and
$F_2$\cite{hep-ex/000904}.  This, however, is just a matter of the
statistics needed for a global fit to all of the unknown structure
functions in $x$ and $Q^2$ bins \cite{hep-ex/9806023}.  With the high
statistics of NuSOnG, precise measurement of $R_L$ will be
possible from neutrino scattering for the first time.

In addition to fitting to the inclusive DIS sample,
neutrino scattering can also probe parton distributions through
exclusive samples.  A unique and important case is the measurement of
the strange sea through opposite sign dimuon production.  When the
neutrino interacts with an $s$ or $d$ quark, it produces a charm quark
that fragments into a charmed hadron. The charmed hadron's
semileptonic decay (with branching ratio $B_c\sim 10\%$ ) produces a second muon of
opposite sign from the first:
\begin{eqnarray}
\nu _\mu \;+\;{\rm N}\;\longrightarrow \;\mu ^{-}\!\! &+&\!c\;+\;{\rm X} \\
&&\!\!\hookrightarrow s\;+\;\mu ^{+}\;+\;\nu _\mu.
\end{eqnarray}
Similarly, with antineutrinos, the 
interaction is with an $\overline{s}$ or $\overline{d}$,
\begin{eqnarray}
\overline{\nu }_\mu \;+\;{\rm N}\;\longrightarrow \;\mu ^{+}\!\! &+&\!%
\overline{c}\;+\;{\rm X} \\
&&\!\!\hookrightarrow \!\overline{s}+\;\mu ^{-}\;+\;\overline{\nu }_\mu .
\end{eqnarray}
The opposite sign of the two muons can be determined for those events
where both muons reach the toroid spectrometer.   Study of these 
events as a function of the kinematic variables allows extraction of 
the strange sea, the charm quark mass, the charmed particle branching ratio
($B_c$), and the Cabibbo-Kobayashi-Maskaka matrix element, $|V_{cd}|$.

For a more in-depth review of precision measurement of parton 
distributions in neutrino scattering, see ref.~\cite{Conrad:1997ne}.

\subsubsection{Nuclear Effects}
\label{subsubnukes}

Historically, neutrino experiments have played a major role in expanding 
our understanding of parton distribution functions through
high statistics experiments such as CCFR~\cite{hep-ex/000904}, 
NuTeV~\cite{nutev}, and CHORUS~\cite{chorus}. 
However, the high statistics extract a price since 
the large event samples require the use of nuclear targets -- iron in the 
case of both CCFR and NuTeV and lead in the case of the Chorus experiment. 
The problem is that if one wants to extract information on {\it nucleon} 
PDFs, then the effects of the nuclear targets must first be removed. 
NuSOnG can provide key measurements which will improve these corrections.

In the case of charged lepton deep inelastic scattering, there are data 
available from nuclear targets covering the range from deuterium through iron 
and beyond. Thus, it has been possible to perform detailed studies of 
the $A$-dependence as a function of $x \ \mbox{and}\  Q^2$ from
both the cross section and the structure function $F_2$. Such is not the 
case in $\nu \ \mbox{and}\ \overline \nu$ interactions where 
the corrections can be different for both cross sections or, equivalently, 
for $F_2\ \mbox{and}\ xF_3$. In this case one must rely on theoretical models 
of the nuclear corrections. This is an unsatisfactory situation since one 
is essentially measuring quantities sensitive to the convolution of the 
the desired PDFs and unknown -- or model dependent -- nuclear corrections. 

It is important to address the question of nuclear effects in neutrino
scattering so that the neutrino data can be used in fits without
bringing in substantial uncertainties.  For example, in a recent
analysis~\cite{CTEQ} the impact of new neutrino data on global fits
for PDFs was assessed. The conclusion reached in this analysis was that the
uncertainties associated with nuclear corrections precluded using the
neutrino data to constrain the nucleon PDFs.  If this uncertainty is
addressed, the neutrino data will be a powerful addition to these
fits.

Furthermore, nuclear effects are interesting in their own right.
Comparison of the charged and neutral lepton scattering data
can provide clues to the sources of the major features which appear
in nuclear effects:  shadowing, antishadowing, and the EMC effect.
There is phenomenological 
evidence which suggests that the nuclear corrections for the $\nu 
\ \mbox{and}\ \overline\nu$ cross sections might be rather similar and, 
in both cases, somewhat smaller than the corresponding corrections in 
charged lepton deep inelastic scattering. These latter two observations 
differ from the pattern suggested by the theoretical model~\cite{kp} 
for nuclear corrections used in the analysis. 

\begin{figure}
\centering
\includegraphics[height=4 in]{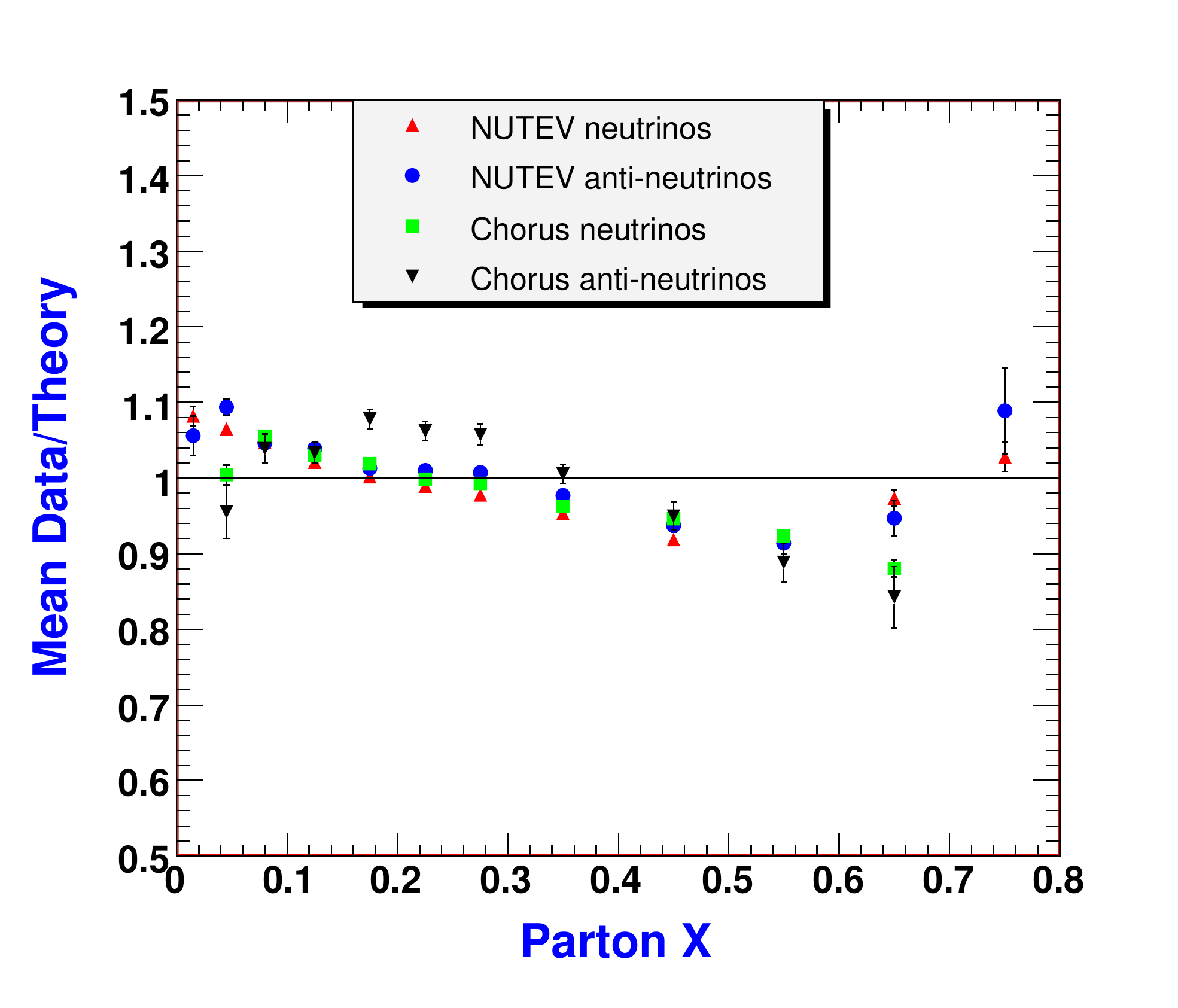}
\caption{Comparison between the reference fit and the unshifted Chorus and
NuTeV neutrino data without any nuclear corrections.}
\label{no_shift_no_cor}
\end{figure}

Fig. \ref{no_shift_no_cor} shows some results from Ref.~\cite{CTEQ} in 
the form of ``data/theory'' averaged over $Q^2$ and presented versus $x$. 
The results are from a global fit but are plotted {\it without} the model-dependent 
nuclear corrections which were used in the fits. What is striking is 
the similarity of the $\nu \ \mbox{and}\ \overline \nu$ results, and the 
overall pattern of deviations, similar to that seen in charged lepton
DIS, although the deviations from unity are somewhat smaller.  
It is interesting to note that 
there is no clear indication of the turnover at low $x$ which is 
observed in charged lepton scattering, 
called shadowing.   However, this may be due
to kinematic limits of the measurements.

To make progress in understanding nuclear corrections in 
neutrino interactions, access to high-statistics data on a variety of nuclear 
targets will be essential. This will allow the $A$-dependence to be 
studied as a function of both $x \ \mbox{and}\ Q^2$, as has been done in 
charged lepton deep inelastic scattering. PDFs from global fits without the 
neutrino data can then be used to make predictions to be compared with the 
$A$-dependent $\nu \ \mbox{and}\ \overline\nu$ cross sections, thereby 
allowing the nuclear corrections to be mapped out for comparison with 
theoretical models.
 
The primary target of NuSOnG will be SiO$_2$. However, we can address
this issue by replacing a few slabs of glass with alternative target
materials: C, Al, Fe, and Pb. This range of nuclear targets would both
extend the results of Miner$\nu$a to the NuSOnG kinematic region, and
provide a check (via the Fe target) against the NuTeV measurement.

Given the NuSOnG neutrino flux, we anticipate $58k$ $\nu$-induced and
$30k$ $\bar\nu$-induced CC DIS events per ton of material.  A single
ton would be sufficient to extract $F_{2}(x)$ and $xF_{3}(x)$ averaged
over all $Q^{2}$; a single 5 m$\times$5 m$\times$2.54 cm slab of any of
the above materials will weigh more than that. The use of additional
slabs would permit further extraction of the structure functions into
separate $(x,Q^{2})$ bins as was done in the NuTeV analysis, at the
potential expense of complicating the shower energy resolution in the
sub-detectors containing the alternative targets; this issue will be
studied via simulation.

Table~\ref{ta:nuclear} shows that two 50-module stacks would be
sufficient to accumulate enough statistics on alternative nuclear
targets for a full structure-function extraction for each
material. However, for basic cross-section ratios in $x$, a single
slab of each would suffice.

\begin{table}
\centering
\begin{tabular}{c c c}
Material & Mass of                      & Number of slabs needed  \\
               & 2.54 cm slab (tons)  &  for NuTeV-equivalent statistics \\
\hline
C & 1.6 & 33 \\
Al & 1.9 & 27 \\
Fe & 5.5 & 10 \\
Pb & 7.9 & 7 \\
\hline

\end{tabular}
\caption{Alternative target materials for cross-section analysis}
\label{ta:nuclear}
\end{table}

\subsubsection{Isospin Violations}
\label{subsubsec:isospin}

When we relate DIS measurements from heavy targets such as
${}_{26}^{56}$Fe or ${}_{82}^{207}$Pb back to a proton or isoscalar
target, we generally make use of isospin symmetry where we assume that
the proton and neutron PDFs can be related via a $u \leftrightarrow d$
interchange.  While  isospin symmetry is elegant and well-motivated, 
the validity of this exact
charge symmetry  must ultimately be established by experimental measurement.
There have been a number of studies investigating 
isospin symmetry violation \cite{Boros:1999fy,Ball:2000qd,Kretzer:2001mb,Martin:2001es};
therefore, it is important to be aware of the magnitude of potential 
violations of isospin symmetry and the consequences on 
the extracted PDF components. 
For example, 
the naive parton model relations are modified if we have a violation
of exact $p\leftrightarrow n$ isospin-symmetry, (or charge symmetry);
\textit{e.g.}, 
$u_{n}(x)\not\equiv d_{p}(x)$ and 
$u_{p}(x)\not\equiv d_{n}(x)$.

Combinations of structure functions can be particularly sensitive
to isospin violations, and NuSOnG is well suited to measure some of 
these observables. 
For example, residual $u,d$-contributions
to $\Delta xF_{3}=xF_{3}^{\nu}-xF_{3}^{\bar{\nu}}$ from charge
symmetry violation (CSV) would be amplified due to enhanced valence
components $\{ u_{v}(x),d_{v}(x)\}$, and because the $d\rightarrow u$
transitions are not subject to slow-rescaling corrections which strongly
suppress the $s\rightarrow c$ contribution to 
$\Delta xF_{3}$. \cite{Kretzer:2001mb}
Here the ability of NuSOnG to separately measure 
$xF_{3}^{\nu}$
and 
$xF_{3}^{\bar{\nu}}$
over a broad kinematic range will provide powerful constraints on the
sensitive  structure function combination  $\Delta xF_{3}$.

There are a wide variety of models that study
CSV \cite{Boros:1999fy,Ball:2000qd,Kretzer:2001mb,Martin:2001es}.  One
method to quantify possible CSV contributions is via a one-parameter
``toy'' model where we characterize the CSV as a rotation in isospin
space: $q_{n}^{{\rm CSV}}=N_{q}\,\sum_{q^{\prime}}\,
R_{qq^{\prime}}(\theta)\, q^{\prime}_{p}$, where $R$ is a rotation
matrix, and $N_{q}$ is the normalization factor.  For example, the
$u$-distribution in the neutron can be related to the proton
distributions via:
\begin{equation}
u_{n}^{{\rm CSV}}(x,Q^{2})=N_{u}^{2}
\left[
\cos^{2}(\theta)\ u_{p}(x,Q^{2}) +
\sin^{2}(\theta)\ d_{p}(x,Q^{2})\right]
\quad . 
\end{equation}
For $\theta=\pi/2$, we recover the symmetric limit 
$u_{p}(x,Q^{2})=d_{n}(x,Q^{2})$.
While this parameterization does not offer any 
explanation for the source of the
CSV, it does provide a simple one-parameter ($\theta$)
characterization which is flexible enough to quantify the range of CSV
effects. (For more details, {\it cf.} Ref.~\cite{Kretzer:2001mb}.)

At present, there are constraints on isospin violation from a number
of experiments which cover different ranges of $x$ and $Q^2$.
  For example, we note that while the above ``toy'' model leaves the
neutron singlet combination $(q+{\bar{q}})$ invariant at the
$\lesssim2\%$ level in the region $x\,\epsilon\,[0.01;0.1]$, it would
lower the NC observable
$\left[\frac{4}{9}(u+{\bar{u}})+\frac{1}{9}(d+{\bar{d}})\right]_{n}$
in this region by about 10\%. An effect of this size would definitely
be visible in the NMC $F_{2}^{n}/F_{2}^{p}$ data which has an
uncertainty of order a few percent.\cite{Arneodo:1996qe}
 The measurement of the lepton charge asymmetry in W decays from the
Tevatron \cite{Abe:1998rv,Bodek:1999bb} places tight constraints on
the up and down quark distributions in the range $0.007<x<0.24$.
While only strictly telling us about parton distributions in the
proton, these data rule out isospin violations at the 
$5\%$ level, as demonstrated in Ref.~\cite{Bodek:1999bb}.
 In addition, there are also fixed-target Drell-Yan experiments such as
NA51  \cite{Baldit:1994jk} and E866 \cite{Hawker:1998ty} which precisely
measure $\bar{d}/\bar{u}$ in the range $0.04<x<0.27$; these are also
sensitive to isospin-violating effects.

NuSOnG will be able to provide high statistics DIS measurements across
a wide $x$ range. Because the target material (SiO$_2$) is very nearly
isoscalar, this will essentially allow a direct extraction of the
isoscalar observables.  Consequently, if isospin violations are
present, they can be measured more precisely than would be the case on a 
highly non-isoscalar target.

\subsubsection{Measurement of the Strange Sea}
\label{subsec:strangesea}

There are several reasons why an improved measurement of the
strange sea is of interest.
First, it contributes to the low-$Q^2$ properties of the nucleon in
the nonperturbative regime -- a question of practical as well as
intellectual interest, since many precision oscillation experiments
are being performed in the 1 to 20 GeV (hence, nonperturbative) range.
It is critical for 
charm production which  provides an important testing ground for NLO QCD
calculations.  In addition, understanding the threshold behavior
associated with the heavy charm mass is of interest to future neutrino
experiments.

Distinguishing
the difference between the $s(x)$ and $\bar s(x)$ distributions,
\begin{equation}
 xs^-(x)\equiv xs(x)-x\overline{s}(x),
\end{equation}
is even more important, and poses additional challenges.  First, it is of intrinsic interest in nucleon
structure
models \cite{Barone:1999yv,Davidson:2001ji,McFarland:2003jw,McFarland:2002rk,Strumia:2003pc}.
Second, the integrated strange sea asymmetry,
\begin{equation}
 S^- \equiv \int_0^1 s^-(x) dx,
\end{equation}
has important implications for the
precision measurement of the weak mixing angle in deep inelastic
scattering of neutrinos ({\it cf.}
Sec.~\ref{nutevsection} and references \cite{Zeller:2002du,Barone:1999yv,McFarland:2003jw,Davidson:2001ji,Zeller:2002du,McFarland:2002rk,Olness:2003wz}).
This was not recognized at the time of the NuTeV $\sin^2 \theta_W$
publication; an error due to $S^-$ will be included in the
NuTeV reanalysis, to be presented in late summer 2007
\cite{SamPrivateComm}.

Historically, information on the $s(x)$ and $\bar{s}(x)$ distributions
was derived from inclusive cross sections for neutral and charged
current DIS via $\Delta(xF_3)$.
These analyses made the implicit assumption that
the $s(x)$ and $\bar{s}(x)$ seas had the same distribution in $x$.
Because the strange sea is relatively small compared to the dominant
$u(x)$ and $d(x)$ processes, the resulting uncertainties on the
strange sea components were large.
For example, the strangeness
content of the nucleon, as measured by the momentum fraction carried
by $s$ or $\overline{s}$, is of order 3\% at $Q=1.5$ GeV.  For this
reason, the strange PDF was typically parametrized using the ansatz
$s(x)=\bar{s}(x)=\kappa(\bar{u}+\bar{d})/2$, where $\kappa$ measured
the deviation from $SU(3)$ flavor symmetry at some low value of $Q$.

Introducing information from opposite sign dimuon production
allows substantial improvement in the strange PDF measurement.
Neutrino-induced dimuon production,
$(\nu/\bar{\nu})N\rightarrow\mu^{+}\mu^{-}X$, proceeds primarily
through the sub-processes $W^{+}s\to c$ and $W^{-}\bar{s}\to\bar{c}$
(respectively), so this provides a mechanism to directly probe the
$s(x)$ and
$\bar{s}(x)$ distributions without being overwhelmed by the larger $u(x)$
and $d(x)$ distributions.
Hence, the recent high-statistics dimuon measurements
\cite{Bazarko:1994tt,Goncharov:2001qe,Tzanov:2003gq,Vilain:1998uw,Astier:2000us}
play an essential role in constraining the strange component of
the proton.

\begin{figure}[t]
\vspace{-0.5in}
\centering
\scalebox{0.4}{\includegraphics{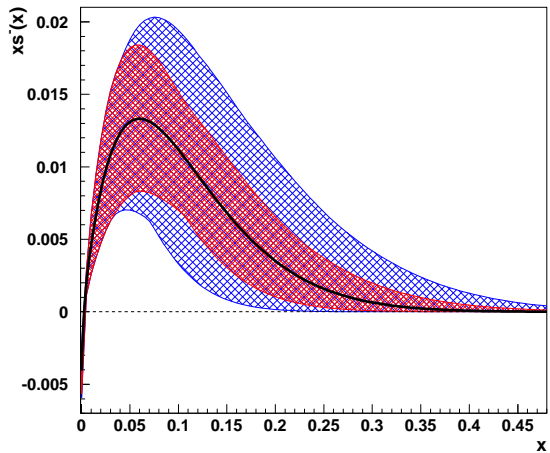}}
\caption{\label{sasym} $xs^-(x)$ vs $x$ at $Q^2=16$ GeV$^2$.  Outer band
is combined errors, inner band is without
$B_c$ uncertainty.}
\end{figure}

The highest precision study of $s^-$ to date is from the NuTeV
experiment \cite{MasonPRL}.  The sign selected beam allowed measurement
of the strange and antistrange seas independently, recording 5163
neutrino-induced dimuons, and 1380 antineutrino-induced dimuon events
in its iron target.  Figure \ref{sasym} shows the measured asymmetry
between the strange and antistrange seas.  With more than 100 times
the statistics of NuTeV, NuSOnG will have substantially finer binning.

\begin{figure}[h]
\centering
\scalebox{0.4}{\includegraphics{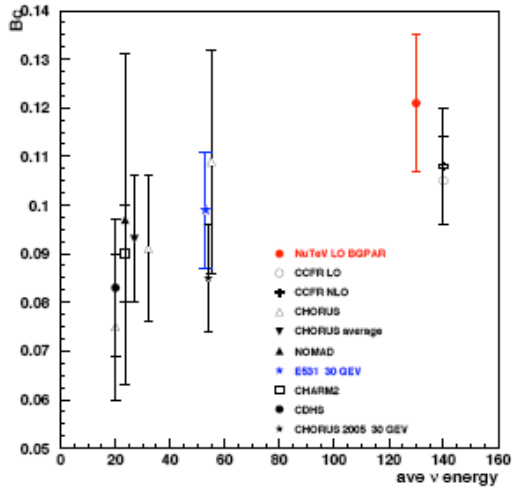}}
\caption{\label{Bc} World measurements of $B_c$.}
\end{figure}

The integrated strange sea asymmetry from NuTeV has a positive central
value: $0.00196 \pm 0.00046$ (stat) $\pm 0.00045$ (syst)
$^{+0.00148}_{-0.00107}$ (external).  The ``external" error on the
measurement is dominated by the error on the average charm
semi-muonic branching ratio, $B_c$ which is determined by
other experiments.  This error currently is about
10\%.  A rescan of Chorus data, which would increase the statistics,
is under consideration \cite{NakamuraPrivateComm}.

The key to an improved result on the strange sea from NuSOnG is in a
more precise measurement of $B_c$ at NuSOnG energies.  This can be
accomplished in two ways.  First, the very high statistics of NuSOnG
allow for an accurate fit to $B_c$ and the $s$ and $\bar s$
distributions simultaneously.  Second, we plan to incorporate an
emulsion detector into the design. The emulsion 
will be scanned by the Nagoya University group.  This group
has substantial expertise, having provided the emulsion and scanning for
Chorus, DoNuT and other emulsion-based experiments.  The goal
will be to obtain $>10$k events in the emulsion during the NuSOnG run.

Beyond this, we will also consider placing a liquid argon TPC of
similar size to microBooNE \cite{muBooNE} (70 tons fiducial volume) or
even Gargamelle (20 tons fiducial volume) in the gap between two of
the NuSOnG subdetectors to directly measure $B_c$.  If one were to,
for example, use a volume comparable to the Gargamelle bubble chamber,
we could observe on the order of one million charged current events
within it for $5 \times 10^{19}$ POT.  This would yield approximately
100,000 events with charm in the final state, and about 10,000 dimuon
events.  

In addition to an improved measurement of $B_c$, the more finely-grained
liquid argon TPC and/or emulsion detectors could be used to aid the calibration of
the four glass detector modules by measuring any differences between
hadron and electron showers from pion and electron beams versus those
within a neutrino induced event.  Coupled with the precision test beam, 
it may also be possible to improve understanding of the background due to muons produced
by pion and kaon decays in the hadron shower.  An improved parameterization of
this background, currently from a CCFR measurement 
\cite{Sandler:1990ck,Sandler:1992wj,Sandler:1992ai}
could help extend the kinematic range of charmed dimuon measurements beyond what
was possible for the NuTeV and CCFR experiments.

\subsubsection{Measurement of the Total Cross Section}
\label{subsubxsec}

Precision measurement of the total neutrino and antineutrino cross
sections at high energies will be valuable to a future neutrino
factory experiment which seeks to make precision measurement of CP
violation.  Because NuSOnG can measure the flux to 0.5\% (see
Sec.~\ref{subprecisionflux}, precise measurements can be made.  Also,
higher accuracy on the ratio of $\sigma^{\bar \nu}/\sigma^{\nu}$
will also improve the electroweak measurement (see
Tab.\ref{NuTeVerrs}).

\begin{figure}[t]
\centering
\vspace{-2.in}
\scalebox{0.4}{\includegraphics{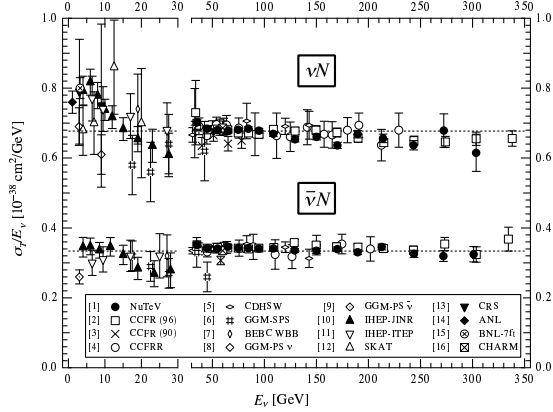}}
\caption{\label{totxsec} World measurements of the total $\nu$ and
  $\bar \nu$ cross sections.  See references \cite{nutev} and
  \cite{xsec2} through \cite{xsec16}.}
\end{figure}

Fig.~\ref{totxsec} shows $\sigma/E_{\nu }$ for the muon neutrino and
antineutrino charged-current total cross-section as a function of
neutrino energy (\cite{nutev} and \cite{xsec2}-\cite{xsec16}).  The
error bars include both statistical and systematic errors.  The
results are from a wide range of target materials, but the experiments
with the smallest errors and largest energy range used iron.  The
straight lines are the isoscalar-corrected total cross-section values
averaged over 30-200 $GeV$ as measured by the experiments in
Refs. \cite{xsec3} to \cite{xsec5}.  The fit \cite{courtesyseligman}
gives: $\sigma ^{\nu \,Iso}/E_{\nu }=(0.677\pm 0.014)\times
10^{-38}cm^{2}/GeV$; $\sigma ^{\bar\nu \,Iso}/E_{\bar\nu }=(0.334\pm
0.008)\times 10^{-38}$ cm$^{2}/$ GeV. The average ratio of the
antineutrino to neutrino cross-section in the energy range 30-200
GeV is $\sigma ^{\bar\nu \,Iso}/\sigma ^{\nu \,Iso}=0.504\pm0.003$
as measured by Refs. \cite{nutev} and \cite{xsec2}-\cite{xsec5}.  Note the change in
the energy scale at 30 GeV.

The most precise measurements are systematics limited.  The largest
contributions to the systematics in recent experiments (CCFR, NuTeV)
come from flux normalization, the model parameterization used in
determination of the flux, and the charm mass used to parameterize
charm threshold.  NuSOnG measures the neutrino flux normalization to
high precision via the IMD events.  Also, as discussed in
Sec.~\ref{subsubfixednu}, NuSOnG's high statistics allow cuts which
substantially improve the model parameterization error.  Lastly, the charm
mass, $m_c$ is expected to be improved from the high statistics fits
to the opposite sign dimuon events described in the previous section.
While more study is needed, it likely that NuSOnG can substantially
improve on the world measurements of the total cross section and the
cross section ratios.

%% file: FluxContent_v1A.tex
\subsection{The Neutrino Flux}
\label{se:fluxcontent}

\begin{figure}
\centering
\includegraphics[width=7in]{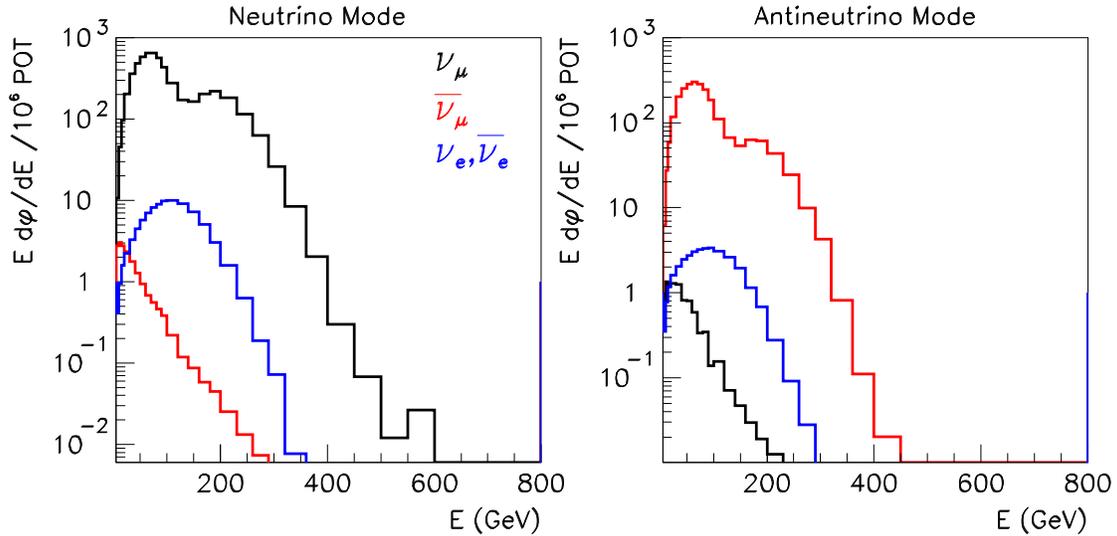}
\vspace{-10cm}
\caption{NuSOnG flux in neutrino mode (left) and antineutrino mode
  (right).  Black: muon neutrino flux, red: muon antineutrino flux,
  blue: electron neutrino and antineutrino flux}
\label{Beam}
\end{figure}

For the purposes of this expression of interest, we assume the same
SSQT design as was used at NuTeV.  The resulting neutrino
(antineutrino) flux \cite{SamThesis} is shown in Fig.~\ref{Beam}, left
(right).  The $\nu_\mu$ flux is shown in black, $\bar \nu_\mu$ in red,
and $\nu_e + \bar \nu_e$ in blue.  The shape of the flux is dominated
by the dichromatic neutrino spectrum from $\pi$ and $K$ two-body decay.

In neutrino mode, 98.2\% of neutrino interactions are due to $\pi^+$
and $K^+$ secondaries, while in antineutrino mode 97.3\% come from
$\pi^-$ and $K^-$.  The ``wrong sign'' content is very low, with an
0.03\% antineutrino contamination in neutrino mode and 0.4\% neutrino
contamination in antineutrino mode.  The electron-flavor content is
1.8\% in neutrino mode and 2.3\% in antineutrino mode.  The major
source of these neutrinos is $K^{\pm}_{e3}$ decay, representing 1.7\%
of the total flux in neutrino mode, and 1.6\% in antineutrino mode.
Other contributions come from $K_{Le3}$, $K_{Se3}$, charmed meson,
muon, $\Lambda_C$, $\Lambda$, and $\Sigma$ decays.

Precise knowledge of the electron-flavor content is crucial for many
NuSOnG analyses.  The largest source of error in the knowledge of the
electron-flavor content in NuTeV was from the $K^{\pm}_{e3}$ branching
ratio, which led to an error on $\nu_e$ content of 1.4\%
\cite{SamThesis}.  While the other sources of $\nu_e$s have large
fractional errors, they constitute a much smaller fraction of the
flux.  An error of 1.5\% for the electron-flavor contamination,
consistent with NuTeV, will be assumed for NuSOnG.

%% file: EventContent_v3A.tex
\subsection{Event Rates}

The {\it approximate} event rates presented here serve to set the
scale for the physics case presented in this document.  They are based
on running the Nuance event generator \cite{Nuance} with the NuTeV
flux, and then scaling to the expectations of NuSOnG assuming a 3 kton
fiducial mass. Some simplifying assumptions, which will be corrected
as the simulation develops, have been made.  For example, C$_2$ is
used as a target rather than SiO$_2$.  Also, note that Nuance is not
yet tuned as a high energy event generator.  Thus, these event rates
are only representative.

For neutrino running, approximate event rates for $5\times 10^{19}$ protons
are:
\begin{eqnarray*}
507\/k &~& {\rm \nu_\mu~CC~quasi-elastic~scatters} \\ 
178\/k &~& {\rm \nu_\mu~NC-elastic~scatters} \\ 
1016\/k &~& {\rm \nu_\mu~CC~\pi^+} \\
302\/k &~& {\rm \nu_\mu~CC~\pi^0} \\
272\/k &~& {\rm \nu_\mu~NC~\pi^0} \\
226\/k &~& {\rm \nu_\mu~NC~\pi^\pm} \\
1379\/k &~& {\rm \nu_\mu~CC~and~NC~Resonance~multi-pion} \\
202\/M &~& {\rm \nu_\mu~CC~Deep~Inelastic~Scattering} \\
63\/M &~& {\rm \nu_\mu~NC~Deep~Inelastic~Scattering} \\
24\/k &~& {\rm \nu_\mu~neutrino-electron~NC~elastic~scatters} \\
235\/k &~& {\rm \nu_\mu~neutrino-electron~CC~quasielastic~scatters} (IMD)
\end{eqnarray*}

For antineutrino running, which assumes $1.5\times 10^{20}$ protons
on target, approximate event rates are:
\begin{eqnarray*}
548\/k &~& {\rm \bar \nu_\mu~CC~quasi-elastic~scatters} \\ 
195\/k &~& {\rm \bar \nu_\mu~NC-elastic~scatters} \\ 
1103\/k &~& {\rm \bar \nu_\mu~CC~\pi^+} \\
321\/k &~& {\rm \bar \nu_\mu~CC~\pi^0} \\
297\/k &~& {\rm \bar \nu_\mu~NC~\pi^0} \\
246\/k &~& {\rm \bar \nu_\mu~NC~\pi^\pm} \\
1516\/k &~& {\rm \bar \nu_\mu~CC~and~NC~Resonance~multi-pion} \\
102\/M &~& {\rm \bar \nu_\mu~CC~Deep~Inelastic~Scattering} \\
36\/M &~& {\rm \bar \nu_\mu~NC~Deep~Inelastic~Scattering} \\
21\/k &~& {\rm \bar \nu_\mu~neutrino-electron~NC~elastic~scatters} \\
0\/k &~& {\rm \bar \nu_\mu~neutrino-electron~CC~quasielastic~scatters}~(IMD)
\end{eqnarray*}

The above were run for $\nu_\mu$ and $\bar \nu_\mu$ beams.  The
relative ratios of event-weighted contents in neutrino mode are:
$\nu_\mu$ -- 98.33\%, $\bar \nu_\mu$ -- 0.08\%, $\nu_e$ -- 1.56\%,
$\bar \nu_e$ 0.03\%.  The relative ratios of event-weighted contents
in antineutrino mode are: $\nu_\mu$ -- 0.42\%, $\bar \nu_\mu$ --
98.07\%, $\nu_e$ -- 0.26\%, $\bar \nu_e$ 1.26\%.

%% file: precisionflux_v3A.tex
\subsection{Precision Measurement of the Flux from Events in the Detector}
\label{subprecisionflux}

Precise knowledge of the neutrino flux is key to many of the physics
goals of the experiment.  The goal, which is ambitious, will be
to measure the neutrino flux as a function of energy 
to a precision better than 0.5\%.  This
goal is a design-driver for the experiment.   In this section, we
outline an analysis plan to achieve this goal using the event
types described in the previous section.

The flux will be determined through the following steps:
\begin{enumerate}
\item The 
inverse muon decay (IMD) events $(\nu_\mu + e^- \rightarrow \mu^- +
\nu_e)$ are, in principle, 
ideal for measuring the total flux because the IMD cross section is well
known in the Standard Model.   Therefore, these events will be used to 
determine the normalization of the flux.
An important background to this measurement, however, comes from the CCQE events
($\nu_\mu + n \rightarrow \mu + p$), 
which must be subtracted.    In this step, the predicted number
of CCQE events is based on external cross section measurements.
The error on the external cross section is likely to be the 
limiting systematic on the normalization determined in this step.
\item The shape of the flux is measured using the 
traditional ``fixed $\nu$'' measurement method, which was applied in
CCFR \cite{hep-ex/000904,xsec2}. and NuTeV \cite{nutev}, and is currently being
used for in the Minos Experiment \cite{DonnaTalk} to measure the shape
of both the neutrino and antineutrino flux.   The flux shape is 
then normalized by the IMD events from step 1 to obtain the initial
flux prediction.
\item The initial flux prediction is used to determine a more precise
CCQE cross section based on the NuSOnG data. 
\item Step 1 is repeated using the more precise cross section determined
in step 3.  This produces the final normalization which is used to
scale the results of step 2, yielding the final flux.
\end{enumerate}
When the analysis is performed, it may be more effective and efficient
to combine the above steps into a single multiparameter fit to the 
IMD and CCQE data, constrained by the external cross section information.
However, for transparency we will consider the stepwise approach below.

Reaching the goal of $\lesssim 0.5\%$ systematic error depends mainly
on the systematics of the IMD total event rate measurement.  To set
the scale of the problem, the best measurement of IMD events to date,
from CHARM II, had a systematic error of 3\% \cite{CharmIIIMD}.  Thus,
we must achieve an order of magnitude improvement in the IMD total
systematic error.  While we present a well-grounded back-of-the-envelope
argument below, this level of measurement has yet to be demonstrated
by simulation.  That is a priority for future work on the development
of NuSOnG.

Two useful cross checks of the measured NuSOnG  flux are possible.
First, one can extract an energy binned neutrino flux from the
IMD events in step 1.  Because of angular resolution, this flux may
have substantial smearing, but it can be used as a compelling
cross-check of the flux shape derived in step 2.  Second, the neutrino
to antineutrino flux ratio can be compared to the first principles
prediction based on secondary production measurements.

\subsubsection{Step 1:  The IMD Measurement for Normalization}

\label{imdnorm}

NuSOnG expects to observe $>200$k IMD events during
neutrino running.  The high statistics is a consequence of both the
high neutrino flux and high neutrino energy.  High energy is required
because the threshold for IMD scattering is $E_\nu \ge E_\mu \ge
{{m_\mu^2}\over{2 m_e}} =10.9$ GeV.  The SSQT beam design for NuSOnG
produces minimal flux below 30 GeV, well within the 
range of IMD production.  This indicates that there will be high statistics
for IMD events in all flux bins.

\begin{figure}
\centering
\scalebox{0.5}{\includegraphics{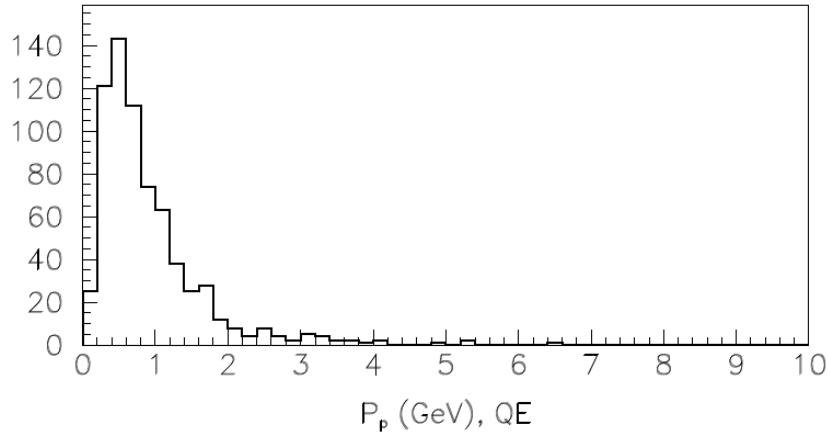}}
\vspace{-0.5in}
\caption{Momentum distribution of protons in $\nu$ CCQE events, from
the Nuance Event generator.}
\label{CCQEprot}
\end{figure}

These events will be used for total flux normalization, with the shape
determined using the Fixed $\nu$ method described in step 2.  This is done
because, while these events can in principle be fully reconstructed
assuming that the incoming neutrino enters parallel to the z-axis, 
the reconstruction in practice suffers substantial smearing.  At
$\sim$100 GeV, IMD events will have scattering angles of $\lesssim$ 1
mrad.  This is similar in magnitude to the expected divergence of the
beam, which was 0.62 mrad in NuTeV.   Angular resolution errors
are expected to be at a similar level.

IMD events must be separated from background, mainly due to
CCQE-like interactions (which include both real CCQE interactions and
single $\pi$ events where the pion was absorbed in the nucleus, and
thus are effectively CCQE events).  IMD events are qualitatively
different from CCQE-like ones in two ways: there is no hadronic energy in the
event, and there is a strict kinematic limit on the transverse momentum
of the outgoing muon, $p_T \le 2 m_e E_\mu$.  It is therefore crucial to
design NuSOnG for observation of very low hadronic energy in the
presence of a muon track, and for excellent angular resolution on the
outgoing muon.  The fine segmentation of NuSOnG should allow hadron
identification in the presence of a muon to substantially lower energy
($\sim$0.75 GeV) compared to 1.5 GeV for Charm II \cite{CharmIIIMD}
and 3 GeV for NuTeV\cite{Formaggio:2001jz}.  One can see from the momentum
distribution of the protons produced in CCQE interactions, shown in
Fig.~\ref{CCQEprot}, that this approach will allow NuSOnG to cut more of the
CCQE background than was possible in the previous
experiments.  NuSOnG also expects better IMD resolution than Charm II,
due to the finer segmentation, which will reduce backgrounds.

Events which produce very low energy pions can also produce a
background, although at a lower level than the CCQE background.
NuSOnG's open trigger will allow many of the low multiplicity DIS
events and CC$\pi^+$ events to be identified and cut due to the
presence of subsequent michel electrons which come from the $\pi^+
\rightarrow \mu^+ \rightarrow e^+$ decay chain.  A 50 MeV michel
electron will traverse 12 cm of glass, producing hits in up to four
chambers in the vicinity of the interaction vertex.

The IMD method for determining the flux proceeds in the following
manner. After cutting on hadronic energy, minimum energy for the
outgoing muon, and no michel electrons near the vertex, the plot of
muon $p_T$ will show a sharp peak at $p_T \sim 0$ superimposed on a
broad continuum of background events extending to high $p_T$.  The
continuum is fit and extrapolated under the IMD peak, to extract the
number of IMD events.  This is divided by the theoretical cross
section to yield the flux.

At the high energies of NuSOnG, the only nuclear effect expected for
CCQE events comes from the Pauli exclusion effect.  This produces an
overall suppression of the cross section across all energies.  Both the
NuTeV and Charm II measurements suffered from the lack of availability
of precise information on the Pauli exclusion effect.  This resulted
in an error on the Charm II measurement from the CCQE model of 2.1\%.

NuSOnG will be in the fortunate position that a number of new
measurements of the CCQE cross section on nuclear targets will be
available as inputs into the CCQE model.  Results from MiniBooNE
\cite{MBCCQE} and SciBooNE \cite{SBprop} will address Pauli
suppression in CCQE interactions on carbon.  Minerva \cite{Minprop} is
studying a series of nuclear targets, and are willing to consider
running a glass target for NuSOnG, if we were to supply the target
panels.  The precision on the CCQE cross section in the NuSOnG era may
be 5\%, which is $\sim$5 times better than the CHARM II era
measurements.  Thus, at this step, the CCQE model error for NuSOnG may
be as low as 0.4\%.

A CCQE model error which was not addressed in the Charm II analysis
was the long-standing discrepancy between models and data at low $Q^2$
\cite{MBCCQE}.    Low $Q^2$ events having small scattering 
angles represents a  significant error on the extrapolation
under the IMD peak.   This discrepancy has recently been resolved
by MiniBooNE under a dipole form-factor model \cite{MBCCQE}.  Minerva
plans to address the $Q^2$ dependence of the form factor in a 
model-independent way \cite{BuddBodek}.     We will assume that the
discrepancy will be fully addressed by the time of the NuSOnG run.

Another 1.5\% systematic error in Charm II came from the model of the
other sources of low hadronic energy CC events, which are dominated by
$\Delta$ resonant production.  As described above, NuSOnG expects a
substantially lower contamination from these sources because of the
lower energy threshold and the michel electron veto.  For those
background events which are not cut, the modeling of these sources
will substantially improved using Minerva data.  In Minerva, the
tracks from CC$\pi^+$ events are well reconstructed, so models $p_T$
distribution of outgoing muons can be tuned.  Similarly, Minerva
offers the opportunity to accurately parametrize CCpi$^0$ events.
Again, the total error on the low hadron multiplicity events in 2015
is expected to be on the order of 5\%, so the modeling of these
backgrounds should not be a limiting systematic.

IMD events can be cut from the sample due to electromagnetic radiation
of the muon near the vertex region which is mistakenly identified as
hadronic energy at the vertex.  The NuTeV IMD analysis
\cite{Formaggio:2001jz} assigned a 1\% systematic error due to
radiative effects.  We address this error in two ways.  First, in the
NuTeV experiment, the photons immediately converted in the 10 cm iron
plates, while in NuSOnG, the 2.5 cm glass plates are 0.25$\lambda_0$,
giving photons a 50\% probability of traversing three plates before
showering.  Fewer IMD radiative events will therefore be misidentified as
events with hadronic energy at the vertex.  Second, because of the
higher segmentation, NuSOnG will employ an improved model of
electromagnetic showers, reducing the systematic error.

\subsubsection{Step 2: The Fixed-$\nu$ Measurement to Determine the Shape}
\label{subsubfixednu}

The central premise of the Fixed $\nu$ method for measuring the flux
is that, for small hadronic energy exchange ($\nu$), the differential
cross section is independent of energy to a good approximation.  The
Fixed-$\nu$ method utilizes this fact to measure the relative flux
between energy bins and the relative flux between neutrino and
antineutrino interactions.  External input is then needed to determine
the overall normalization.

To motivate the premise, consider the differential cross section at a
fixed $\nu$ integrated over all $x$:
\begin{equation}
\label{fixednuxsec}
{{d\sigma}\over{d\nu}} = A(1+{B\over A}{\nu \over E_\nu} - {C\over A}{\nu^2 \over {2E_\nu^2}}).  
\end{equation}
In this equation,
\begin{eqnarray}
A & = & k \int F_2(x,Q^2) dx, \\
B & = & -k \int [F_2(x,Q^2) \mp xF_3(x,Q^2)] dx, \\
C & = & B-k \int F_2(x,Q^2) \big( {{1+{2Mx}\over{\nu}}\over{1+R(x,Q^2)}} -
{{Mx}\over{\nu}} -1 \big) dx, \\
\end{eqnarray}
where $k= (G_F^2 M)/\pi$, and $\mp$ refers to neutrinos ($-$) or antineutrinos ($+$).
For simplicity, first consider $\nu \rightarrow 0$.  The cross section 
becomes equivalent to $A$, which is a constant. Since it is impossible to measure
scattering for $\nu=0$, consider scattering for $\nu=\nu_0$ where $\nu_0 \ll E_\nu$.
As long as $\nu_0$ is small enough, the terms which depend on $\nu_0/E_\nu$ will have 
negligible contribution.  Thus for a fixed, low value of $\nu$,
$d\sigma/d\nu \rightarrow A$, independent of beam energy.   Note that terms $B$ and $C$ 
differ for neutrinos and antineutrinos.   However, as long as $\nu_0/E_\nu$ is negligible,
these terms do not contribute and the cross section for antineutrinos is equal to the 
cross section for neutrinos.

From this, one can see how to measure the relative fluxes.
If one measures the number of events at a given $\nu_0$ in bins of $E_\nu$,
one can solve for the flux:
\begin{equation}
\Phi(E_\nu) = N(E_\nu, \nu_0)/A \quad .
\end{equation}
The relative change of flux between two energy bins is independent of $A$:
\begin{equation}
\Phi(E_\nu^{bin 1})/\Phi(E_\nu^{bin 2}) = N^{bin 1}(E_\nu, \nu_0)/N(E_\nu^{bin 2}, \nu_0) \quad .
\end{equation}
Since the neutrino and antineutrino cross sections are equal, this method also 
allows the relative fluxes to be extracted, independent of $A$.
\begin{equation}
\Phi(E_\nu)/\Phi(E_{\bar\nu}) = N(E_\nu, \nu_0)/N(E_{\bar\nu}, \nu_0) \quad .
\end{equation}
Thus one can extract the relative bin-to-bin and neutrino-to-antineutrino fluxes 
strictly from the data, with no theoretical 
input on the value of $A$.    

In practice one uses a low $\nu$ region, defined by $\nu<\nu_0$ where $\nu_0$ is some appropriate 
upper limit.  CCFR and NuTeV, used $\nu<\nu_0=20$ GeV, which allowed high statistical precision
for the measurement.   From the theoretical point of view, however,
this was not optimal since the goal was to measure the flux down to $E_\nu=30$ GeV, thus at 
$\nu = 20$ GeV, the $\nu/E_\nu$ terms were not negligible.
The flux is then given by:
\begin{equation}
\Phi(E_\nu) = \int_0^{\nu_0} {{{dN}\over{d\nu}} \over {1+{B\over A}{\nu \over E_\nu} - {C\over A}{\nu^2 \over {2E_\nu^2}}}} d\nu \quad .
\end{equation}
A fit to $dN/d\nu$ determines $B/A$ and $C/A$.   
One can test the quality of the bin-to-bin result by fitting $\sigma/E$ to a line.
A good fit results in small slope, due to QCD effects on the order of a few percent (somewhat smaller in antineutrino mode), with small error.  
NuTeV found values consistent with expectation \cite{nutev}:
\begin{eqnarray}
{{\Delta({{\sigma^\nu}\over{E}})\over{\Delta E}}}& = & (-2.2 \pm 0.8)\%/100{\rm GeV}, \\
{{\Delta ({{\sigma^{\bar \nu}}\over{E}})\over{\Delta E}}}& = & (-0.2 \pm 0.8)\%/100{\rm GeV}. 
\end{eqnarray}
The NuTeV analysis indicated a good fit to a straight line,
as expected.
The extracted shape of the flux was obtained to very high precision across the full
energy range by this approach.

NuSOnG has an important advantage over NuTeV when implementing this
method, in that the high statistics and good segmentation will all us
to reduce this range of the low $\nu$ substantially, perhaps to as low
as $\nu<\nu_0=10$ GeV.  This should allow an even more precise measure
of the shape than was obtained by past experiments, since the
contribution of the fit to the $B$ and $C$ terms will be reduced.  In
particular, the systematic error contribution from the charm mass will
be substantially reduced.  

NuTeV also required $\nu>5$ GeV to cut the
resonance region.   NuSOnG is also likely to introduce such a cut.
However, this should be revisited in light of the expected 
new data from Minerva in the resonance region.

The most important detector systematic to this measurement is likely to be the
muon energy scale.  NuTeV achieved knowledge of the muon energy
scale to 0.7\%, although the absolute
calibration beam was known to 0.3\%.   The difficulty was mapping 
across the full area of the toroids.    For NuSOnG to achieve its 
goal of measuring the flux with $\lesssim 0.5\%$ total error, the
muon energy scale will need to be known to about 0.25\%.    Careful
thought must be put in to understand how to achieve this.

In past experiments, the next step was to obtain the absolute flux by
normalizing to the world's total, which is $\sigma/E_\nu = 0.667 \pm
0.014 \times 10^-{38} cm^2/{\rm GeV}$.  It necessarily introduces a
2\% normalization error into this method.  NuSOnG will use the IMD
events to perform the absolute normalization, rather than relying on
the world average neutrino cross section measurement.  This is done by
scaling the total flux measured in neutrino mode with the Fixed $\nu$
method to $\sum_i N^{IMD}(E_i)\int \sigma^{IMD} dE$.  At the end of
this step, the predicted flux is expected to be known to $\sim 1\%$.

\subsubsection{Step 3: A Precise Measurement of the CCQE Cross Section }
\label{subsubccqe}

At this point in the procedure, the limiting systematic is likely to
be the CCQE cross section model error in the IMD normalization.  In
this step, this cross section is further constrained using the CCQE
data in NuSOnG.

The background to the CCQE cross section analysis will be the low
hadronic energy events.  These can be reduced using the michel veto
method discussed in Step 1.  Beyond this, because CCQE scatters extend
to higher angles, excess hits due to the presence of charged pions and
photons from $\pi^0$ decay should be more easily resolved from the
photon track.  NuSOnG expects $\sim 500$k CCQE events, and thus
stringent cuts can be applied to remove backgrounds without
substantial statistical error, assuming the efficiency of the cuts can
be well-understood.

The goal will be for NuSOnG to measure the CCQE cross section to
$\lesssim$2\%.  This would be a very valuable measurement in its own
right, as well as allowing for improvement in the flux extraction in
the following steps.  This result can be used to constrain the
normalization for a glass-target measurement in Minerva.  Ratios to
the other nuclear target cross section measurements by Minerva then
allow precisely determined measurements at low E across a wide range
of nuclei.  This will be useful input to future precision neutrino
oscillation measurements.

\subsubsection{Step 4: The Final NuSOnG Flux}

Once the CCQE cross section has been determined at the $\lesssim$2\%
level, one can iterate the IMD analysis of step 2 and then renormalize
the distributions in step 3.  The resulting flux is expected to have
errors of $\lesssim 0.5$\%.

\subsubsection{Cross Checks}

Two useful cross checks of the flux are possible.  The first takes the
measured flux and compares it to the IMD event rate in energy bins.
The second uses external data to cross check the shape and normalization
of the antineutrino flux.

The first cross check compares the shape of the neutrino flux
determined at step 1 to the shape determined through step 2.  This
will be done by running the final flux through the MC and using it to
predict the IMD rate in energy bins.  We will then extract the
predicted flux in energy bins to compared to the measurement performed
in step 1.  This provides a powerful consistency check.

We can also cross check the fluxes obtained by the above method using a
first-principles prediction based on external secondary production
measurements.  The absolute predictions in neutrino and antineutrino
mode are unlikely to be an effective cross check because of large
errors in the secondary production predictions, which vary from 5 to
10\%.  However, the prediction of the ratio of the neutrino to
antineutrino fluxes may be possible to high precision.   This requires
some investigation.

Reference \cite{Sacha} provides a compendium of secondary production
experiments in Table 3.  None extend up to 800 GeV.  The most relevant
experiment was NA56/SPY at 450 GeV, which took data on beryllium
targets\cite{SPY}.  This experiment published yields of
$\pi^+$,$\pi^-$,$K^+$ and $K^-$ with errors on each measurement of
$\sim 5\%$.  However, because many of the systematics cancel in ratio,
the $\pi^-/pi^+$, $K^-/K^+$ and $\pi/K$ ratios are each determined to
$\sim 2.5\%$.  This data should allow a good cross check of the
individual $\pi$ and $K$ shape contributions.  We may choose to run
for a short period at 450 GeV in order to have an exact
cross-comparison.

\subsubsection{The Electron Neutrino Flux}
\label{eflux}

We will begin by tuning the NuSOnG Beam Monte Carlo using the recent
secondary meson production measurements described above.  The new $K$
production results will improve the first principles prediction for
electron neutrinos beyond those of NuTeV.  The electron neutrino
contamination then can be further constrained through the precision
measurement of the $\nu_\mu$ flux, which can be tied to the $\nu_e$
flux, and through the measurement of $\nu_e$ CCQE events.

Once the muon neutrino flux is measured to high precision, it can be
used to constrain the electron neutrino flux.  This is because the
$\nu_e$ ($\bar \nu_e$) background is largely due to $K^+$ ($K^-$)
decays in neutrino (antineutrino mode).  Using the measured $\nu_\mu$
peak from $K^+$ events, the beam Monte Carlo can be precisely tuned.
Having measured the CCQE cross section precisely in the process of
determining the $\nu_\mu$ flux, this result can then be applied to
$\nu_e$ CCQE events to cross check the $\nu_e$ flux prediction.

%% file: ProtondeliveryMJSA.tex
\subsection{Proton Delivery to NuSOnG}

Our goal is to obtain $2 \times 10^{20}$ protons on target during
a 5-year run.   This section outlines
how we might achieve this goal.

Proton delivery occurs via the following lines:
\begin{itemize}
\item The Linac
\item The Booster
\item The Main Injector
\item The Tevatron 
\item Extraction to targeting
\end{itemize}

The existing Linac and the Booster should perform to the
level needed by NuSOnG without problems.
The Booster fills the MI in
batches of $5\times 10^{12}$ protons and will operate between 9 and 15
Hz by 2015.  The Proton Plan projects $7\times 10^{13}$ protons in
each MI fill by 2010 \cite{Protplan}.  Two pulses from the MI are used 
to fill the Tevatron.  In principle, therefore, it is conceivable that the 
Tevatron could receive nearly $1.5\times 10^{14}$ protons per fill under
this scenario. 

Let's suppose that with care the Tevatron can accelerate $8\times 10^{13}$ 
ppp to 800 GeV using two pulses from the Main Injector at $4\times 10^{13}$ 
each pulse, similar to today's MI operation.
To date, the highest intensities extracted from the Tevatron in a single
pulse at 800 GeV were around 2.5 to $3\times 10^{13}$.  The limiting
issue was longitudinal instabilities for energies above 600 GeV at
high intensities, as the bunch length shrank.  ``Bunch spreaders''
were used to compensate.  A better method to compensate will be
required for NuSOnG.   However,
advances in rf techniques and technology and in damper systems make
finding a satisfactory solution conceivable.   More detailed study is
needed.

Our proposal is for a Tevatron cycle time of 40 s, with a 1 s 
flattop at 800 GeV.    Since the MI cycle time will be 2.2 s,
and we need two injections, 
our impact on NuMI is  4.4/40 = 11\% of their run time.

If the uptime for the Tevatron is 66\%, then we will receive
$5\times10^5$ cycles per year.  At $8\times 10^{13}$ ppp, this gives
$4\times 10^{19}$ protons per year.  We then achieve our goal in five
years of running.

%% file: SsqtA.tex
\subsubsection{ Target}

Beryllium oxide was the target material in NuTeV and prior high energy
neutrino beamlines \cite{Bernstein, Yu}.  Beryllium is efficient at 
producing secondary mesons,
and BeO has good structural and thermal properties.  The NuTeV target
consisted of two 30 cm long, 2.5 cm diameter segmented rods in a
copper cooling block, mounted on a movable drive that could select
between centering the beam on either of the two targets or no
target. This target was designed to accept up to 1$\times 10^{13}$ protons per
pulse (ppp).  A similar target will be acceptable for NuSOnG, but it may
be a challenge to provide adequate cooling at our design intensity of
$8\times 10^{13}$ protons per cycle. The NuTeV protons were delivered in five 4
msec "pings" separated by 0.5 sec; we intend to have one pulse
of about 200 msec.  This means that our instantaneous heating rate will be
somewhat lower than NuTeV's, but the total number of protons per cycle
is eight times higher.  In NuTeV the beam width was 0.6 mm, which was
significantly smaller than necessary; a wider more diffuse beam would
help relieve the localized heating problem.  Careful design of the
target support and cooling system will be a necessity.

\subsubsection{ SSQT}

A Sign Selecting Quadrupole Train (SSQT) can be used to provide beams
of either neutrinos or antineutrinos with very low contamination from
either wrong-sign muon neutrinos or electron neutrinos from neutral
kaons.  The NuTeV SSQT utilized two dipoles and six quadrupoles, with
two dumps \cite{Bernstein, Yu}.  The first dipole provided a 6.1 mrad bend for 250 GeV
daughter mesons of the selected sign. In antineutrino mode the
unreacted protons are bent in the opposite direction and are absorbed
in the first dump.  In neutrino mode the protons are absorbed in the
second dump. The first two quadrupoles capture the secondary beam.  A
second dipole then bends the beam by another 1.6 mrad, enhancing the
sign separation and sweeping out low energy particles produced by
scraping in upstream magnets.  Neutral particles are not bent and
therefore travel away from the detector.  NuSOnG will use a similar
SSQT. The only challenge will be designing the proton dumps for our
significantly higher intensity.  In the NuTeV upstream dump in
antineutrino mode, the dump temperature approached 100 C at $1.3\times10^{13}$
protons per pulse; the temperature limit was 110 C. The NuSOnG dumps
will need to be water-cooled.

\subsubsection{Monitoring}

Primary beam monitoring in NuTeV was accomplished with four beam
position monitors (BPMs), four vacuum segmented wire ionization
chambers (SWICs), four secondary emission electron detectors
(SEEDs), a beam current toroid, and a thin foil secondary emission
monitor (SEM) \cite{Bernstein, Yu}. The toroid, SEM, SEEDs and BPMs measured proton
intensity; the BPMs, SWICs, and SEEDs monitored position. It was
found that the SEM degraded over the course of the run, so the beam
toroid was used as the primary measure of intensity. The BPMs and
SEEDs gave closely correllated position measurements, and the SWICs
and SEEDs gave beam profiles that agreed well except in the tails;
the SEED tails dropped more rapidly than those from the SWICs.  With
the exception of the SEMs, which would suffer even more radiation
damage at our higher intensities, a combination of any of these
monitoring devices could be used by NuSOnG.

%% file: NusOnGDetector_v2A.tex
This section details our first ideas about the detector configuration; these are 
summarized in Tab.~\ref{ta:det}.
In thinking about NuSOnG, we have drawn on previous large, high energy
neutrino detectors whose characteristics are summarized in
Table~\ref{ta:comparison}.  NuSOnG represents a natural evolution of
these designs and we believe this makes construction low
risk.  Of particular note regarding Table~\ref{ta:comparison} is the
excellent performance achieved by CHARM II using digital proportional
tubes, a glass target, and fine granularity.

The primary event signatures NuSOnG will need to identify are: 
\begin{itemize}
\item charged current deep inelastic scattering, characterized by a
  hadronic shower and a high energy muon
\item neutral current deep inelastic scattering, characterized by a
  hadronic shower
\item inverse muon decay, $\nu_{\mu}+e^- \rightarrow
  \mu^-+\overline{\nu}_e$, which is characterized by a high energy
  muon accompanied by no hadronic activity.
\item neutrino and antineutrino electron scattering, characterized by
  an electromagnetic shower with no hadronic activity
\item stopped muon decay, which results in an electromagnetic shower
  with energy up to 50 MeV.  These events will be used to reject low
  hadronic energy events which are tagged through the $\pi \rightarrow
  \mu \rightarrow e$ decay chain (see sec.~\ref{imdnorm}).
\end{itemize}

In order to achieve the rates and carry out the measurements given in
Section~\ref{se:rates}, NuSOnG consists of a 3.5 kton (3 kton fiducial
volume) isoscalar target with high segmentation resulting in good
separation between electromagnetic and hadronic showers and muon tracks
with good energy resolution for each.  Good separation between hadronic
and electromagnetic showers and good muon identification are necessary
for separation of neutral and charged current events, and for low
systematic errors on the measurements of the neutrino and
antineutrino electron scattering cross sections.  Finally, good muon
identification is critical for detecting inverse muon decay events for
a precise flux measurement.

\begin{table}[t]
\centering
\begin{tabular}{l r}
Parameter & Value \\
\hline
Total target mass & 3.492 \\
Fiducial mass & 2.975 kt \\
Total length & 192 m \\
Number of glass planes & 2500 \\
Number of toroid washers & 96 \\
Number of muon detector wire planes & 60 \\
(two coordinates each) & \\
\end{tabular}
\caption{Summary of NuSOnG detector parameters.}
\label{ta:det}
\end{table}

\begin{table}[t]

\centering
\begin{tabular}{l c c c c c c c}
 &  \multicolumn{3}{c}{Resolution} & Sampling & Absorber \\
 &   EM & Hadronic & Muon & & \\
 &    ($\sigma_{E}/E$) & ($\sigma_{E}/E$) & ($\sigma_p/p$)  & & \\
 \hline
 FMMF   & 1.04/$\sqrt{E}$ & 0.72/$\sqrt{E}$ & 8\% & 0.11 X$_o$ & sand/shot \\
(Flash tubes,       &                                    &                             &         &     &              \\
digital)                  &                                    &                             &         &     &              \\
CDHS                 &   0.80/$\sqrt{E} $& -                            &  5\% & 2.8/8.3 X$_o$ & steel \\
(Scintillator)       &                                        &                           &         &      &             \\
CHARM II          & 0.52/$\sqrt{E}$+0.02  & 0.24/$\sqrt{E}$+ 0.34 & 5\% & 0.5 X$_o$ & glass \\
(Prop. tubes,     &                                         &                           &         &       &            \\
digital)                &                                         &                           &         &       &            \\
NuTeV               &  0.86/$\sqrt{E}$+0.022 & 0.5/$\sqrt{E}$ +0.042 & 10\% & 5.8 X$_o$ & steel \\
(Scintillator)     &                                          &                            &         &        &            \\
\hline

\end{tabular}
\caption{Comparison of high energy neutrino detectors.}
\label{ta:comparison}
\end{table}

Given the large size of the detector, ease of construction and low
cost technologies are important.  The long running time requires high
stability and robust operation.

\begin{figure}
\centering
\vspace{1.5cm}
\includegraphics[width=6.in]{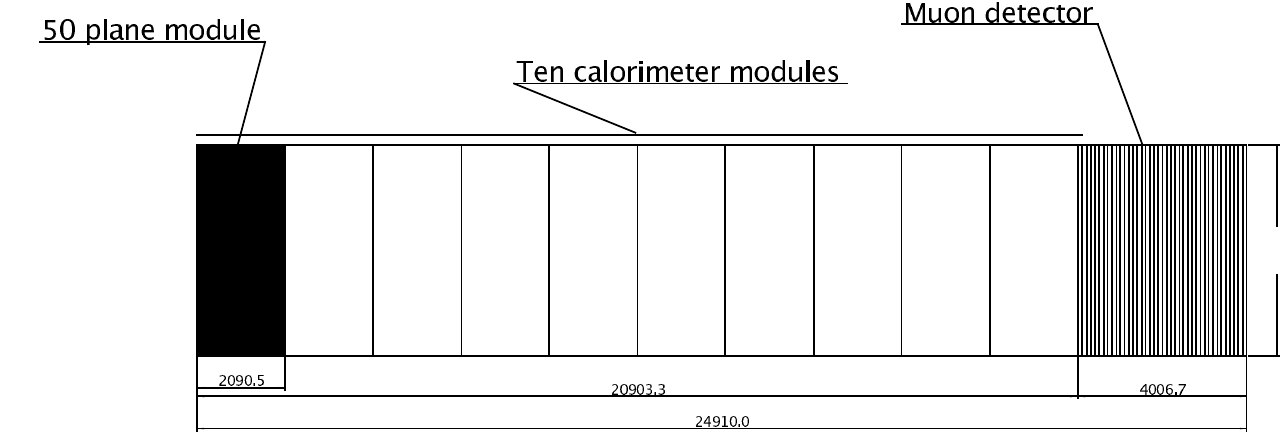}
\vspace{0.5cm}
\caption{NuSOnG detector showing calorimeter modules and muon detector.}
\label{fi:module}
\end{figure}

\begin{figure}
\centering
\vspace{0.5cm}
\includegraphics[width=6.in]{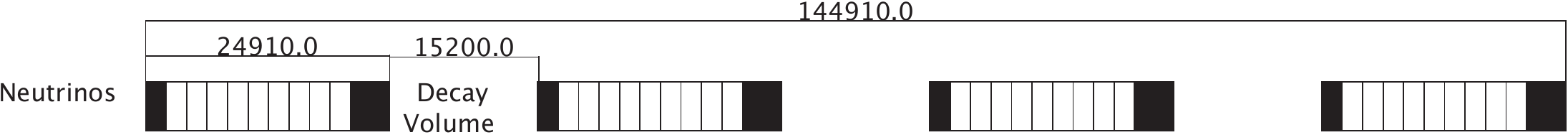}
\vspace{0.5cm}
\caption{The full NuSOnG experiment showing four detectors separated by a decay volume.}
\label{fi:full}
\end{figure}

Our first design is shown in Figs.~\ref{fi:module} and \ref{fi:full}
and summarized in Table~\ref{ta:det}.  NuSOnG consists of four
calorimeters each with a muon spectrometer.  15 m decay volumes
separate the four detector elements.  Interspersing the decay volumes
between the detectors will allow a calibration beam to be brought to
each of the four detector regions.

\begin{figure}
\centering
\vspace{0.5cm}
\includegraphics[width=6.in]{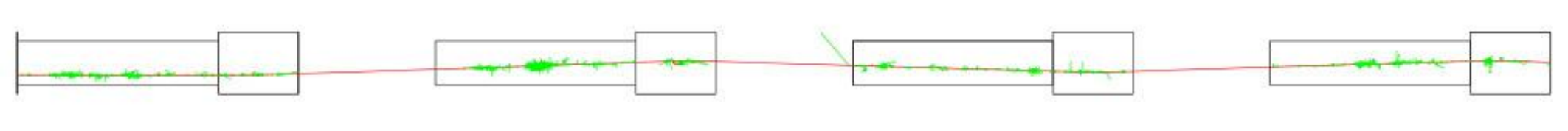}
\vspace{0.0cm}
\caption{A 100 GeV muon traversing the detector from the NuSOnG initial
GEANT4 Monte Carlo.}
\label{GoLittleMuonGo}
\end{figure}

Each calorimeter has 500 SiO$_2$ 2.5 cm (X$_o$/4) glass target planes
interleaved with active detectors with two dimensional readout.  The
active detectors could be proportional tubes, scintillator panels, or a
combination of both.  These three options are discussed below.
Neutrinos interact in the target planes, creating secondary particles;
the active detector determines the total energies of the hadronic
and electromagnetic secondaries.  The muon detector measures the
momentum of muon secondaries and serves to identify them.  The
pattern of the shower serves to identify the shower type: showers in
which all the energy resides in ten of fifteen planes will be
electromagnetic, and more extended showers will be hadronic.  The
lateral extent of the shower also resolves electromagnetic from
hadronic showers.

We have chosen an SiO$_2$ target.  This material provides a balance
between longer radiation length, important to particle ID issues, and
shorter detector length, important for acceptance and calibration
issues.  The target could be commercial glass or thin walled plastic
boxes filled with sand.  Glass planes have the advantage of being easy
to install and require no construction.  Sand-filled boxes could be
much less expensive.  We will investigate both possibilities.  Either
way, SiO$_2$ has the advantage of being isoscalar ($\langle
N_u\rangle>/\langle N_d\rangle$=0.998). SiO$_2$ has a density of 2.2
g/cm$^3$; a high energy muon will lose 10 MeV per plane, which gives 5
GeV across all 500 planes in one calorimeter.  Energy loss will also
occur through electromagnetic showers.  An example straight-through
muon event from our initial GEANT4 detector simulation is shown in
Fig.~\ref{GoLittleMuonGo}.  A michel electron with 30 MeV energy
should be clearly visible across three planes.  Each calorimeter is
followed by a toroidal muon spectrometer consisting of magnetized iron
plates interleaved with drift chambers.

Other target materials, including emulsion, are under consideration,
as has been discussed in previous sections of this document.  These
materials are not yet incorporated into the preliminary design presented
here, but should be straightforward to include in the future.

The design must address beam correlated backgrounds.  These include
backgrounds arising from debris (muons, remnants of hadronic showers)
from neutrino interactions in the earth surrounding the detector.  We
plan for a forward veto consisting of a three layer scintillator
hodoscope.  Since our detector is so long, we may also need a veto
system along the sides, top, and bottom of the calorimeters.  We plan a
Monte Carlo study of veto requirements in the coming months.  Cosmic
rays muons and their attendant showers present a beam-uncorrelated
background which we will need to eliminate.  We envisage a counter on
the top of the detector similar to that used by the MINOS experiment.

NuTeV showed the value of continuous beam calibration and this will be
discussed in a separate section.  

While our detector is quite large, the robust, simple design will make
the cost and construction manageable.  The modules design makes
upgrades and improvements straightforward.  While we are designing
with an initial four  to five  year run in mind, this detector can be
put to other uses should the physics warrant.

\subsubsection{Active detector options}

The active detector performs two roles: first, it tracks the particles
emerging from a neutrino interaction;  second, it
samples the particle's energy loss along the trajectory giving an
measurement of the total energy.  Simplicity, robustness, and high
efficiency are essential, as is low cost.

Two technologies immediately present themselves: gas-filled
proportional tubes and plastic scintillator read out by phototubes.
Both have been used in several experiments (see
Table~\ref{ta:comparison}).  At this point, it is not clear to us which
is the best approach for NuSOnG.  We are also considering a design
with both proportional tubes and scintillator.  In the coming months,
we plan to study the performance of each via simulation, develop
preliminary design prototypes, and carry out a detailed cost estimate.
We describe each detector concept below.

\paragraph{Proportional Tubes}

~~~\\
~~~

\begin{figure}
\centering
\includegraphics[width=5.9in]{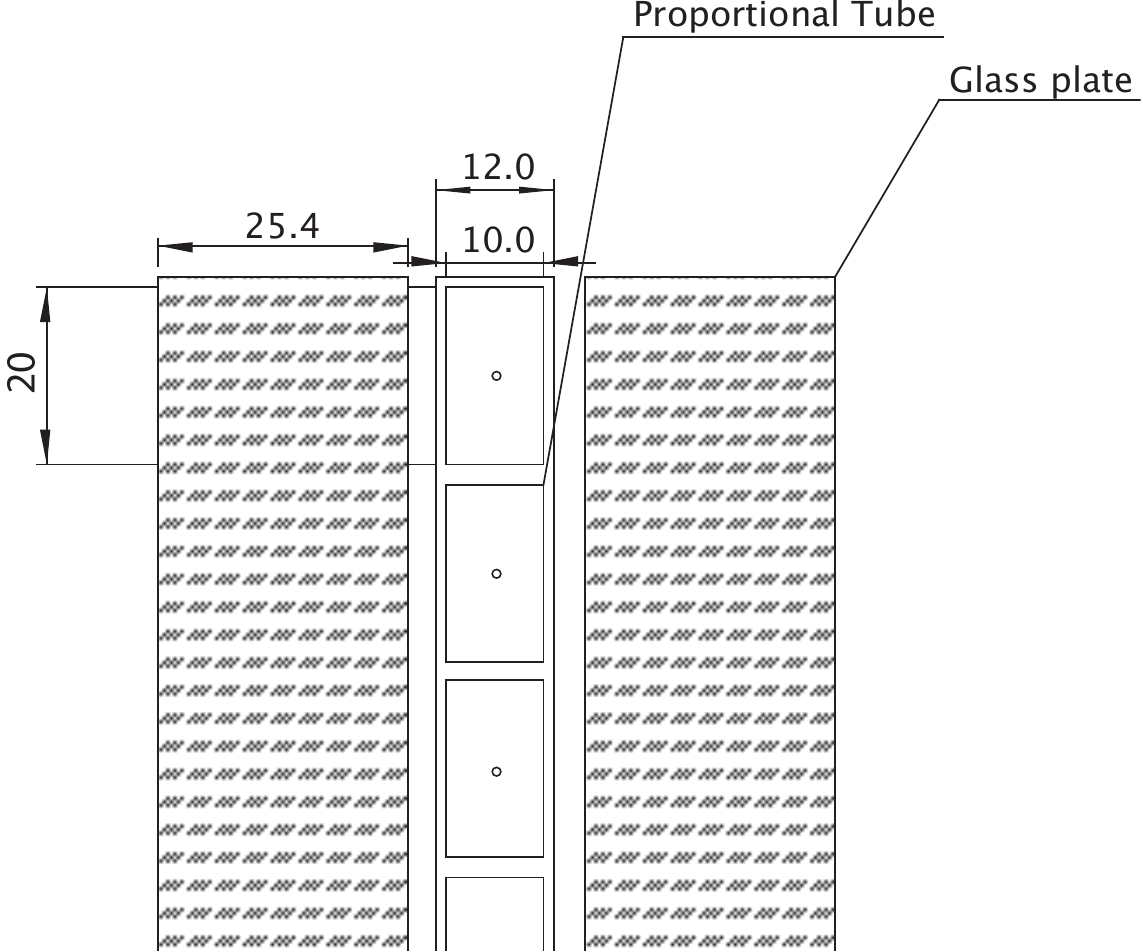}
\vspace{-8pt}
\caption{Proportional tubes design.}
\label{fi:prop}
\end{figure}

A first design for a proportional tube active detector is shown in
Fig.~\ref{fi:prop}.  Each active detector plane is made from five 1 m
$\times$ 5 m extruded aluminum panels.  Each panel contains fifty 1 cm
$\times$ 2 cm drift cells.  A 50 $\mu$m wire is strung down the
center of each tube, and the applied high voltage produces both drift
and proportional amplification fields.  Ar:CO$_2$ (80:20) provides a
good candidate for a fill gas; with 1.8 - 2 kV applied to the wire,
the drift field will give a drift velocity of about 50 $\mu$m/ns and a
gain of 3000.  A minimum ionizing particle crossing the 1 cm cell will
deposit 2.7 keV of energy, liberating about 160 drift electrons in ten
or so clusters.  The drift time across the cell will be about 500 ns
and proportional multiplication will give a collected charge of 80 fC
over a time of 250 ns.

As an example of a readout scheme, we look to the ATLAS Transition
Radiation Tracker (TRT) ASIC chips.  The TRT readout has a peaking
time of 7.5 ns and a charge threshold of 2 fC, making them well
matched to our proportional tubes.  Each chip set reads out sixteen
channels and can be configured to provide trigger information.  The
TRT system, which is based around 6 mm straw tubes, has achieved a
spatial resolution of 127 $\mu$m, albeit with higher energy deposition
resulting from the use of a xenon mixture.  Scaling by the energy deposition
gives a resolution of 200$\mu$m for our argon-filled tubes.  The TRT
readout chip set has sufficient charge sensitivity to allow us to use
charge division; this should give position resolution of 5-10 cm along
the wire.

\paragraph{Scintillating Strips}

~~~\\
~~~

The second option uses planes of scintillator strips read out with
green wavelength-shifting fibers fed into multi-anode photomultipliers.
This option would be similar to that used for the SciBar detector in
K2K, the Minos neutrino detector, and the Opera neutrino detector.
NuSOnG would have 2500 5 m by 5 m planes with each plane made up of
128 3.9 cm $\times$ 1.3 cm strips. Each 64 strip plane will be
separately wrapped in an Al skin that will provide the light seal and
strength for the module.

The scintillator strips will be coextruded with a TiO$_2$ reflective coating 
and have a 1.8 mm diameter hole in the middle.  A 1.5 mm diameter green 
wavelength-shifting fiber will be put in the hole and routed to multianode 
photomultipliers for readout.  The 64 wavelength-shifting fibers on one side 
of a plane will be coupled to a Hamamatsu M64 multianode photomultiplier tube.  
The readout side will alternate between subsequent planes to improve uniformity.  
The fiber end opposite to the tubes will be polished and mirrored to increase the 
light output and uniformity. Planes will alternate between horizontal and vertical 
strips to provide two view tracking; readout tubes will alternate.  

The readout would be based on a custom ASIC combined with a standard FPGA.  
One example is the 64 channel MAROC2 custom integrated circuit, designed at 
LAL (Orsay) for the ATLAS luminosity monitor.  This chip allows adjustment of the 
electronic gain of each of the 64 channels, which will be needed to correct for 
the expected factor of 3 pixel-to-pixel gain variation of the M64 tubes.  The system 
provides a self-triggering analog readout into an external flash ADC.  A fast discriminator 
signal for triggering is also available for each strip with a common threshold.

Based on the performance of the SciBar detector, a minimum ionizing particle 
traversing a strip will yields about 20 photoelectons close to the tube, and 
the strip/fiber system will have an attenuation length of 3.5 m.  This would then produce 
about 10 photoelectrons at the center of the detector per plane.  

\paragraph{Hybrid Design}

~~~\\
~~~

Our initial estimates indicate the scintillator option may cost more
than the proportional tube option.  However, the scintillator system
described above does provide a stable, easy to characterize active
detector.  In particular, scintillating strips offer very stable
response that does not vary with pressure or temperature.  We will
investigate a hybrid system in which every fourth or eighth plane (one
or two radiation lengths) would be a scintillator panel.  The high
granularity of the proportional tube design would give good pattern
recognition, and the excellent energy resolution of the scintillator
would give a better energy measurement.  Reducing the fiber spacing in
the scintillator may be possible; this would reduce the cost.

One issue with adding 12-25\% scintillator would be the change in the
fraction of protons in the detector.  The precise change depends on
the scintillator used, but for CH$_4$ and one scintillator panel every
quarter radiation length, the proton-neutron ratio changes from 0.998
to 0.940.  The impact of this change will have to be balanced against
the cost reduction and stability improvement.  This will be part of
our Monte Carlo effort in the coming months.

%% file: mag_v2A.tex
\subsubsection{\bf Toroid Spectrometers}

High energy muons produced in charged-current interactions will be
momentum analyzed in three iron toroid spectrometers
downstream of each subdetector (set of ten ``stacks'').
Each spectrometer will be composed of layers of magnetized iron
instrumented with drift chambers for tracking.

Since NuSOnG will see muons of the same energies as NuTeV/CCFR a
similar arrangement for measuring muon momenta would be suitable. 
CCFR used sections of 8'' thick steel washers instrumented with scintillator
hodoscopes for calorimeter.\footnote{The CCFR arrangement used 
two C-shaped sections with a horizontal
crack at the center to allow placement of hall probes 
for field calibration. This crack would be eliminated in NuSOnG
and instead small slots could be included for this purpose.}
Tracking was performed
using four views of each x and y chambers (0.5 mm coordinate resolution)
in three gaps located after each 1.6 m of steel. 
The magnetic field was produced by four coils carrying approximately
1500 A each which passed through the center hole. 
The field was nearly radially symmetric and pointed in the azimuthal direction
with magnitude ranging from 1.9 T near the center hole to 1.55 T near the
outer edge (at R=1.8 m). Details can be found in reference \cite{nim:king}.

\begin{figure}[tbp]
\begin{center}
\centerline{\includegraphics[width=4.5in,height=3.0in]{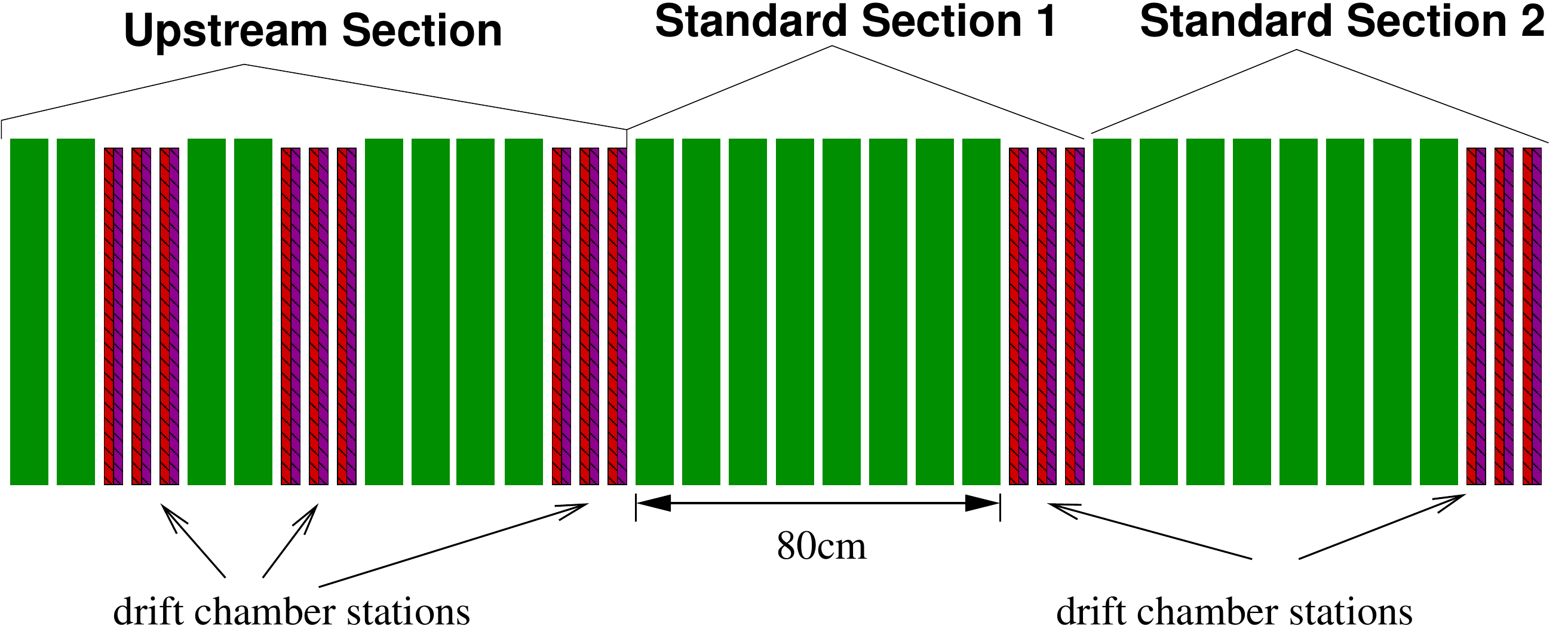}}
\end{center}
\vspace{-.3in} 
\caption[]{Conceptual Schematic for a NuSOnG toroid element. The three sections contain the same amount of steel (eight washers of 8'' each). The upstream most section has additional drift chamber stations to improve acceptance for low energy muons. Each of the five drift chamber stations has 3x and 3y view chambers.}
\label{fig:tor}
\end{figure}

Figure \ref{fig:tor} shows a possible arrangement for a NuSOnG toroid 
spectrometer. One ``Upstream section'' and two downstream 
``Standard sections'' are shown. The downstream sections contain
eight 8'' washers with 
one drift chamber station with 3x and 3y view chambers each.
The most upstream section of a spectrometer unit has two additional 
drift chamber stations 
to improve acceptance for low energy muons. To pass the coil through this
arrangement the upstream chamber stations would be half size (the same 
chambers but rotated for each view). 
Each of the three sections contain the same amount of steel.
 Hodoscope paddles could be added in each chamber
station for triggering purposes.
Resolution of this arrangement would be dominated by multiple 
Coulomb scattering and would be $\sim$ 11\% independent of momentum.

The NuSOnG arrangement will provide good acceptance
for high energy primary muons of both signs since in 
a sign-selected beam the can be routinely operated with 
the polarity set to focus the primary muon.
Very high energy particles can be tracked into the downstream
target sections with a long lever arm and their momentum analyzed.
(resolution for very high energy muons ($>150$GeV was limited
in NuTeV and CCFR; this resulted in large uncertainties in
measuring flux in the high energy tail of the beam).
Improving flux measurements in this region may help constrain
kaon fluxes and therefore electron neutrino beam contamination.

%% file: cal_v2A.tex
\subsubsection{ Detector Calibration }

\label{detcalib}

A thorough and precise calibration of the entire detector will
be required to achieve the physics goals of NuSOnG. Some of the
response features of the detector can be understood using beam 
and cosmic ray muon samples, but a dedicated calibration 
effort will be required to study the hadronic and electromagnetic
response of the detector and to measure the absolute energy scales.
Precise calibration of a detector of this size will require
a dedicated {\it in situ} calibration beam such as was used in NuTeV for
this purpose \cite{ref:nutevnim}.   

The requirements for NuSOnG calibration beam would be similar to
those of NuTeV. Tagged beams of hadrons, electrons, and muons over
a wide energy range (5-200 GeV) would be required. 
The calibration beam should have the ability to be steered over the transverse
face of the detector in order to map the magnetic field of each toroid
with muons. This could be accomplished in several ways; for example,
gaps of a few meters in front of each toroid could be incorporated into the design,
and the beam could be steered into each toroid in turn; or the 
toroids could each be moved into the test beam for these
calibration runs.  
Steering for hadrons and electrons
would be less crucial than it was in NuTeV's case but would still be useful.

The calibration beam can be constructed with a similar design to 
NuTeV.
Upstream elements were used to select hadrons, electrons, or
muons. An enhanced beam of electrons was produced by introducing a
thin lead radiator into the beam and detuning the portion of the beam
downstream of the radiator.  A radiator was also used in the nominal
beam tune to remove electrons.  Particle ID (a threshold cerenkov and
TRDs) was incorporated in the spectrometer and used to tag electrons
when running at low energy.  A pure muon beam was produced by
introducing a 7~m long beryllium filter in the beam as an absorber.

\begin{figure}[tbp]
\begin{center}
\centerline{\includegraphics[height=0.40\textheight]{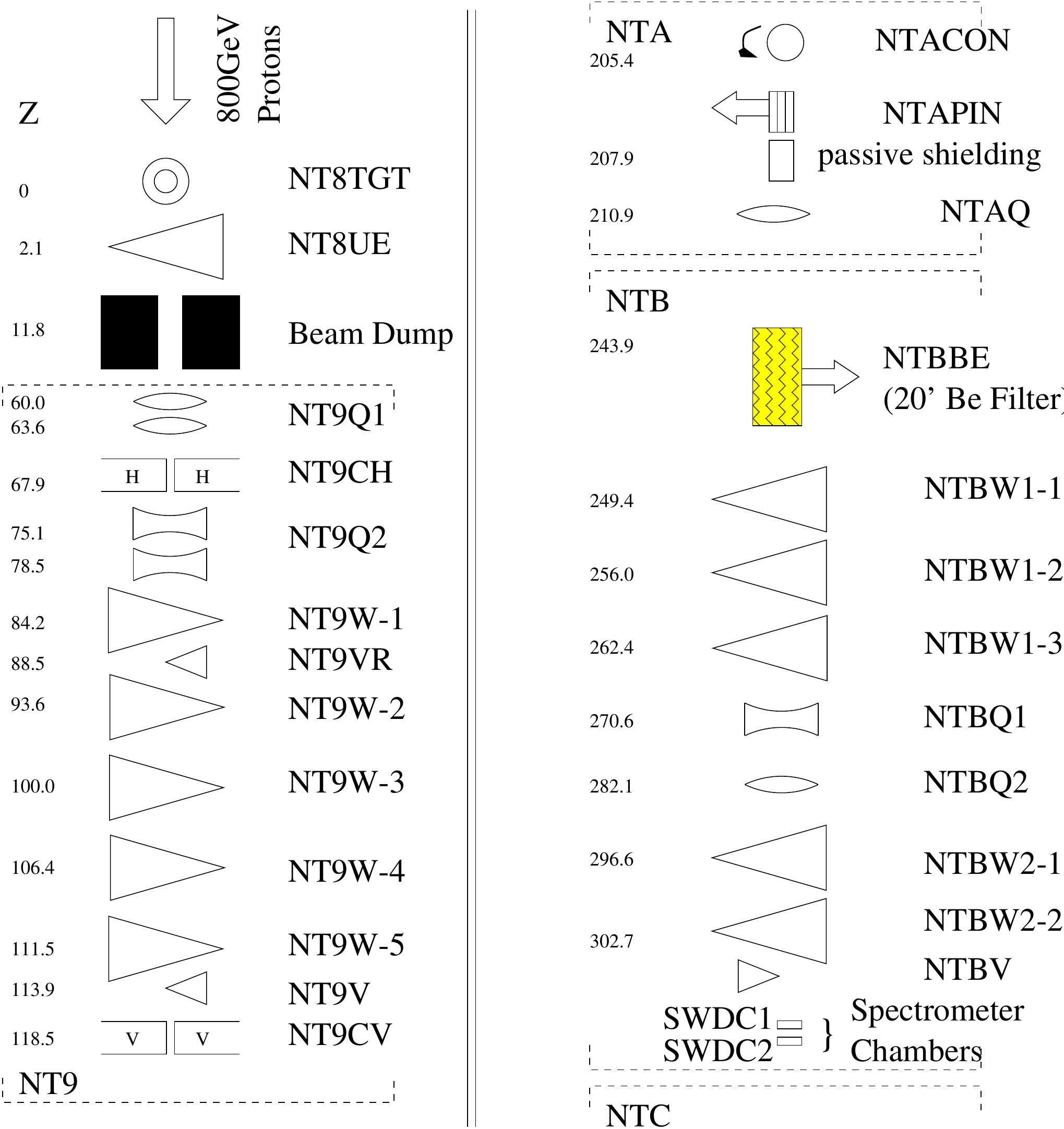}}
\centerline{\includegraphics[height=0.300\textheight]{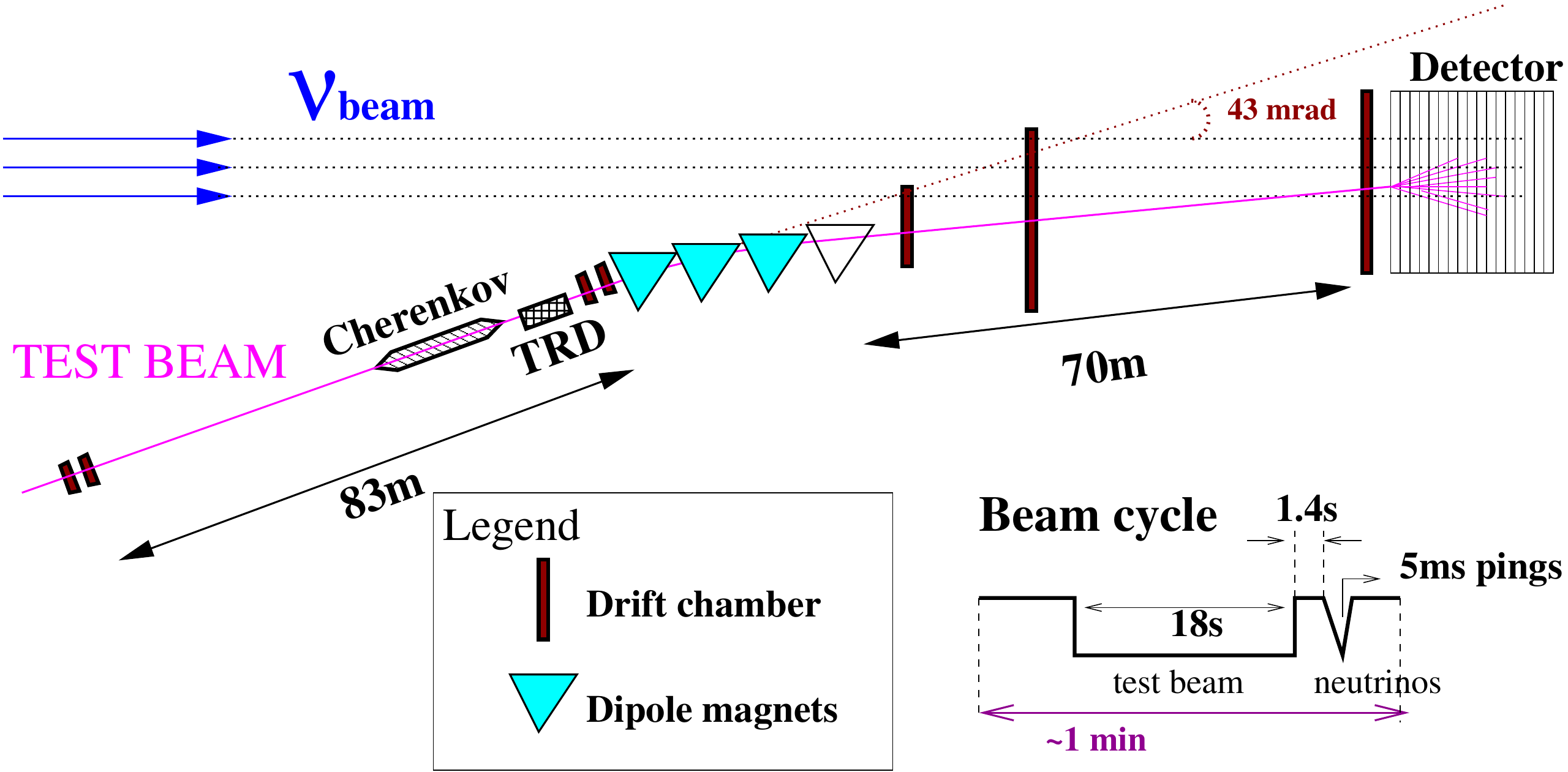}}
\end{center}
\vspace{-.3in} 
\caption[]{(Top) Components of the NTEST beamline used
to calibrate the NuTeV detector. Four 
different thicknesses of converter material at NTACON were used to select 
pure hadrons or electrons.
The 7~m long Be filter(NTBBE) was used to select pure muons.
The numbers on the left-hand-side of each component indicate 
	the relative distance of the component to the primary 
	target (NT8TGT) in meters.
(Bottom) NuTeV's long lever arm spectrometer. The four dipole
bend magnets were located in an enclosure approximately 70~m upstream of
the Lab E detector. The spectrometer spanned over 150~m in length; this
allowed precision measurement of the bending angle.}
\label{fig:tb}
\end{figure}

The NuTeV calibration spectrometer was able to determine incoming particle
momenta with a precision of better than 0.3\% absolute. 
This was accomplished by two means.  First, precisely calibrated dipole 
spectrometer magnets were used, with $\int Bd\ell$ known to better 
than 0.1\% in the region traversed by the beam. Secondly, 
the bend angle was determined to
better than 0.1\% using drift chambers positioned over the 
150~m spectrometer. This long lever arm allowed 
a modest alignment uncertainty of a few mm to translate
into only a 0.1\% uncertainty in the absolute momentum scale. 
The event-by-event resolution of the spectrometer, dominated by
multiple scattering in the drift chamber walls, was better 
than 0.3\% for most energies.
(Helium in the region between the last dipole and the 
upstream part of the detector reduced the scattering in air).

Figure \ref{fig:tb} shows the NuTeV calibration beam configuration and
the long lever arm spectrometer used to tag particle momenta with an
absolute precision of better than 0.3\%.  The most downstream dipole
was mounted on a rotating stand which gave the ability to steer the
beam out of the plane.

The NuSOnG goal of the calibration precision would be to measure 
energy scales to a precision of about 0.5\%.
NuTeV achieved 0.43\% precision on absolute hadronic
energy scale and 0.7\% on absolute muon energy scale (dominated
by the ability to accurately determine the toroid map). 
Precise knowledge of the
muon energy scale is especially important in order to achieve
high measurement accuracy on the neutrino fluxes using the 
low-$\nu$ method. For example a 0.5\% precision
on muon energy scale translates into about a 1\% precision on the flux.
Both energy scales
are important for precision structure function measurements 
and were the largest contributions to structure function measurement
uncertainties in NuTeV \cite{nutev}.

%% file: where_is_itA.tex
\subsection{Possible Locations}

\begin{figure}[tp]
\centering
\includegraphics[width=6.in]{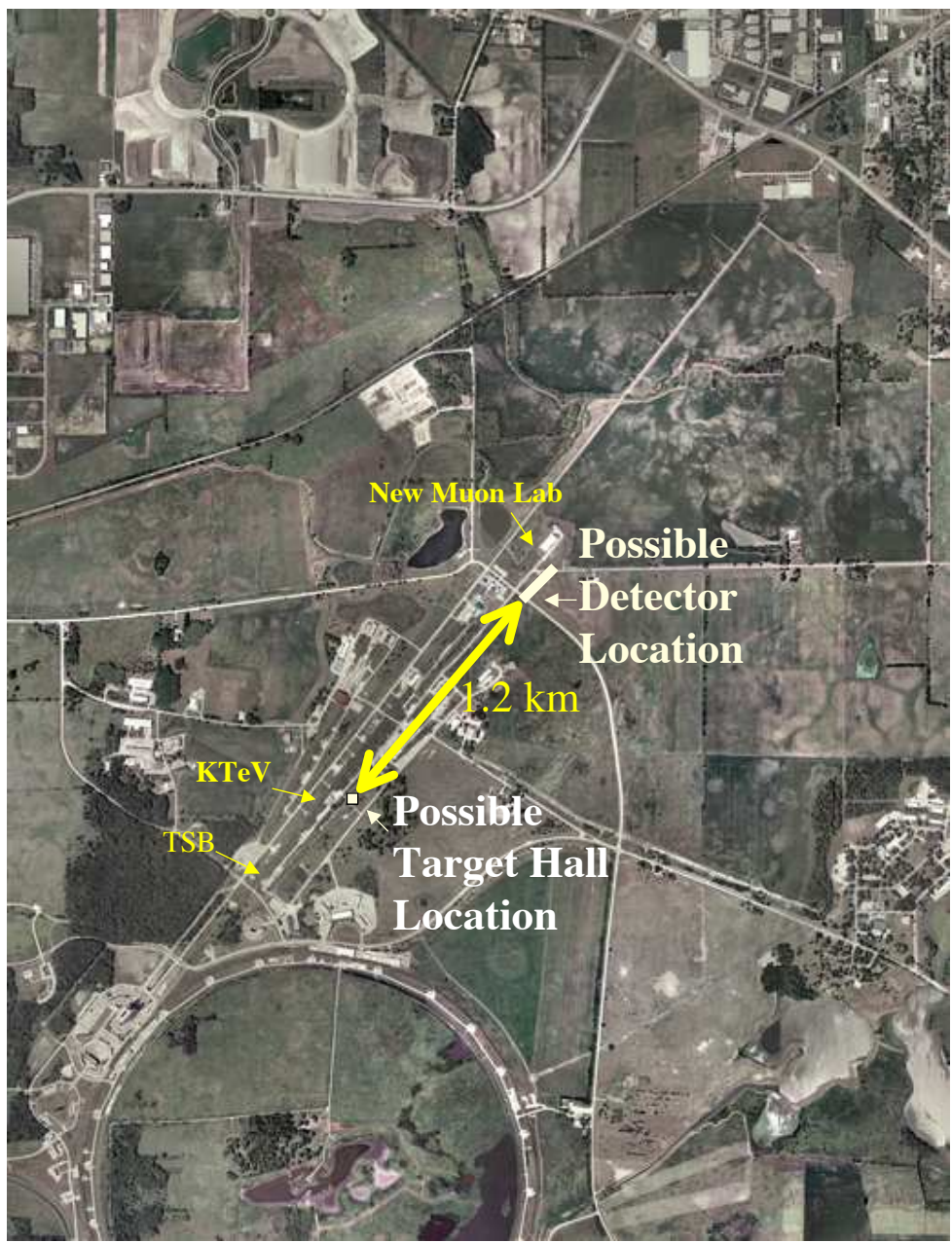}
\caption{Aerial view of Fermilab showing the Tevatron, external beam lines and potential site for NuSOnG target and detector halls.}
\label{fi:site}
\end{figure}

Fig.~\ref{fi:site} shows a possible location for the NuSOnG beam target
hall and detector.  Other layouts are possible; this is just meant
to provide an example.

This layout assumes that the beam is extracted at A0 from the TeVatron
and directed through the Switchyard Complex to a new targeting hall.
This location allows low luminosity extraction down existing beamlines
for the calibration beam.

The detector is located near the New Muon Lab.
This is a region with more than 200 m of clear length, with roads and
utilities nearby.

The calibration beam could be delivered to the NuSOnG hall using a
scheme similar to that used in NuTeV with the NTest beamline. 
Only a short extension of the existing NTest line
would be required to reach a detector located near the New Muon Lab.
The beam
was split off from the same beamline (Ncenter) and then bent around to
impinge on the detector at a 43~mrad angle.

%% file: summaryA.tex
NuSOnG is an experimental program with high discovery potential. The
precision neutrino scattering measurements probe terascale physics and
will complement discoveries at the LHC.  Through precision electroweak
measurements, NuSOnG will be sensitive to such new phenomena as extra
$Z$ bosons with masses beyond the 1 TeV (depending on the model) and
compositeness scales above 5 TeV.  The NuSOnG measurement of the
coupling to the $Z$, when combined with the LEP measurement of the
invisible width, is a more sensitive method to search for new physics
than this same measurement at the ILC.  NuSOnG can also probe the
existence of neutrissimos, moderately heavy neutral heavy leptons which
may be produced at the LHC, but which could be difficult to reconstruct and
identify.  A wide range of direct searches for new particles and
interactions can be accomplished.  The high neutrino flux and
isoscalar target will make allow measurements which probe deeper into
nuclear structure.

The high energy neutrino facility, which uses 800 GeV protons from the
TeVatron, has been endorsed by the 2007 Fermilab Steering Group.
While NuSOnG is the first to propose an experiment for this facility,
a wide range of interesting measurements can be made on this line.

The proposed 3 kton (fiducial) NuSOnG detector design, which is
opitmized for the physics goals, is based largely on the experiences
of NuTeV and CHARM II.  The basic technology is straightforward,
although challenges exist because of the high precision demanded by
the physics goals.  Detailed simulations of the detector are now
underway

Our plan is to develop these ideas over the coming months.  We plan to
submit a proposal to the Fermilab Directorate in the near
future.